\documentclass[applsci,article,accept,moreauthors,pdftex,10pt,a4paper]{Definitions/mdpi}

\setitemize{parsep=6pt,itemsep=0pt,leftmargin=*,labelsep=5.5mm}
\setenumerate{parsep=6pt,itemsep=0pt,leftmargin=*,labelsep=5.5mm}
\setlist[description]{itemsep=0mm}

\firstpage{1}
\pubvolume{10}
\issuenum{5}
\articlenumber{1699}
\pubyear{2020}
\copyrightyear{2020}
\history{Received: 7 January 2020; Accepted: 21 February 2020; Published: 2 March 2020}
\updates{yes} 

\usepackage{listings}
\usepackage{color}
\usepackage{algorithm}
\usepackage{algorithmic}

\Title{Combining Machine Learning and Social Network Analysis to~Reveal the Organizational Structures} 

\Author{Mateusz Nurek 
\orcidA{} and Radosław Michalski *\href{https://orcid.org/0000-0002-0106-655X}{\orcidicon}}

\AuthorNames{Mateusz Nurek, Radosław Michalski}

\address[1]{%
Department of Computational Intelligence, Faculty of Computer Science and Management,\\Wrocław University of Science and Technology, 50-370 Wrocław, Poland; mateusz.nurek@pwr.edu.pl}

\corres{\hangafter=1 \hangindent=1.05em \hspace{-0.82em} Correspondence: radoslaw.michalski@pwr.edu.pl}

\abstract{Formation of a hierarchy within an organization is a natural way of assigning the duties, delegating responsibilities and optimizing the flow of information. Only for the smallest companies the lack of the hierarchy, that is, a flat one, is possible. Yet, if they grow, the introduction of a hierarchy is inevitable. Most often, its existence results in different nature of the tasks and duties of its members located at various organizational levels or in distant parts of it. On the other hand, employees often send dozens of emails each day, and by doing so, and also by being engaged in other activities, they~naturally form an informal social network where nodes are individuals and edges are the actions linking them. At first, such a social network seems distinct from the organizational one. However, the analysis of this network may lead to reproducing the organizational hierarchy of companies. This~is due to the fact that that people holding a similar position in the hierarchy possibly share also a similar way of behaving and communicating attributed to their role. The key concept of this work is to evaluate how well social network measures when combined with other features gained from the feature engineering align with the classification of the members of organizational social network. As a technique for answering this research question, machine learning apparatus was employed. Here, for the classification task, Decision Trees, Random Forest, {Neural Networks and Support Vector Machines} have been evaluated, as well as a collective classification algorithm, which is also proposed in this paper. The used approach allowed to compare how traditional methods of machine learning classification, while supported by social network analysis, performed in comparison to a typical graph algorithm. The results demonstrate that the social network built using the metadata on communication highly exposes the organizational structure.}


\begin{document}


\section{Introduction}
\label{sec:introduction}
People around the world send hundreds of emails to exchange information within organizations. As~an implicit result of that, each of these interactions forms a link in a social network. This network can be a valuable source of knowledge about human behaviors and what is more, conducting the analysis can reveal groups of employees with similar communication patterns. These groups usually coincide with different levels of the organization’s hierarchy and additionally, employees who work in the same position generally have a comparable scope of duties. It is common for organizations to observe some hierarchy because, formally, an~organized structure helps with better management of employees and gaining an advantage within the market. Therefore, the~analysis of the network created from a set of emails could retrieve valuable data about inner corporation processes and recreate an organizational structure. An~interesting and promising idea seems to be the combination of network measures and additional features extracted from messages for classification tasks. Social network analysis (SNA) has the potential to boost machine learning algorithms in a field of organization structure detection, thus capturing relations between data seems to be very important for this kind of~dataset.

The reverse engineering of the corporate structure of an organization can be perceived two-way. On~the one hand, if~successful, it could reveal company structure by having only meta-data and this imposes a risk in the case when the structure is intentionally kept secret, for example,~for keeping the competitive edge or protecting the employees from takeovers by other companies. On~the other hand, this could lead to reconstructing the structure of malicious organizations when only partial information about them is~available.

In the literature, there are several works describing the detection of organizational structures. However, most of them use the Enron dataset~\cite{McCallum2007Topic} or focus rather on a network approach and omit standard supervised classification algorithms. It should be noted that each organization is managed in a slightly different way which means that communication patterns could differ within each of them. These differences imply that some solutions may give better or worse results depending on the network's specificity; it is important that studies on an organization hierarchy should not be limited to only one~dataset.

The authors of Reference~\cite{PalusEvaluation,Michalski2011Matching} introduced a concept of matching a formal organizational structure and a social network created from email communications. Experiments were carried out on messages coming from a manufacturing company located in Poland as well as the well-known Enron dataset. The~research results showed that in both cases, some network metrics were able to reveal organizational hierarchy better than others. This work also touched on the problem that a formal structure sometimes may significantly differ and will not converge with the daily~reality.

The idea of combining network metrics and other features extracted from email dataset to reveal corporate hierarchy is introduced in Reference~\cite{Creamer2009Segmentation}. The~authors presented their own metric named “social score” which defines the importance of each employee in the network. This metric is defined as a weighted average of all features and is used in a grouping algorithm. The~grouping method is a simple straight scale level division algorithm which assigns employees to defined intervals by the social~score.

The study on the usage of network measures as input features for classification algorithms was presented in Reference~\cite{Fire2015Organization}. The~basic concept of this work focused on retrieving company hierarchy based on the network created from social media accounts of the employees. The~authors presented that centrality measures and clustering coefficients in combination with other features extracted from social media can detect leaders in a corporate structure. However, this research used individual features of a person like a gender, hometown or number of friends instead of features gained from job activities and interactions among employees. Other articles describing the combination of SNA and standard classification methods are References~\cite{Namata2006Inferring,Zhang2009Analyzing}. They both work using the Enron dataset and features based on the number of sent/received messages. In Reference~\cite{Wang2013Analyzing} the usage of some network metrics as input for classification and clustering algorithms has been described. Furthermore, the~results were compared to a novel measure called Human Rank (improvement of Page Rank). {However, the~use of classification based on social network features is not limited only to the corporate environment. For~instance, following the ideas of studying the social networks of criminalists~\cite{coles2001s}, the authors of Reference~\cite{shaabani2015early} used features of a social network of co-arrestees for predicting the possibility of future violent crimes. A~similar concept was also used in Reference~\cite{tayebi2014spatially} for analyzing co-offending networks. In~that work, a~co-offence prediction algorithm using supervised learning has been developed. Yet, the~classification in social networks based on communication or behaviour in social media, can relate to completely different areas, such as poverty detection~\cite{sundsoy2016deep}, personality traits discovery~\cite{kosinski2013private} or occupation~\cite{huang2015multi}. All that is possible because our digital traces do differ depending on our role or status.}

In the area of the problem being solved also many solutions concentrated mainly on classification from a graph perspective. For~instance, identification of key players of social network based on entropy~\cite{Ortiz-Arroyo2009Discovering}, applying graphical models~\cite{McCallum2007Topic} or factor graph models~\cite{Dong2015Inferring}.

The following work focuses on the organizational structure detection based on nine-months of e-mail communication between employees of a manufacturing company located in Poland as well as the Enron dataset. The~research used Decision Tree, Random Forest, {Neural Network and Support Vector Machine (SVM) algorithm} for classification, moreover influence of minimum employee activity was examined. The~obtained results were compared with the simple graph algorithm of collective classification also proposed in this paper. The~weakness of this approach is the fact that an independent and identical distribution (IID) condition is difficult to meet due to network measures which were calculated once before splitting data on training and test set. In~social network analysis, full satisfaction of the IID condition is hard to achieve because if we had built independent networks for training and test data, we would get totally different network measures and the importance of each node could be biased. However, network measures could be valuable features for machine learning algorithms in sight of capturing connections between data. The~results showed that the combination of classification algorithm and social network analysis can reveal organizational structures, however, small changes in the network can change the efficiency of the algorithms. Furthermore, a~graph approach, such as collective classification, is able to classify well even with limited knowledge about node~labels.

\section{Materials and~Methods}

In this section, after~introductory sections to supervised learning and social network analysis, a~proposed solution is described in detail, as~well as the used datasets. The~presented solution is created with Python language, as~well as {NetworkX} library for a social network creation and {Scikit-learn} for all machine learning~tasks.

\subsection{{Supervised~Learning}}
\label{sec:supervised}
{Machine learning can be considered an application of artificial intelligence that provides systems the ability to automatically learn and improve from experience without being explicitly programmed~\cite{jordan2015machine}. The~set of tools derived from the field of statistics enables the possibility to perform multiple tasks far exceeding human capabilities or simple algorithms in variety of disciplines, ranging from text analysis, computer vision, medicine and others. In~machine learning, classification is a supervised learning approach in which the algorithm learns from the data input given to it and then uses this learning to classify new observation. Here, by~supervised we mean providing the algorithm instances of objects that have been categorized as belonging to certain class and requiring it to develop a method to adequately classify other objects without known class. In~order to fulfill this goal, numerous algorithms have been developed and tuned over last decades, such as logistic regression~\cite{hosmer2013applied}, naive Bayes classifier~\cite{maron1961automatic}, nearest neighbor~\cite{altman1992introduction}, Support Vector Machines~\cite{cortes1995support}, decision trees~\cite{utgoff1989incremental}, random forests~\cite{breiman2001random} or neural networks~\cite{ayodele2010types}. Each of these methods takes a different perspective to the task. Regarding the methods used in this work, decision trees build a tree consisting of the tests of features: each branch represents the outcome of the test, and~each leaf node represents a class label. Random forest extends the concept of decision trees by building a multitude of them and outputs the class that is the mode of the classes of the individual trees. What one can say about these two methods is that the rules of classification are transparent and highly interpretable. Regarding two other methods evaluated in this work, Support Vector Machine constructs a hyperplane or a set of hyperplanes in a high- or infinite-dimensional space, which can be used for classification. Neural networks try to mimic human brain in terms of how information is being passed and analysed. Here, a~neural network is based on a collection of connected units or nodes called neurons. Each connection, like the synapses in a biological brain, can transmit a signal to other neurons. An~artificial neuron that receives a signal then processes it and can signal neurons connected to it. Neurons are grouped in layers and a signal passing the layers is being converted by neurons into the one that at the final (output layer) will be decided upon the class membership. This creates an architecture of neural network. Contrary to decision trees and random forests, Support Vector Machines and neural networks are not that easily interpretable in terms of the importance of features~\cite{molnar2019interpretable}.}

{What links all of these methods is that they usually require that the samples are required to follow IID principle, namely they have to be independent and identically distributed. Unfortunately, this~is not always the case, especially when we consider any network-related data. For~instance in social networks, people tend to cluster in groups of similar interests~\cite{mcpherson2001birds} or change their opinion based on others' opinion~\cite{friedkin1990social}. In~this case it is hard to consider the samples as IID, so another set of approaches has been developed: collective classification that tries to make the use of the networked structure of the data~\cite{sen2008collective}.}

{Another problem in classification is that rarely the instances are equally distributed over all classes to be classified. This problem is referred to as imbalanced data and there are multiple techniques that allow to tackle it, mainly belonging to one of two groups: under-sampling and over-sampling~\cite{sun2009classification, krawczyk2016learning}. In~under-sampling, the~dominant groups are being reduced to be equally represented as previously under-represented classes. Contrary, over-sampling generates the synthetic instances that belong to under-represented class leading to more balanced data. One of the most prominent over-sampling techniques is SMOTE that bases on nearest neighbors judged by Euclidean Distance between data points in feature space and perform vector operations to generate new data points~\cite{chawla2002smote, fernandez2018smote}. Using this technique provides more representative and less biased samples compared to random over-sampling.}

{In Section~\ref{sec:classification} one can find information on which classification algorithms have been used in this work and Section~\ref{sec:collective} contains more information on how we used collective classification for discovering the organizational structure from social network.}

\subsection{{Social Network~Analysis}}
\label{sec:sna}
{The field of social network analysis can be understood as a set of techniques deriving knowledge about human relationships based on the relations they form---usually by being members of social networks of different kinds. These networks can relate to family, friends, companies or organizations they are employees members of, or~social media they participate in. More formally, social network consists of a finite set or sets of actors and the relation or relations defined on them~\cite{wasserman1994social}. To~help understanding this definition of a social network, some other concepts that are fundamental in this case should be explained. An~actor is a discrete individual, corporate or collective social unit~\cite{wasserman1994social}. This can be a person in a group of people, a~department within a company or a nation in the world system. Actors are linked to each other by social ties and these relations are the core of the social network approach. Social networks are presented using graph structures, where nodes are actors and edges are connections between them. Hence, all graph theory methods and measured can be applied. A~graph may be undirected, which means that there is no distinction between the two vertices associated with each edge, or~its edges may be directed from one vertex to another. A~graph is being defined usually as an ordered pair $G:=(V,E)$, where $V$ are vertices or nodes and $E$ are edges. On~top of that multiple methods are being used in order to derive knowledge about network members or network itself, such as centrality measures~\cite{friedkin1991theoretical}, community detection~\cite{pujol2006clustering}, modelling evolution~\cite{michalski2011modelling}, or~detection of influential nodes~\cite{michalski2015maximizing}. As~described in Section~\ref{sec:introduction}, social network analysis also became present in organizations where it is often being referred to as organizational network analysis~\cite{Michalski2014}.}

\subsection{Datasets}
{In this work we evaluated two datasets containing both: metadata on communication and organizational structure in companies. This allowed to use the features extracted from the social network built using the communication data as features for classifiers. These classifiers have been then distinguishing the level of an employee in a corporate hierarchy. Detailed description of the datasets is presented below.}

\subsubsection{Manufacturing~Company}
\begin{figure}[H]
\centering
\includegraphics[angle=180, scale=0.9]{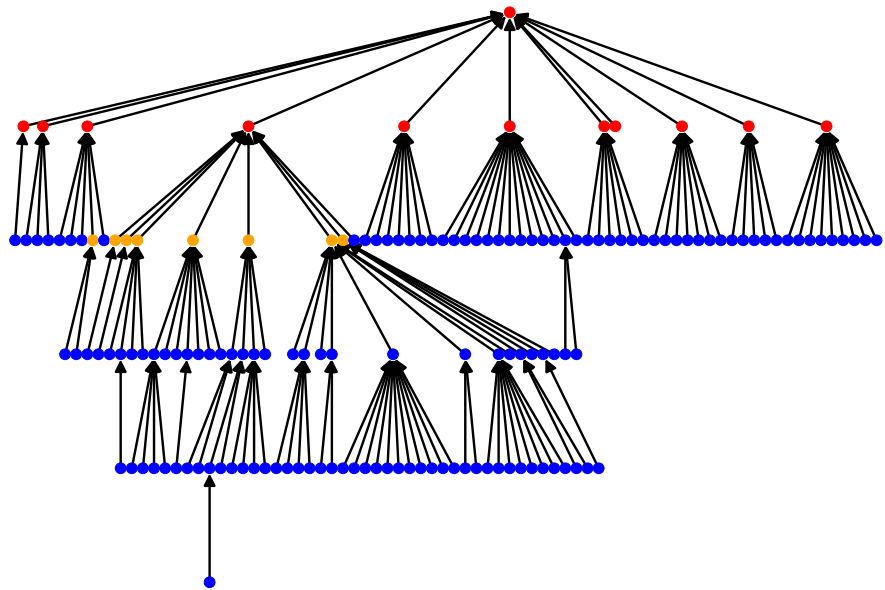}
\caption{Organizational hierarchy after removal of former employees and technical accounts. Red nodes---first level management. Orange nodes---second level management. Blue nodes---regular~employees.}
\label{fig:hierarchy}
\end{figure}

The analyzed dataset contains a nine-month exchange of messages among clerical employees of a manufacturing company located in Poland~\cite{Michalski2011Matching}. The~dataset consists of two files---the first contains the company hierarchy, the~second stores the communication history. The~analyzed company contains the three level hierarchy: the first management level, the~second management level and regular employees. The~file with emails consists of senders and recipients, as~well as the date and time of sent messages. Moreover, emails from former employees and technician accounts are also included in this file. Due to the lack of data about supervisors, former employees and technical accounts were removed from the further research. The~final organizational structure to be analyzed is shown in~Figure~\ref{fig:hierarchy} and in~Table~\ref{tab:hierarchy}. It is important to note that the dataset does not contain any correspondence with anyone outside of the company, moreover, the~company structure has been consistently stable within the period of time being considered and has not undergone any~changes.

\unskip
\begin{table}[H]
\caption{Organizational structure after removal of former employees and technical~accounts.}
\centering
\begin{tabular}{cc}
\toprule
\textbf{Hierarchy Level}    &   \textbf{Number}\\
\midrule
The first management level & 12\\
The second management level & 8\\
Regular employees & 134\\
\bottomrule
\end{tabular}
\label{tab:hierarchy}
\end{table}
\unskip

\subsubsection{Enron}
\label{sec:enron}
The second analyzed dataset comes from the Enron company~\cite{McCallum2007Topic}. Enron was a large American energy establishment founded in 1985 subsequently became famous at the end of 2001 due to financial fraud. During~the investigation, the~dataset has been made public, however the organizational hierarchy has never been officially confirmed. Despite of this limitation, the~Enron email corpus has become the subject of many studies, which allowed to partially reconstruct the company's structure. The~authors of this paper decided to use processed version of this dataset which already include positions assigned to the employees. There is a seven-level hierarchy in this data set, however, to~reduce the complexity of this structure the authors proposed more generic three-level hierarchy showed in Table~\ref{tab:enron_hierarchy}, the~same as in the manufacturing company dataset. The~applied approach allowed for a better distinction of managerial and executive positions from regular employees. The~analyzed period contains messages from over 3 years and due to limited knowledge about inner company processes the authors assumed that the organizational structure was stable during~it.

\begin{table}[H]
\caption{Enron~hierarchy.}
\centering
\begin{tabular}{ccc}
\toprule
\textbf{Flattened} & \textbf{Original} & \textbf{Number}\\
\midrule
\multirow{3}{*}{The first management level} & CEO & \\
& President & {40}\\
& Vice President & \\\midrule
\multirow{3}{*}{The second management level} & Director  & \\
& Managing Director & {37}\\
& Manager & \\\midrule
\multirow{3}{*}{Regular employee} & Employee & \\
& In House Lawyer & {53}\\
& Trader & \\
\bottomrule
\end{tabular}
\label{tab:enron_hierarchy}
\end{table}

Both datasets are available for evaluation or further research, see Supplementary~Materials.

\subsection{Network}
{The network was built using the email exchanges of its members, where the nodes were employees and the edges were the messages. It was decided to use a directed graph defined as follows:
Social network is a tuple $ SN~=~(V, E) $, where $ V~=~\{v_1, \ldots, v_n \}, n \in \mathbb{N}_+ $ is the set of vertices and $ E~=~\{ e_1, \ldots, e_{k^e} \}, k^e \in \mathbb{N}_+ $ is the set of edges between them. Each vertex $ v_i \in V $ represents an individual $ v_i^e $ and each edge $ e_{ij} $ corresponds to the directed social relationship from $ v_i $ to $ v_j $, such that $ E~=~\{ (v_i, v_j, w_{ij}) : v_i \in V, v_j \in V, v_i~=~v^{e}_i, v_j~=~v^{e}_j $ and $ w_{ij} \in [0,1] \} $.}
The edge weights defined according to the following formula:
\begin{align*}
w_{ij}~=~\dfrac{\sum e_{ij}}{\sum e_i},
\tag{1}
\end{align*}
where $\sum e_{ij}$ is the sum of messages sent from node $i$ to node $j$ and $\sum e_i$ is the total sum of messages sent from node $i$. All self loops were~removed.

\begin{figure}[H]
\centering
\includegraphics[width=\textwidth]{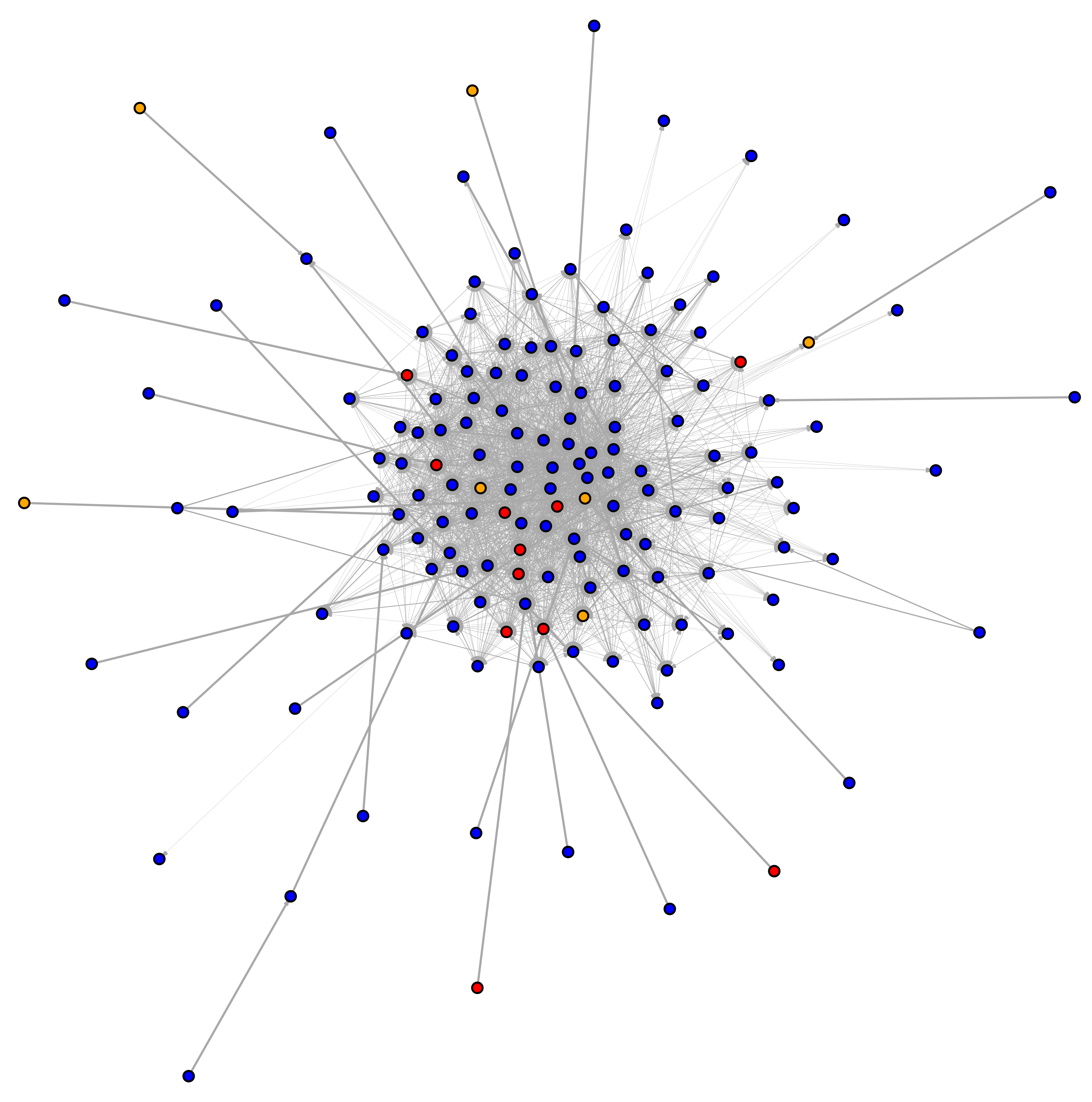}
\caption{{Weighted and directed social network in a manufacturing company built upon e-mail communication. The~colouring scheme is the same as for Figure~\ref{fig:hierarchy}: red nodes---first level management, orange nodes---second level management, blue nodes---regular employees. }{The algorithm used for visualisation is a force-directed large graph layout~\cite{adai2004lgl} with the root node with the highest betweenness.}}
\label{fig:manufacturing-network}
\end{figure}

{In Figure~\ref{fig:manufacturing-network} the weighted directed network built using e-mail communication in the manufacturing company is depicted. What can be noticed is that the position of the first level and second level management is not always central in this network. As~a result of that, using centrality measures would not be enough for detecting positions in organizational hierarchy. The~reasons why there is no direct correlation between position in the social network and the organizational networks can be of many kind, for example,~management positions do not require intense communication, using different forms of communication or having supporting personnel to communicate on behalf.}

\unskip

\subsection{Features}
\label{sec:features}
In the created social network, the~centrality measures presented below have been calculated as input features for classification algorithms. These measures are also briefly described in Table~\ref{tab:features}.
\begin{itemize}
\item indegree centrality:
{
\begin{equation}
\label{eq:indegree}
C_{IN}(v_i)~=~|e_{ji} \in E|, j \neq i,
\end{equation}
where $ e_{ji} $ is the edge going from every node $ v_j $ to evaluated node $ v_i$. \\
}
\item outdegree centrality:
{
\begin{equation}
\label{eq:outdegree}
C_{OUT}(v_i)~=~|e_{ij} \in E|, i \neq j,
\end{equation}
where $ e_{ij} $ is the edge going from evaluated node $ v_i$ to every other node $ v_j $ in the~network. \\
}
\item betweenness centrality:
{
\begin{equation}
\label{eq:betweenness}
C_B(v_i)~=~\sum_{v_s \neq v_i \neq v_d} \frac{\sigma_{v_{s}v_{d}}(v_i)}{\sigma_{v_{s}v_{d}}},
\end{equation}
where $ \sigma_{v_{s}v_{d}}(v_i) $ is the number of shortest paths between nodes $v_s$ and $v_d$ passing through node $v_i$ and $\sigma_{v_{s}v_{d}}$ is the number of all shortest paths between $v_s$ and $v_d$.\\
}
\item closeness centrality:
{
\begin{equation}
\label{eq:closeness}
C_C(v_i)~=~\frac{N}{\sum_{v_y} d(v_y, v_i)},
\end{equation}
where $ N $ is the number of vertices in the network and $ d(v_y, v_i) $ is a distance between vertices $ v_y$ and $ v_i $.\\
}
\item eigenvector centrality:
{
\begin{equation}
\label{eq:eigenvector}
C_E(v_i)~=~\frac{1}{\lambda} \sum_k a_{v_k,v_i} \, C_E(v_k),
\end{equation}
where $A~=~(a_{i,j})$ is the adjacency matrix of a graph and $\lambda \neq 0$ is a~constant.\\
}
\item page rank:
{
\begin{equation}
\label{eq:pagerank}
C_{PR}(v_i)~=~\alpha \sum_k \frac{a_{v_k,v_i}}{d_k} \, C_{PR}(v_k) + \beta,
\end{equation}
where $\alpha$ and $\beta$ are constants and $d_k$ is the out-degree of node $v_k$ if such degree is positive, or~$d_k~=~1$ if the out-degree of node $v_k$ is null. Again, $A~=~(a_{i,j})$ is the adjacency matrix of a~graph.\\
}
\item hub centrality:
{
\begin{equation}
\label{eq:hub}
C_{HUB}(v_i)~=~\beta \sum_k a_{v_i,v_k} \, C_{AUT}(v_k),
\end{equation}
where $A~=~(a_{i,j})$ is the adjacency matrix of a graph and $C_{AUT}(v_k)$ is the authority centrality of a node (see Equation~(\ref{eq:auth})), $\beta$ is a~constant.\\
}
\item authority centrality:
{
\begin{equation}
\label{eq:auth}
C_{AUT}(v_i)~=~\alpha \sum_k a_{v_k,v_i} \, C_{HUB}(v_k),
\end{equation}
where $A~=~(a_{i,j})$ is the adjacency matrix of a graph and $C_{HUB}(v_k)$ is the hub centrality of a node (see Equation~(\ref{eq:hub})), $\alpha$ is a constant.
}
\end{itemize}

Moreover, a~local clustering coefficient was calculated for each node, which allows capturing density of connections between neighbors, as~well as two additional features related to cliques:

{
\begin{equation}
\label{eq:clustering}
C_{CC}(v_i)~=~\frac{2 m_{v_i}}{k_i (k_i - 1)},
\end{equation}
where $m_{v_i}$ is the number of pairs of neighbors of a node $v_i$ that are connected. In~the formula it is linked with the the number of possible pairs of neighbors of node $v_i$, which is {$\frac{k_{v_i} (k_{v_i} - 1)}{2}$}, where $k_{v_i}$ is the degree of a node $v_i$.
}

A clique is defined as a fully connected subgraph which means that each node has directed links to all other nodes in the clique. The~first feature is the total numbers of cliques in which an employee is assigned, furthermore the second is the size of the biggest clique for the specific node. {Reference~\cite{barabasi2016network} contains more details on all the measures introduced above.}

The next features were based on neighborhood variability, which is determined in three ways: sent neighborhood variability, received neighborhood variability and general neighborhood variability. Overall, neighborhood variability is defined as the difference between sets of neighbors which the specific node communicates in the previous and the next month. Sent neighborhood variability considers a set of neighbors to which the given node was sending messages. Received neighborhood variability looks at a set of neighbors from which the given node had been receiving messages. General neighborhood variability uses a set of neighbors with which the node communicates, without~distinguishing between sending and receiving messages. The~Jaccard coefficient was used for calculating the difference between sets, so the coefficient takes values between 0 and 1 where 0 means totally different sets and 1 means identical sets. The~Jaccard coefficient was calculated for each pair of alternating months. Moreover, if~the employee had not been active in a directly following month, the~nearest next month would be considered. For~example: the employee was active in January, but~not in February and again was active in March; therefore, coefficient would be calculated between sets of neighbors in January and March. Furthermore, a~neighborhood variability was calculated as an average Jaccard coefficient for each node based on previous partial coefficients. {Formally, sent~variability measure can be defined as following:}

{
\begin{equation}
\label{eq:variability}
VAR_{SNT}(v_i)~=~\frac{|N_{snt_{v_i,m-1}}\cap{N_{snt_{v_i,m}}}|}{|N_{snt_{v_i,m-1}}\cup{N_{snt_{v_i,m}}}|},
\end{equation}
where $N_{snt_{v_i,m-1}}$ is the set of neighbours that a certain node $v_i$ sent messages to in the month $m-1$ and $N_{snt_{v_i,m}}$ is the is the set of neighbours that a certain node $v_i$ sent messages to in month $m$ or, if~no messages have been sent in $m$, then $m+1, m+2, \ldots, m_{max}$ are considered. Similarly, received and general neighbourhood variability measures can be defined by substituting the sets of neighbours to the neighbours that either sent messages to node $v_i$ (received variability) or the set of neighbours the contact occurred with in any direction and involved $v_i$ (general variability).}

Furthermore, features such as the number of weekends worked and the amount of overtime taken were taken into account. For~overtime, work between 4:00 PM and 6:00 AM were considered. It should be mentioned that overtime was only calculated for the manufacturing company dataset. Due to the limited knowledge about the Enron dataset, it was impossible to know whether different timezones should be considered because the dates were given in the POSIX format and Enron had branches located in different~timezones.

{As a summary, all used features used for classification are presented in Table~\ref{tab:features}.}
\begin{table}[H]
\centering
\caption{Features.}
{\begin{tabular}{cp{2cm}p{6cm}}
\toprule
\textbf{Feature Name} & \textbf{Defined in} & \textbf{Brief Description} \\
\midrule
indegree centrality & Equation~(\ref{eq:indegree}) & a number of incoming links to a given node\\
\midrule
outdegree centrality & Equation~(\ref{eq:outdegree}) & a number of outgoing links from a given node\\
\midrule
betweenness centrality & Equation~(\ref{eq:betweenness}) & the frequency of a node appearing in shortest paths in the network \\
\midrule
closeness centrality& Equation~(\ref{eq:closeness}) & the length of the shortest paths between the node and all other nodes in the graph\\
\midrule
eigenvector centrality & Equation~(\ref{eq:eigenvector}) & a relative measure of importance dependent on the importance of neighbouring nodes in the network \\
\midrule
page rank centrality & Equation~(\ref{eq:pagerank}) & relative measure of importance also based on eigenvectors of an adjacency matrix, more tunable\\
\midrule
hubs centrality & Equation~(\ref{eq:hub}) & indication of position in relevance to important nodes---authorities \\
\bottomrule
\end{tabular}}
\label{tab:features}
\end{table}

\begin{table}[H]\ContinuedFloat
\centering
\caption{\textit{Cont}.}
{\begin{tabular}{cp{2cm}p{6cm}}
\toprule
\textbf{Feature Name} & \textbf{Defined in} & \textbf{Brief Description} \\
\midrule
authorities centrality & Equation~(\ref{eq:auth}) & importance of node based on being referred to by hubs \\
\midrule
clustering coefficient & Equation~(\ref{eq:clustering}) &  degree to which nodes in a graph tend to cluster together \\
\midrule
the total numbers of cliques & Section~\ref{sec:features} &  total numbers of cliques in which an employee is assigned\\
\midrule
the biggest clique & Section~\ref{sec:features} & size of the biggest clique for the specific node\\
\midrule
sent neighborhood variability & Equation~(\ref{eq:variability}) & difference between sets of neighbours a node sends emails to in consecutive months \\
\midrule
received neighborhood variability & Equation~(\ref{eq:variability}) & difference between sets of neighbours a node receives emails from in consecutive months \\
\midrule
general neighborhood variability & Equation~(\ref{eq:variability}) & difference between sets of neighbours a node communicates with in consecutive months \\
\midrule
overtime & Section~\ref{sec:features} & a number of days an employee worked overtime (only for manufacturing company) \\
\midrule
the number of weekends worked & Section~\ref{sec:features} & how many times an employee worked over weekends \\
\bottomrule
\end{tabular}}
\end{table}

\subsection{Classification}
\label{sec:classification}
The classification task was carried out using the Decision Tree~\cite{utgoff1989incremental}, Random Forest, {Neural Network (multi-layer perceptron) with L-BFGS 
solver and SVM algorithm with the polynomial kernel} for different set up of following parameters of the experiment: number of recognized employee groups, minimum number of active months as well as the percentage of used~features.

The first parameter refers to the previously mentioned three-level hierarchy of employees, which~can also be flattened to only two levels---management level and regular employees. The~experiment was run with two values of this parameter to see how the performance of the algorithms vary with recognizing two and three groups of~employees.

The meaning of the second parameter is checked to see how the activity of a person may have influence on the result of the classification and therefore was examined to see if higher minimum months of employee activity correspond with better results. There is an assumption that some patterns of behavior required more time to be revealed, so the classification was run five times starting with one month minimum activity and ending with 5 months minimum activity. For~each value, the~network had to be recreated and features calculated again as some nodes were eliminated from the~network.

The third parameter examines the impact of the elimination of the most significant features. For~this parameter, the~experiment was carried out nine times, starting from all features to only ten percent of features with a continual decrease of ten percent. {The importance of features for Decision Tree and Random Forest algorithms was determined based on Gini importance parameter from previously learned model. The~Neural Network and SVM algorithm are not so easily interpretable and importance of the features cannot be obtained from the outcome of the model. In~this case, the~importance of the features must be determined before learning a model. For~this purpose, the~univariate feature selection method based on the chi-squared test was used.}

In the analyzed manufacturing company dataset, there was a problem with the unbalanced size of classes, which is common for a company structure where the group of regular employees is the most numerous and the management level has fewer members. However, the~Enron dataset is much more balanced in each group, which may indicate a different management model in this company. To~handle this problem the technique of oversampling was used to solve it, therefore to match the size of all minor classes to the size of the majority class of regular employees, SMOTE algorithm was used. {To prevent data leakage, oversampling was performed only on a training set.}

{In general, for~each combination of the above parameters, a~model was trained with the usage of the grid search algorithm with 5-fold cross validation, so all the possible combinations from the range of given values were tested, and~the best one {with respect to the f-score macro average} was returned. The~hyperparameters search space is shown in Table~\ref{tab:hp_space} as well as the best ones for each model in Tables~\ref{tab:mc_best_hp_2l}--\ref{tab:enron_best_hp_3l}.}

\subsection{Collective~Classification}
\label{sec:collective}
Collective classification is a different way of revealing company hierarchy from a graph perspective. This approach uses the connection between nodes to propagate labels within the whole network. Loopy belief propagation is an example of collective classification described in detail in References~\cite{sen2008collective,Kajdanowicz2016learning}. Therefore, in~this paper, a~simplified version of this algorithm is introduced to compare with standard classification~algorithms.

{{The proposed collective classification method is presented as Algorithm~\ref{alg:cc}.} The first step of this algorithm is choosing a utility score and sorting all nodes according to it {(line~\ref{alg_utility_score})}. The~utility score can be one of the calculated features from the previous section. The~next step is to reveal the labels for the given percentage of nodes of each class $l_i \in L$ with the highest utility score {(line~\ref{alg_reveal_top_nodes})}. These nodes are marked as known (their labels are constant) and labeled $V^L$ , whereas the other nodes are treated as unkown $V^{UK}$ and unlabeled. Furthermore, the~propagation of labels begins in a loop until the stop condition is met or the number of iterations exceeds the given maximum number of iterations. In~one iteration, each labeled node sends a message to all of its neighbors by treating edges as undirected {(line~\ref{alg_message_passing})}; moreover, all received labels in a given iteration are saved for each node $v_i$ in a counter $c_{v_i}$ {(line~\ref{alg_counter})}. The~labels update begins after all nodes sent a message to their neighbors, so the sending order does not affect the result. If~the node $v_i$ has received one label more often than others {(line~\ref{alg_only_one})}, this label will be assigned to it {(line~\ref{alg_assign_label})} and the node will be additionally treated as labeled $v_i \in V^L$ {(line~\ref{alg_assign_v_to_vl})}, otherwise for this node counter $u_{v_i}$ will be increased {(line~\ref{alg_counter_increased})}. If~$u_{v_i}$ exceeds the maximum value {(line~\ref{alg_max_value})}, it will be reset {(line~\ref{alg_reset})} and the node will be assigned the label with the highest position in the company hierarchy among the labels with the highest count {(lines~\ref{alg_highest_count_1}~to~\ref{alg_highest_count_2})}. At~the end of iteration the stop conditions is always checked and it is determined as a difference between sets of previous and current labels, therefore if the Jaccard coefficient is bigger than the given minimum Jaccard value and all nodes have assigned label then the algorithm will end {(line~\ref{alg_stop_condition})}. Additionally, in~the case of unbalanced classes, the~algorithm allows defining a $threshold$. During~the phase of counting how many times each label was received by the node, the~result for the majority class will be divided by this threshold to prevent domination of this class. {(line~\ref{alg_threshold})}}

The collective classification algorithm was run with three parameters: number of recognized employee groups, minimum number of active months, percentage of known nodes. The~first two are identical to the parameters from the previous section, but~the last one determines percentage of the known (labeled) nodes. Nine values of this parameter were used from 90\% to 10\% with a decrease of ten percent. The~manufacturing company dataset required setting threshold on the contrary to the Enron dataset where it was not necessary. Additionally, to~find the best utility score experiment was carried out for all calculated features, as~well as the best Jaccard value and threshold were chosen from a range of different values. {The hyperparameters search space is shown in Table~\ref{tab:hp_space_cc} as well as the best ones for each model in Tables~\ref{tab:mc_best_hp_cc_1_2l}--\ref{tab:enron_best_hp_cc_3l}.}
\vspace{12pt}

\begin{algorithm}[H]
\caption{Collective Classification~Algorithm.}
\setstretch{1.5}
\begin{algorithmic}[1]
\STATE sort nodes descending by utility score \label{alg_utility_score}
\STATE assign given percentage of top $v_i \in V$ to $V^L$ for each label \label{alg_reveal_top_nodes}
\REPEAT
\STATE //perform message passing
\FOR{each edge $(v_i,v_j) \in E, v_i \in V^{L},v_j \in V^{UK}$} \label{alg_message_passing}
\STATE $c_{V_j}(l_{V_i}) \gets c_{V_j}(l_{V_i}) + 1$ \label{alg_counter}
\ENDFOR
\STATE //perform label update
\FOR{each node $v_i \in V^{UK}$}
\IF{$l_{v_i}$ is a majority class}
\STATE $c_{V_i}(l_{V_i}) \gets c_{V_i}(l_{V_i}) / threshold$ \label{alg_threshold}
\ENDIF
\IF{exists only one label with highest count for the node $v_i$} \label{alg_only_one}
\STATE $l_{v_i} \gets l: \max_{l \in L}c_{v_i}(l)$ \label{alg_assign_label}
\STATE assign $v_i$ to $V^L$ \label{alg_assign_v_to_vl}
\STATE $u_{v_i} \gets 0$
\ELSE
\STATE $u_{v_i} \gets u_{v_i} + 1$ \label{alg_counter_increased}
\ENDIF
\IF{the maximum value of $u_{v_i}$ has been reached} \label{alg_max_value}
\STATE //get set of labels with the highest count
\STATE $L_{max} \gets l: \max_{l \in L}c_{v_i}(l)$ \label{alg_highest_count_1}
\STATE //get label with the highest position in the hierarchy (smaller is higher)
\STATE $l_{v_i} \gets l: \min L_{max}$ \label{alg_highest_count_2}
\STATE assign $v_i$ to $V^L$
\STATE $u_{v_i} \gets 0$ \label{alg_reset}
\ENDIF
\ENDFOR
\UNTIL{stop condition} \label{alg_stop_condition}
\end{algorithmic}
\label{alg:cc}
\end{algorithm}

\section{Results}

The problem that is tackled can be considered as a binary classification for two groups of employees and multiclass classification for three groups. Therefore, f-score macro average measure was used to evaluate the solution in sight of the one metric which was needed to compare both results. This measure can handle the above cases, moreover, as~was written in Reference~\cite{Ozgur2005text} it copes well with the problem of unbalanced classes. The~biggest advantage of this measure is the equal treatment of all classes which means that a result is not dominated by a majority~class.

The results for the manufacturing company dataset are shown in Figures~\ref{fig:manufacturing_company_classification_2_groups},~\ref{fig:manufacturing_company_classification_3_groups}~and~\ref{fig:manufacturing_company_collective_classification}. {The f-score macro average for the randomly assigned labels is around 0.42 for two levels of the hierarchy and 0.24 for three levels of the hierarchy, in~comparing the best result for the two levels was 0.7768 obtained by Random Forest and for the three levels 0.4737 achieved by Decision Tree. The~much higher score obtained for two groups of classification can be explained by unbalanced classes.} The~classification of the three groups got worse results because of the small number of samples which was insufficient for the distinction between the two levels of management, even when oversampling was used. Furthermore, Random Forest got a slightly better results especially for two groups of employees. A~strange phenomenon can be observed when a reduction of the most important features occasionally concludes with a better result; meaning that there could be some noise among the features which may affect the decision boundary. The~potentially explanation of this phenomenon might be related to the problem described in Reference~\cite{PalusEvaluation}, so the observed alteration could be a result that the hierarchy, which arises from daily duties does not converge with company structure on paper. This inconsistency could be the source of some noise in the used features which has an influence on the obtained result; therefore, changing the network structure by eliminating some nodes, as~well as removing the most important features, could result in moving a decision boundary. {It is also noticeable that the parameter of a minimum employee activity also has impact on the classification but it is difficult to indicate the best value because no clear pattern is visible; however, most of the best results are obtained for a minimum activity greater than one month.} The best results for two and three groups of employees was obtained by the collective classification algorithm which was able to classify nodes even if more than half of the labels were~unknown.

\begin{figure}[H]
\centering
\includegraphics[scale=0.45]{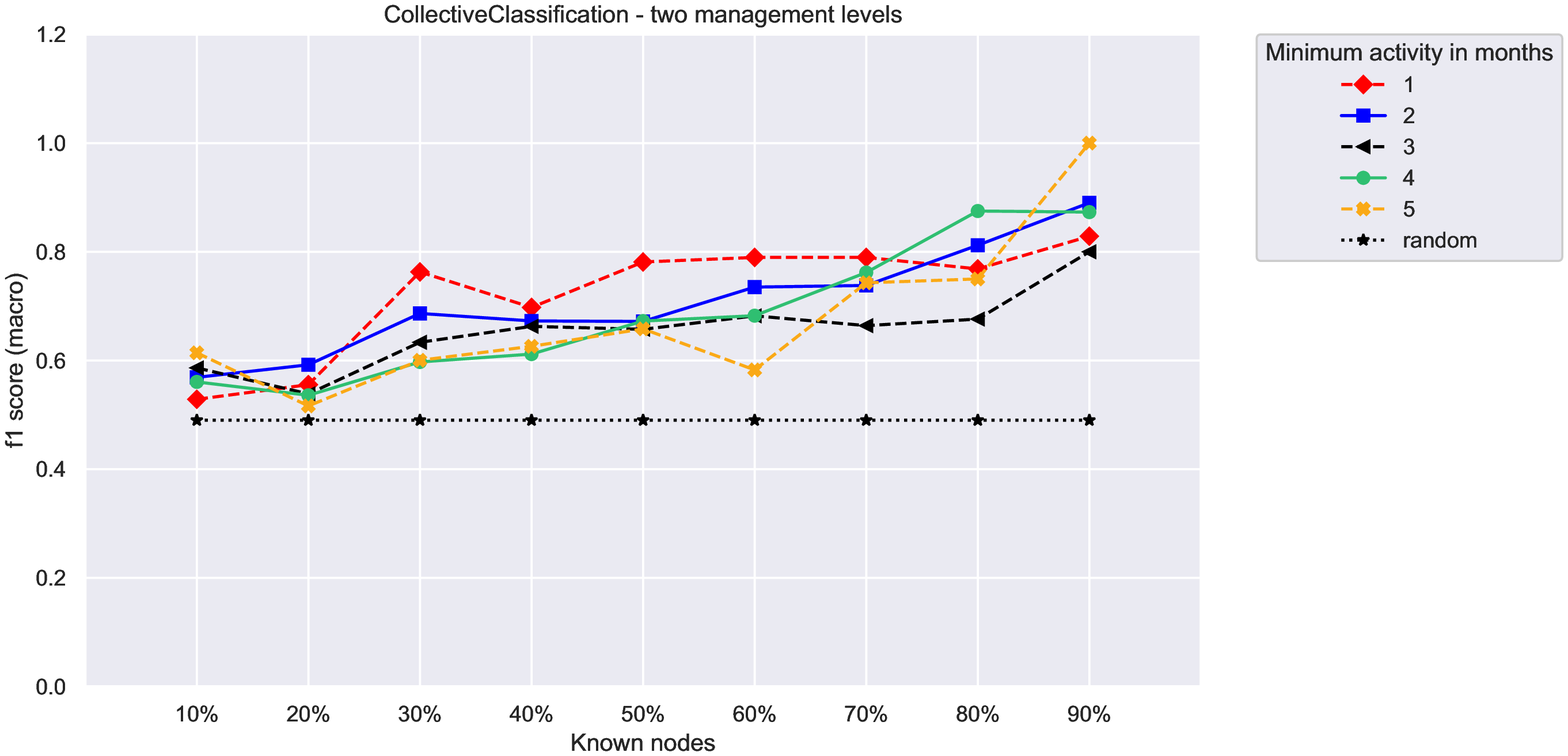}
\includegraphics[scale=0.45]{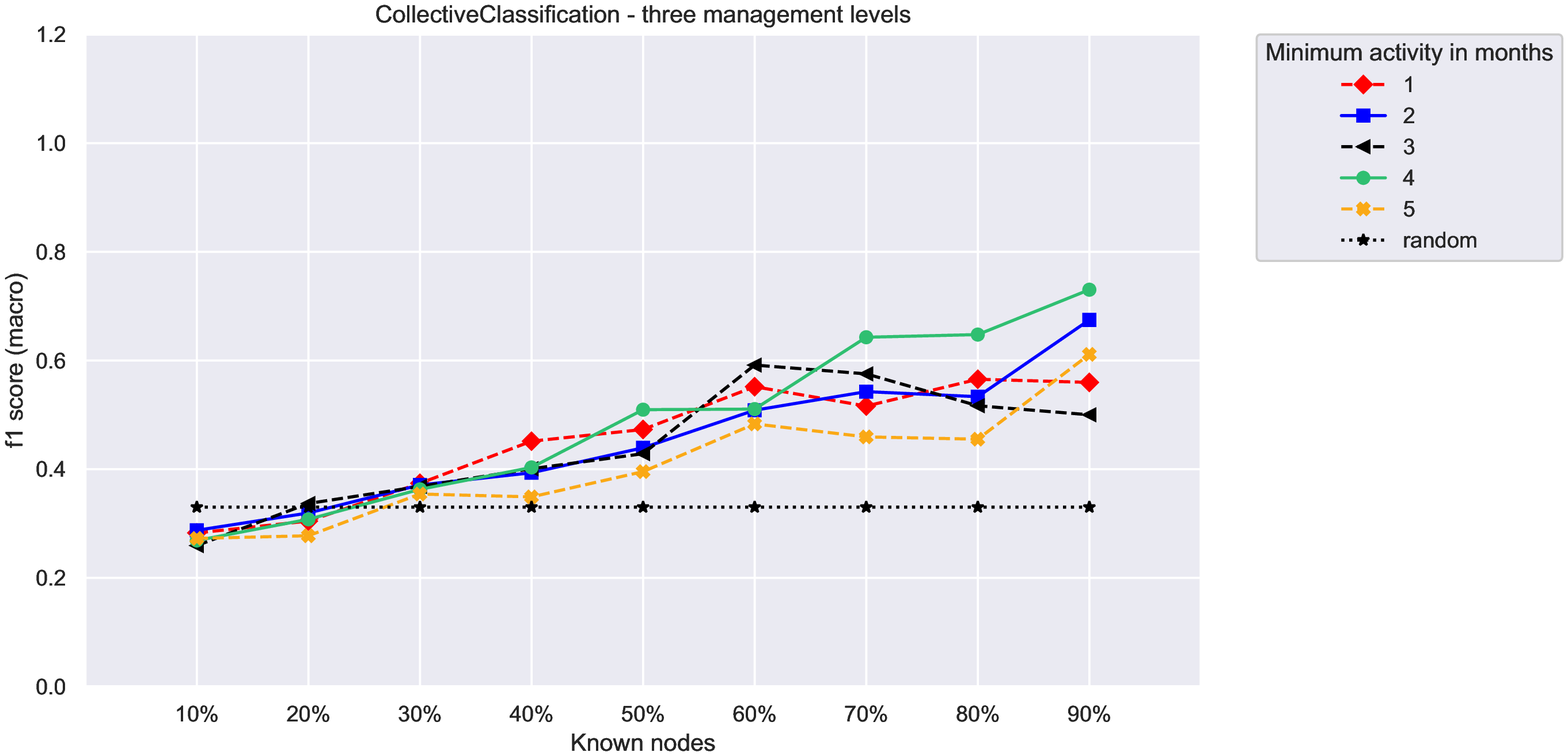}
\caption{The result of the collective classification of three groups for the Enron company~dataset.}
\label{fig:enron_collective_classification}
\end{figure}

Figures~\ref{fig:enron_classification_2_groups},~\ref{fig:enron_classification_3_groups},~and~\ref{fig:enron_collective_classification} present the results for the Enron dataset which are similar to the results of the previous dataset. {The result obtained by random labels was equal 0.49 for the two levels of the hierarchy and 0.33 for the three levels. The~best f1-score for the supervised learning methods was achieved by Random Forest algorithm, and~it was 0.8198 for the two hierarchy levels and 0.6423 for the three levels. The~results of the collective classification algorithm where higher than the results of the standard classification if the knowledge of the node labels was over 70\%.} Below this value, the~results were similar to standard classification, moreover, for~three groups, if~the knowledge of nodes fell below 40\%, the~results significantly deteriorated. It is visible that excessive reduction of features or known nodes leads to the results close to randomness. Furthermore, similarity of the results is important because shows that the presented solution works well for various organizational management models. In~the manufacturing company the majority of the employees are regular clerical workers in contrast to the small management group. In~the Enron dataset the situation is opposite, so the ratio between the first and second management level and regular employees is~balanced.

As a summary, the~best results obtained by supervised learning algorithms are presented in Table~\ref{tab:the_best_models}. Moreover, all numerical results can be found in Appendix in Tables~\ref{tab:mc_dt_2l}--\ref{tab:enron_cc_3l}.

\begin{table}[H]
\centering
\caption{{The best results obtained by supervised methods.}}
{\scalebox{0.95}[0.95]{
\begin{tabular}{cccccc}
\toprule
\textbf{Dataset} & \textbf{Number of Levels} & \textbf{Algorithm} & \textbf{F1-Score} & \textbf{Min. Activity} & \textbf{\% of Features} \\
\midrule
\multirow{8}{*}{manufacturing company} &\multirow{4}{*}{2} & Decision Tree & 0.7039 & 2 months & 80\% \\
& & Random Forest & 0.7768 & 5 months & 90\% \\
& & Neural Network & 0.6247 & 3 months & 100\% \\
& & SVM & 0.6517 & 3 months & 100\% \\ \cmidrule(lr){2-6}
&\multirow{4}{*}{3} &  Decision Tree & 0.4737 & 1 month & 80\% \\
& & Random Forest & 0.4575 & 2 months & 60\% \\
& & Neural Network & 0.4149 & 1 month & 100\% \\
& & SVM & 0.4622 & 4 months & 50\% \\ \midrule
\multirow{8}{*}{Enron} &\multirow{3}{*}{2} & Decision Tree & 0.7849 & 5 months & 100\% \\
& & Random Forest & 0.8198 & 5 months & 100\% \\
& & Neural Network & 0.7835 & 4 months & 80\% \\
& & SVM & 0.7855 & 5 months & 50\% \\  \cmidrule(lr){2-6}
&\multirow{4}{*}{3} &  Decision Tree & 0.5827 & 5 months & 50\% \\
& & Random Forest & 0.6423 & 3 months & 100\% \\
& & Neural Network & 0.6107 & 4 months & 70\% \\
& & SVM & 0.6137 & 3 months & 70\% \\
\bottomrule
\end{tabular}}
}
\label{tab:the_best_models}
\end{table}

{{Interesting conclusions about trained models can be drawn from the Figures \ref{fig:mc_fi_dt_2l}--\ref{fig:enron_fi_chi_3l}, which~presents the importance of the features for models that used a set of all features.} First of all, it should be noticed that the Decision Tree and Random Forest use many features; however, none~of the features stand out significantly in terms of importance. Nevertheless, the~clustering coefficient could be highlighted {for the manufacturing company and indegree centrality for the Enron dataset} because in many cases these features have slightly bigger importance than the others. For~the Neural Network and SVM algorithm the total number of cliques is visibly the best one {for the both datasets}. Moreover, sent and received neighborhood variability are also in some cases significant, therefore, it~shows that not only network centrality measures but also features created from employees' behavior can be important in the classification task. A~common element for all algorithms is the fact that for all parameters of the experiment, the~worst feature is always the eigenvector centrality. Furthermore, Tables~\ref{tab:mc_best_hp_cc_1_2l}--\ref{tab:enron_best_hp_cc_3l} show the best utility score depending on different combinations of experiment parameters. Unlike supervised methods, it is difficult to identify the most discriminating feature for collective classification because a wide range of them is used as a utility score.}

\begin{figure}[H]
\centering
\includegraphics[scale=0.45]{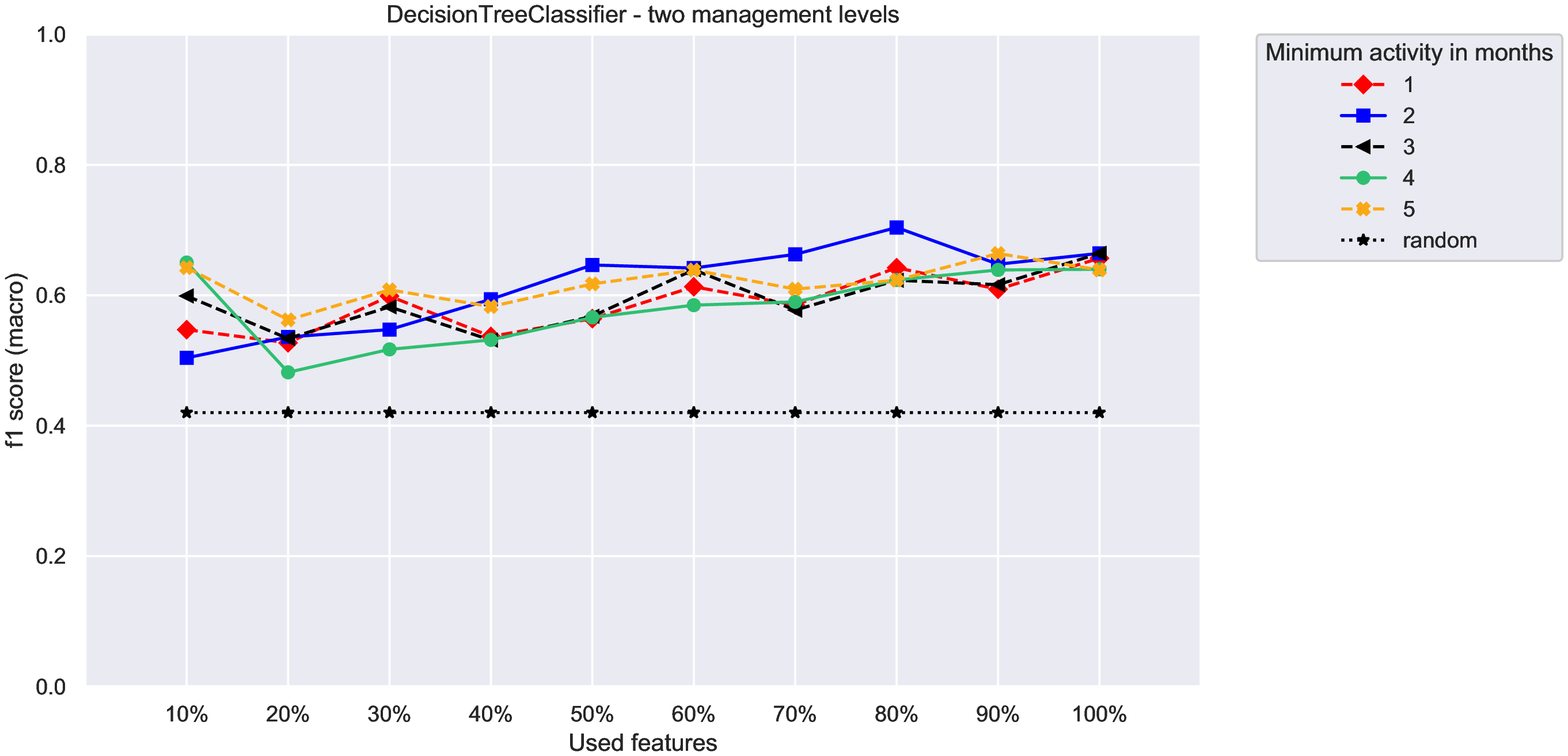}
\includegraphics[scale=0.45]{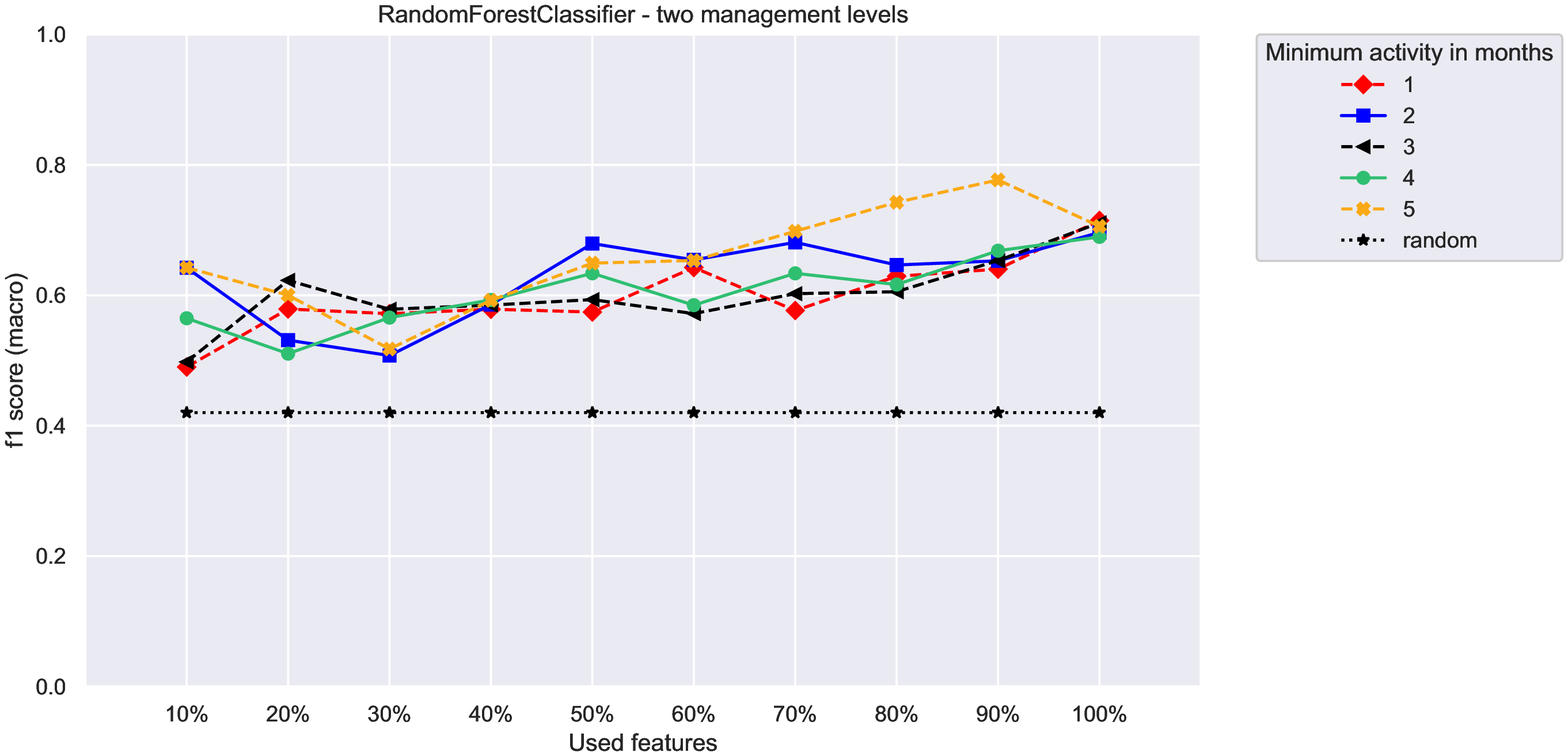}
\includegraphics[scale=0.45]{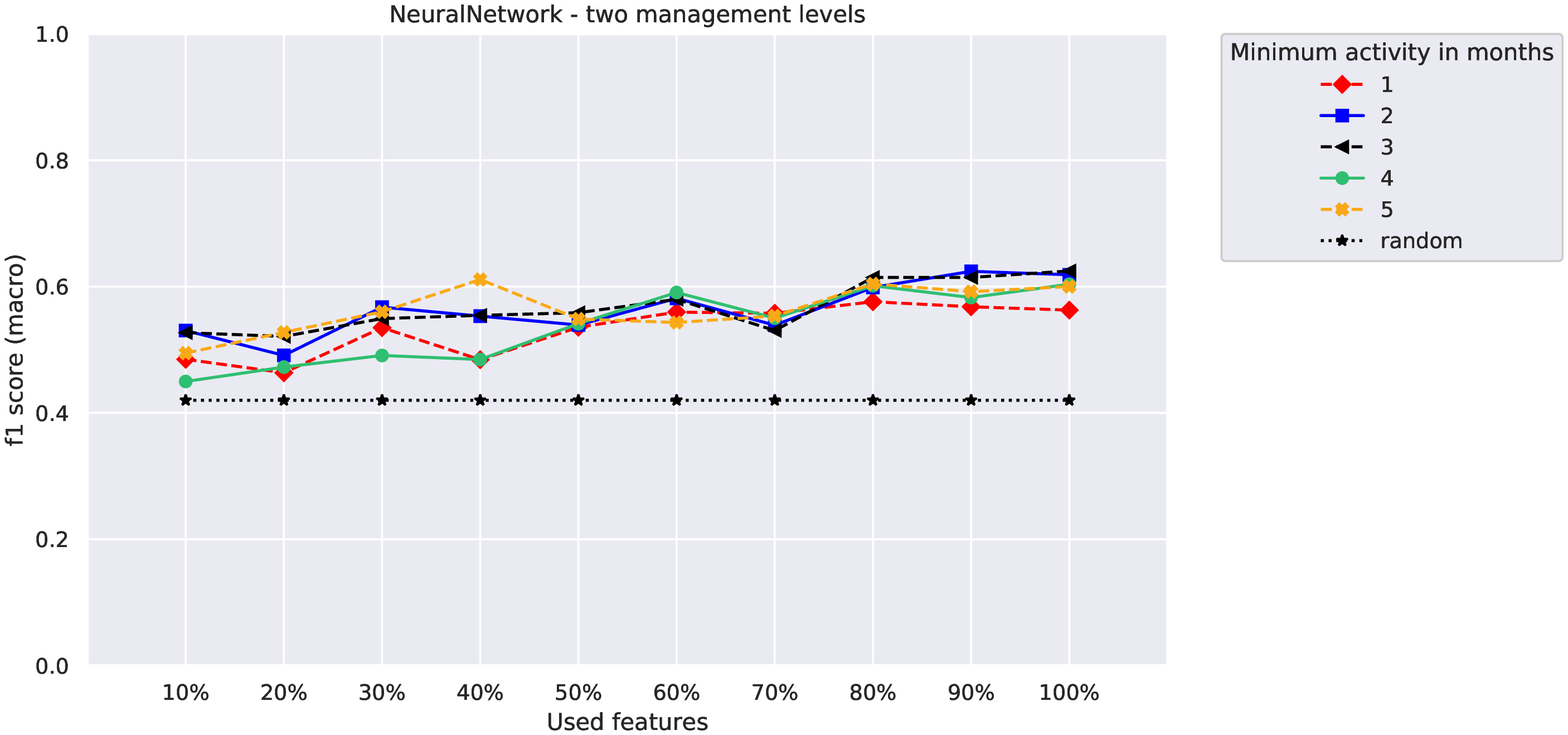}
\caption{\textit{Cont}.}

\end{figure}

\begin{figure}[H]\ContinuedFloat
\centering
\includegraphics[scale=0.45]{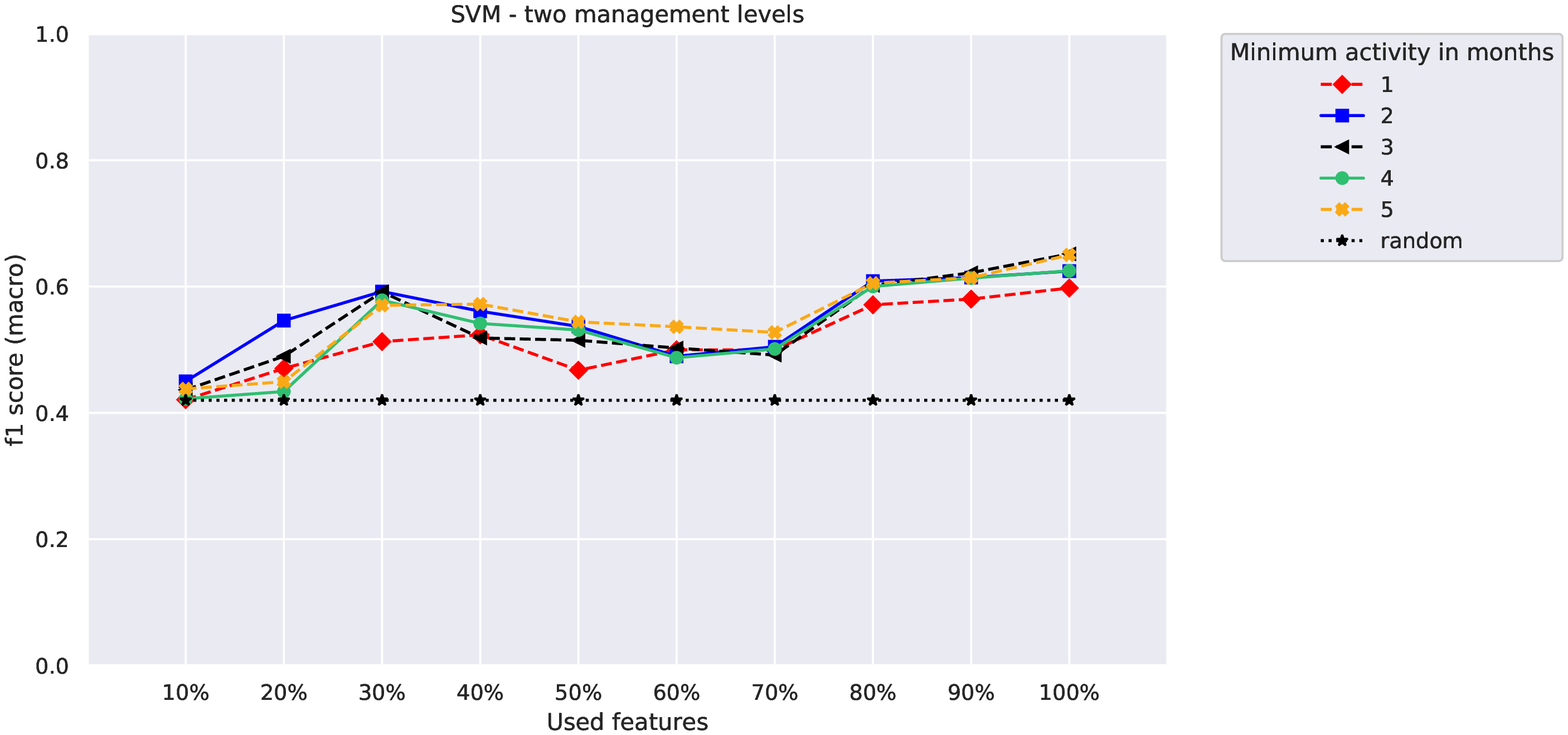}
\caption{The result of the classification of two groups for the manufacturing company~dataset.}
\label{fig:manufacturing_company_classification_2_groups}
\end{figure}
\unskip

\begin{figure}[H]
\centering
\includegraphics[scale=0.45]{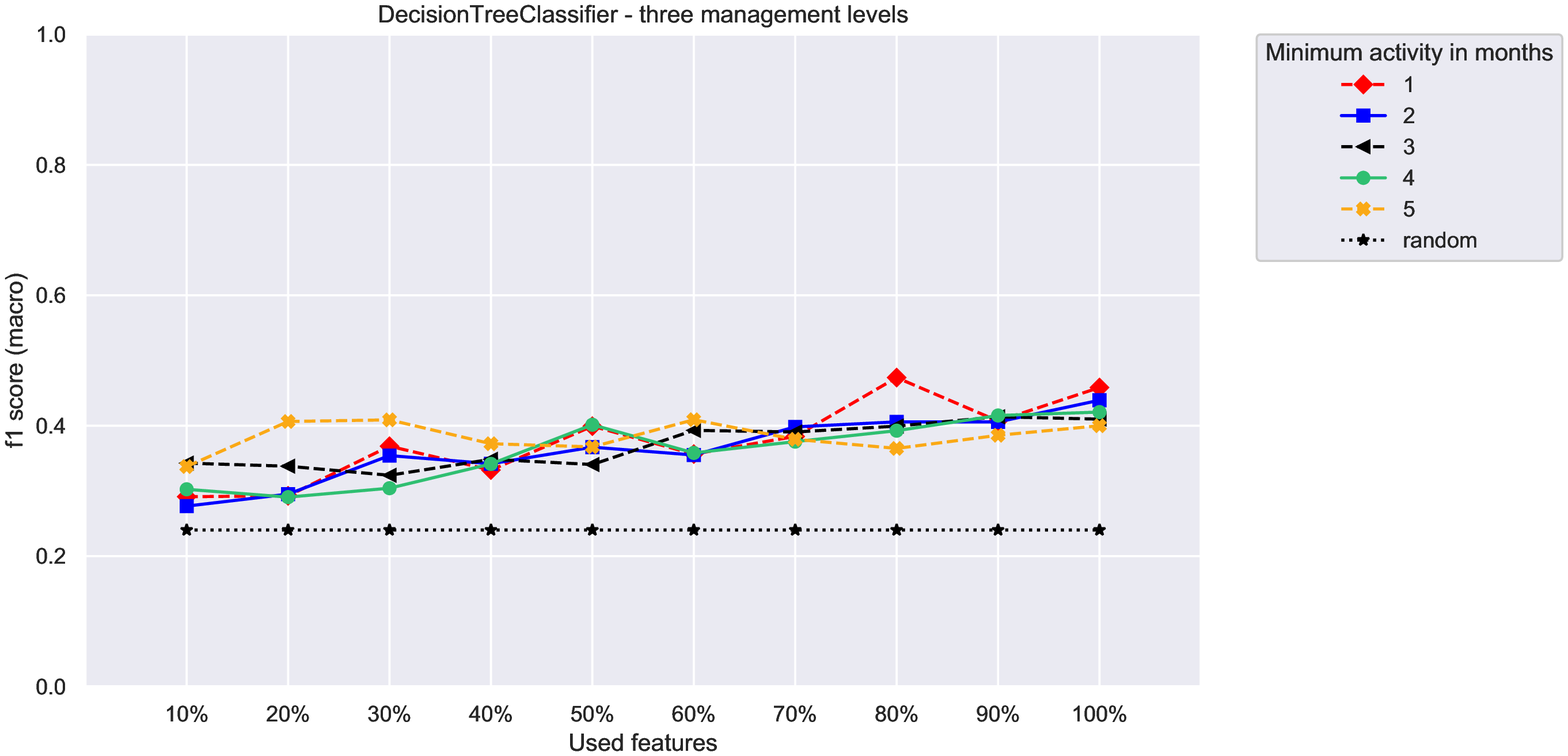}
\includegraphics[scale=0.45]{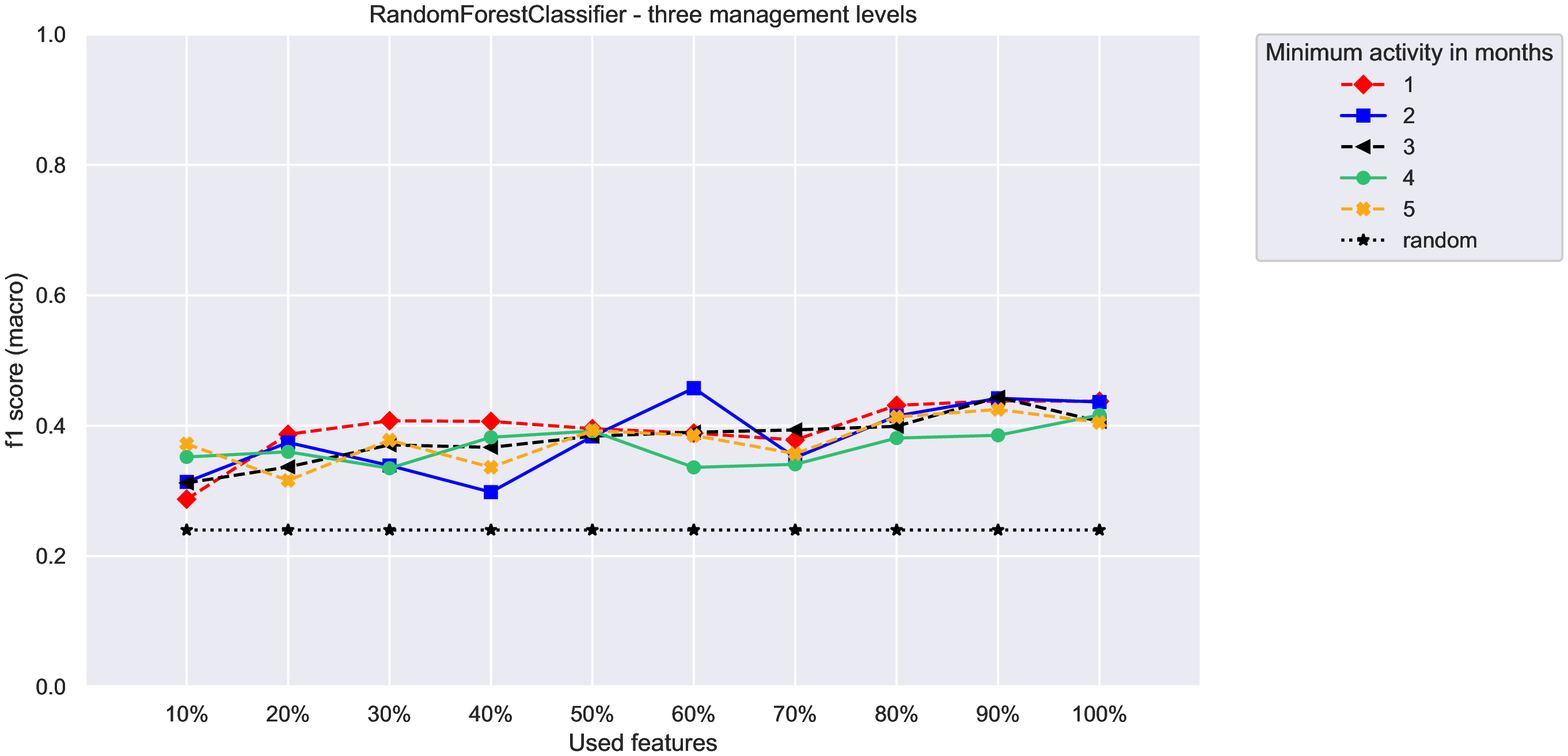}

\caption{\textit{Cont}.}

\end{figure}

\begin{figure}[H]\ContinuedFloat
\centering
\includegraphics[scale=0.45]{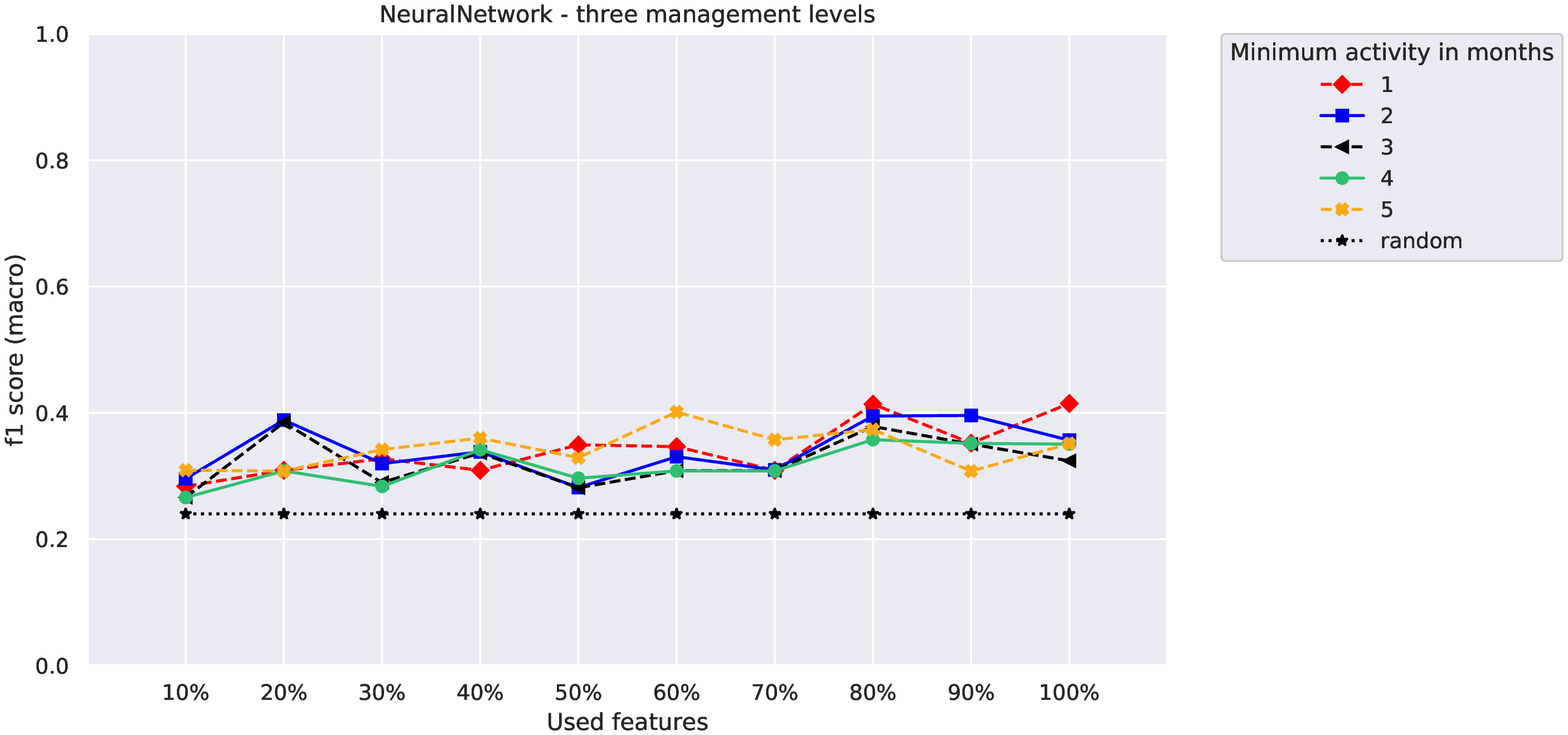}
\includegraphics[scale=0.45]{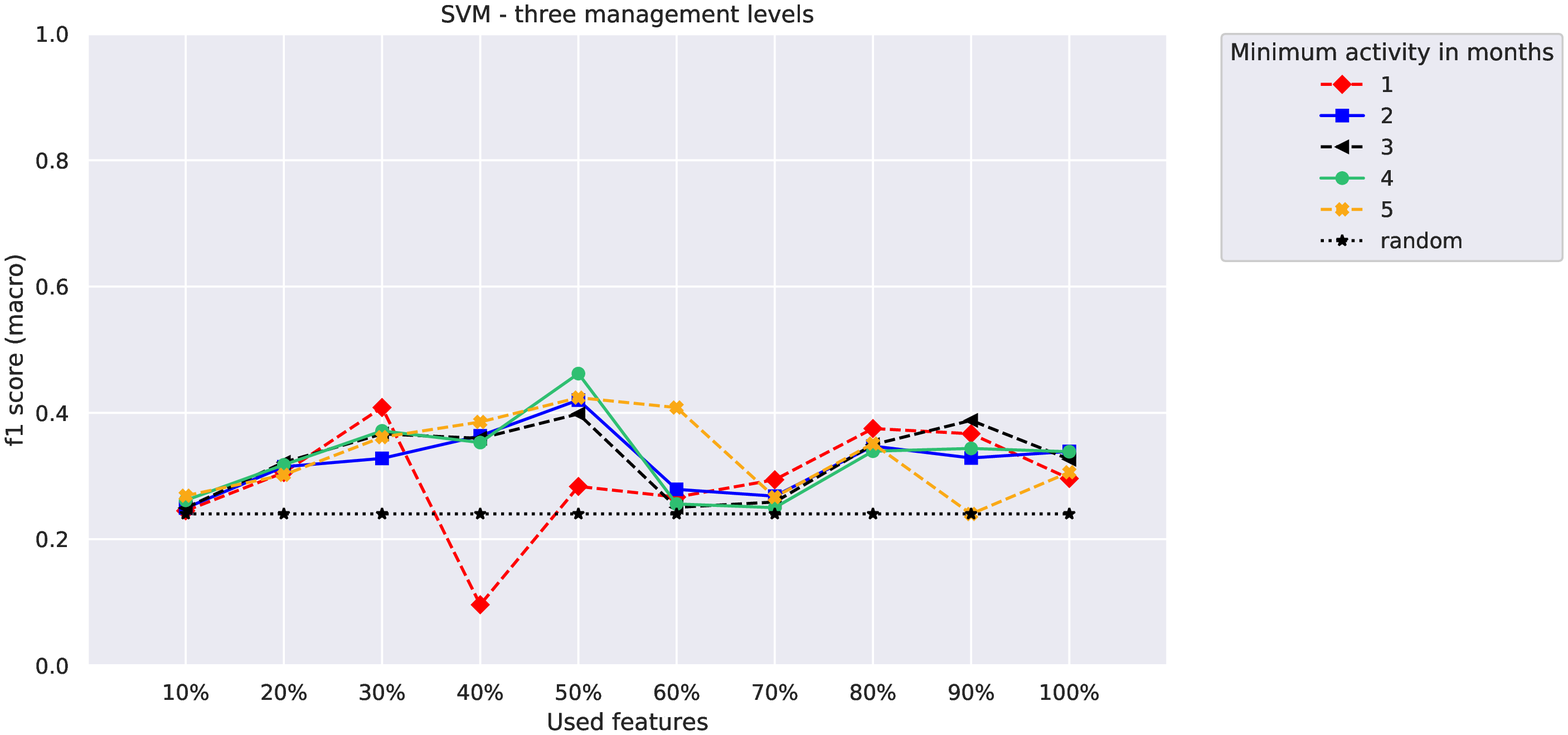}
\caption{The result of the classification of three groups for the manufacturing company~dataset.}
\label{fig:manufacturing_company_classification_3_groups}
\end{figure}
\unskip

\begin{figure}[H]
\centering
\includegraphics[scale=0.45]{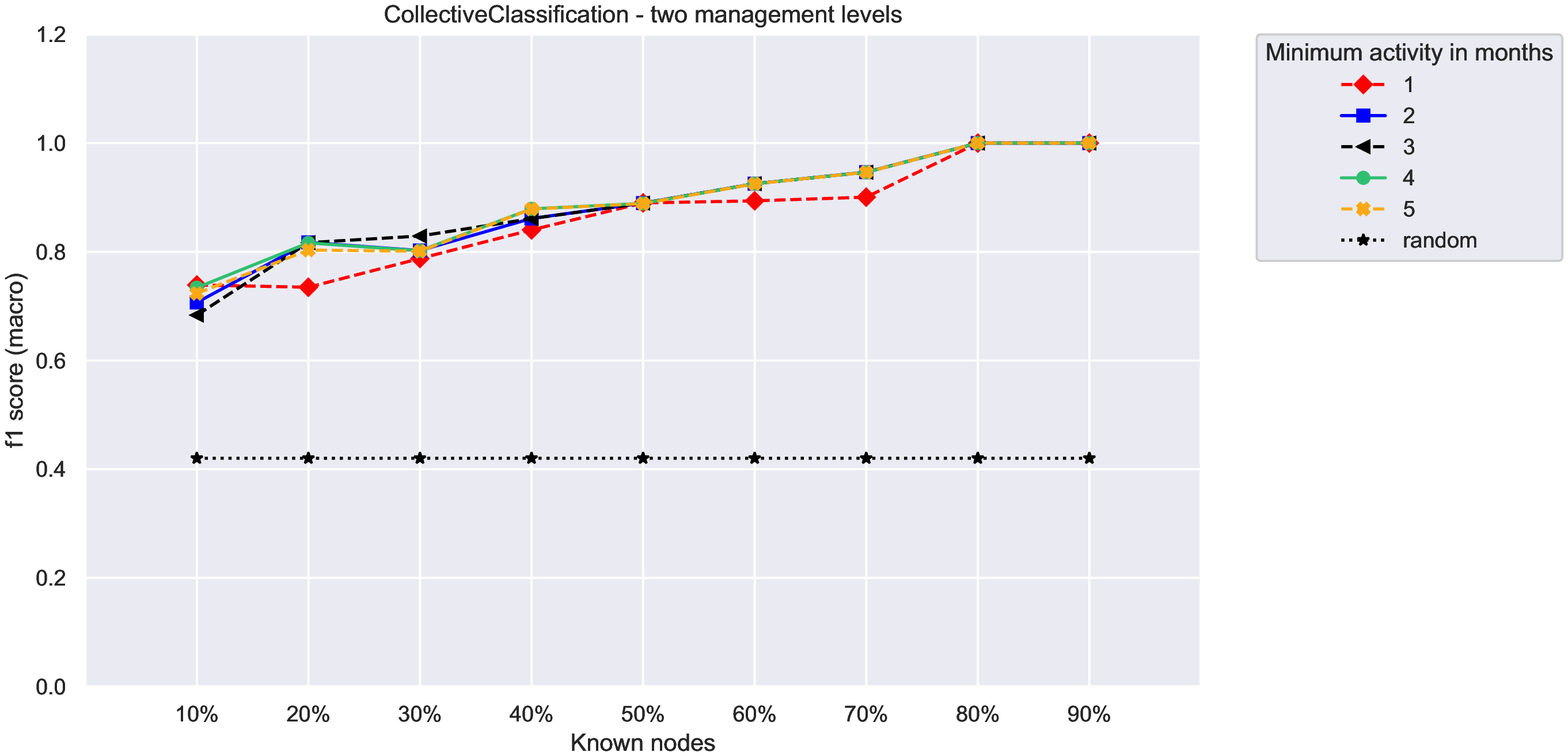}

\caption{\textit{Cont}.}

\end{figure}

\begin{figure}[H]\ContinuedFloat
\centering
\includegraphics[scale=0.45]{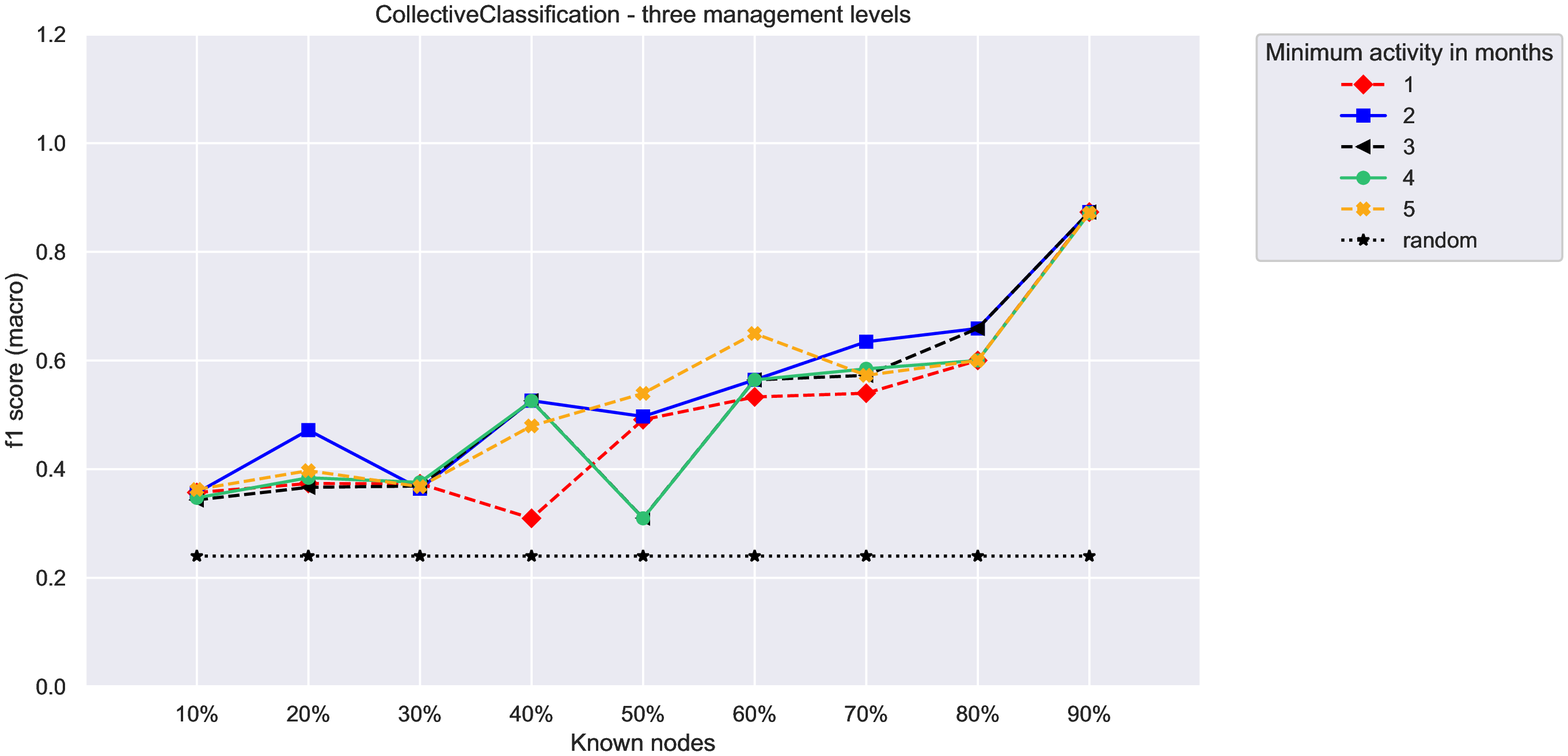}
\caption{The result of the collective classification of three groups for the manufacturing company~dataset.}
\label{fig:manufacturing_company_collective_classification}
\end{figure}
\unskip

\begin{figure}[H]
\centering
\includegraphics[scale=0.45]{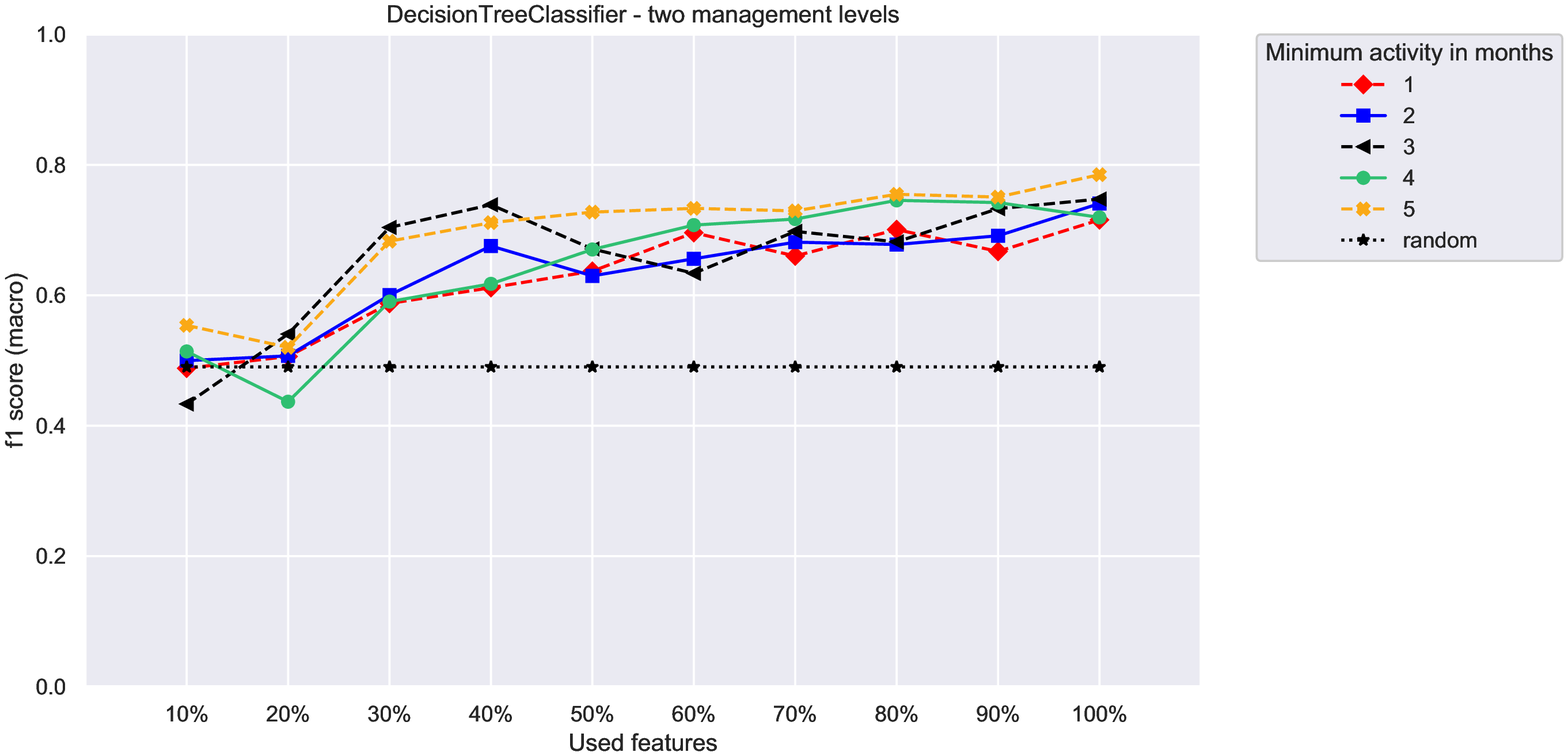}
\includegraphics[scale=0.45]{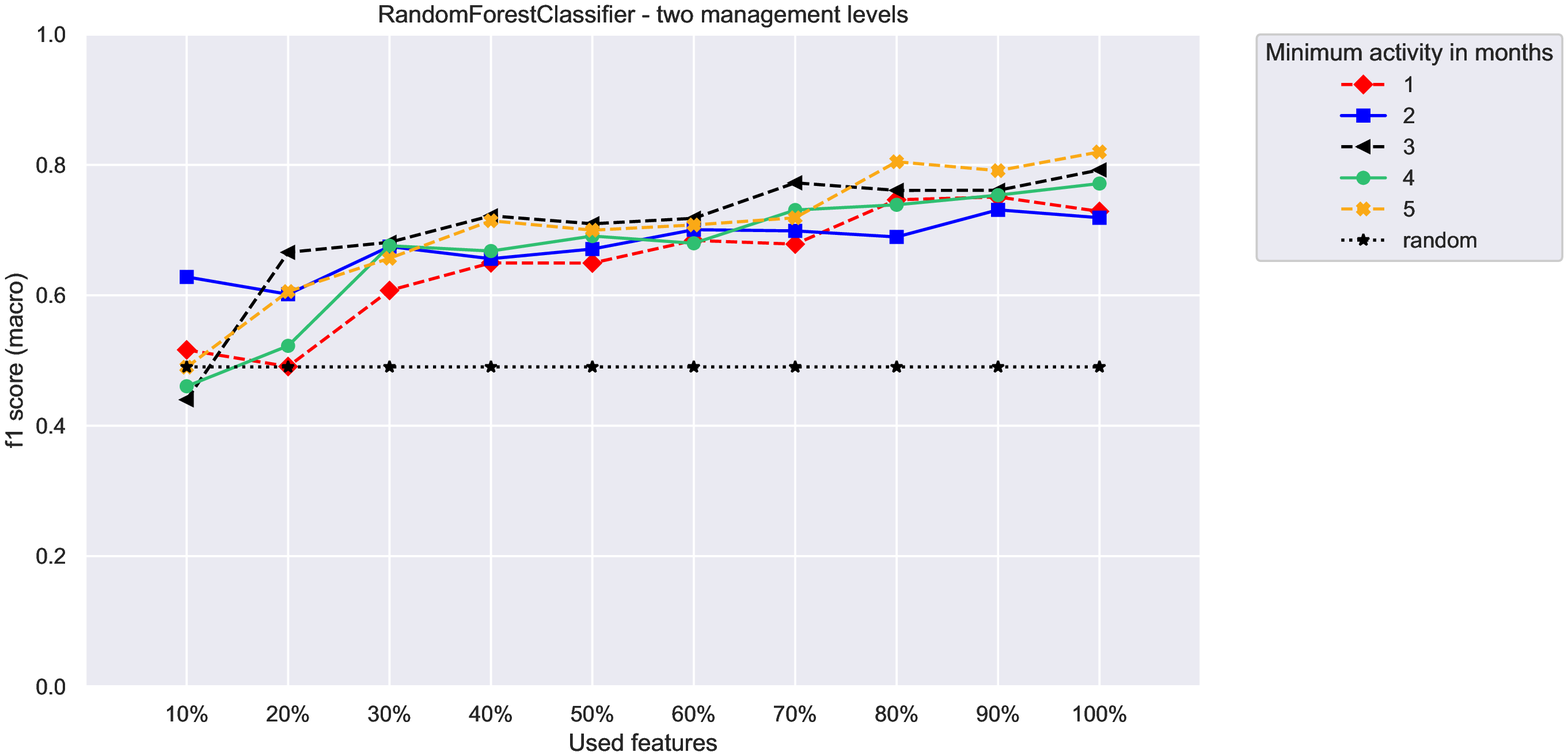}
\caption{\textit{Cont}.}

\end{figure}

\begin{figure}[H]\ContinuedFloat
\centering
\includegraphics[scale=0.45]{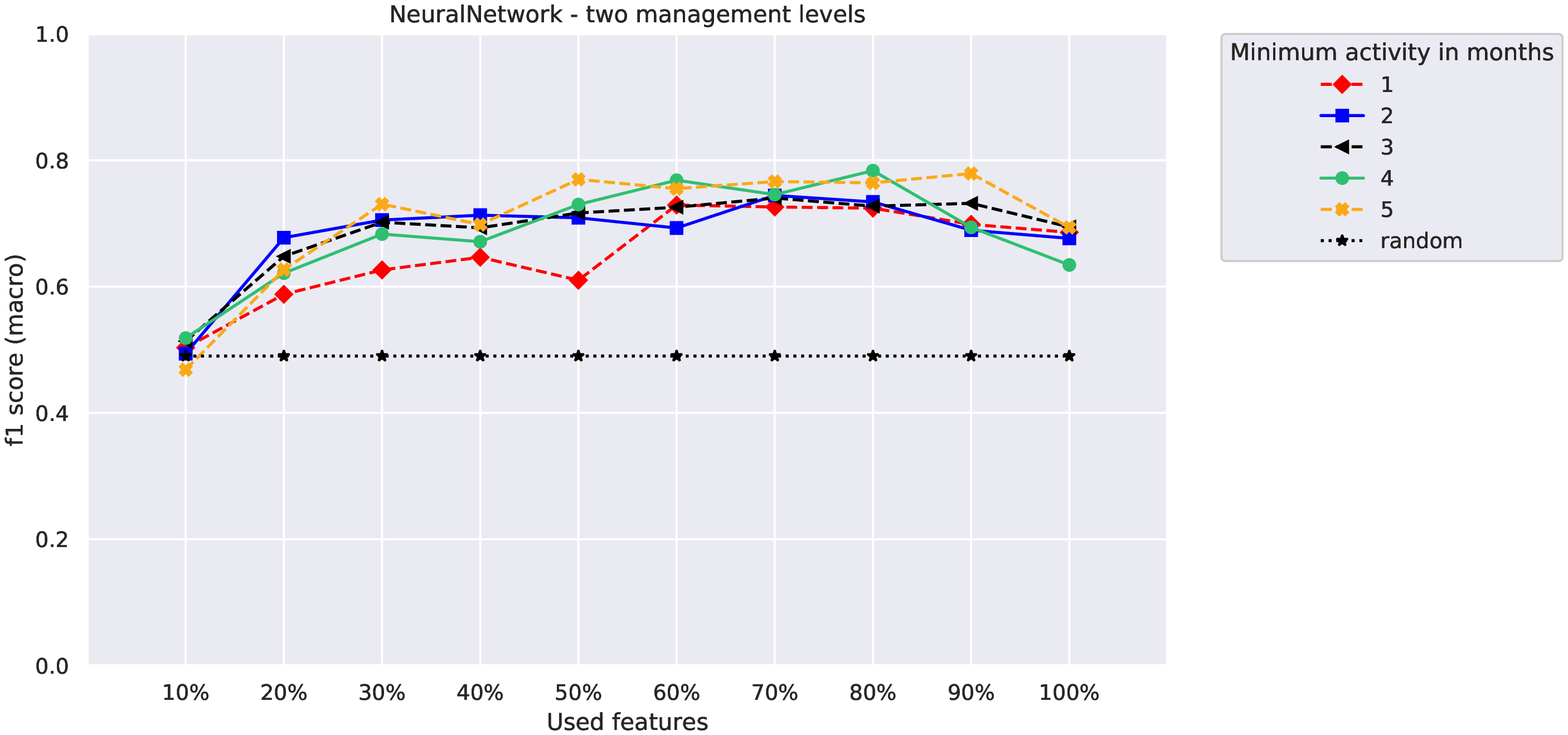}
\includegraphics[scale=0.45]{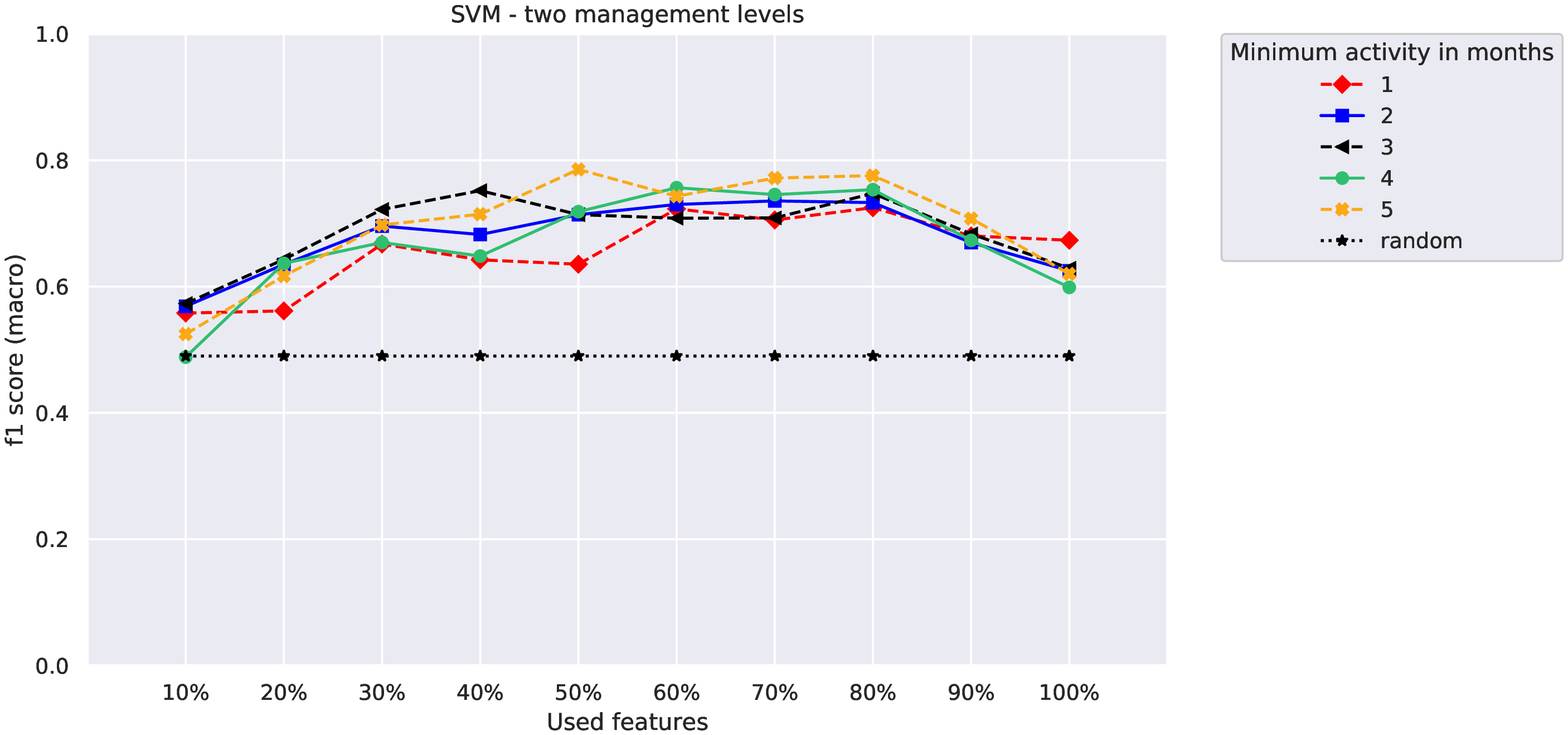}
\caption{The result of the classification of two groups for the Enron~dataset.}
\label{fig:enron_classification_2_groups}
\end{figure}
\unskip

\begin{figure}[H]
\centering
\includegraphics[scale=0.45]{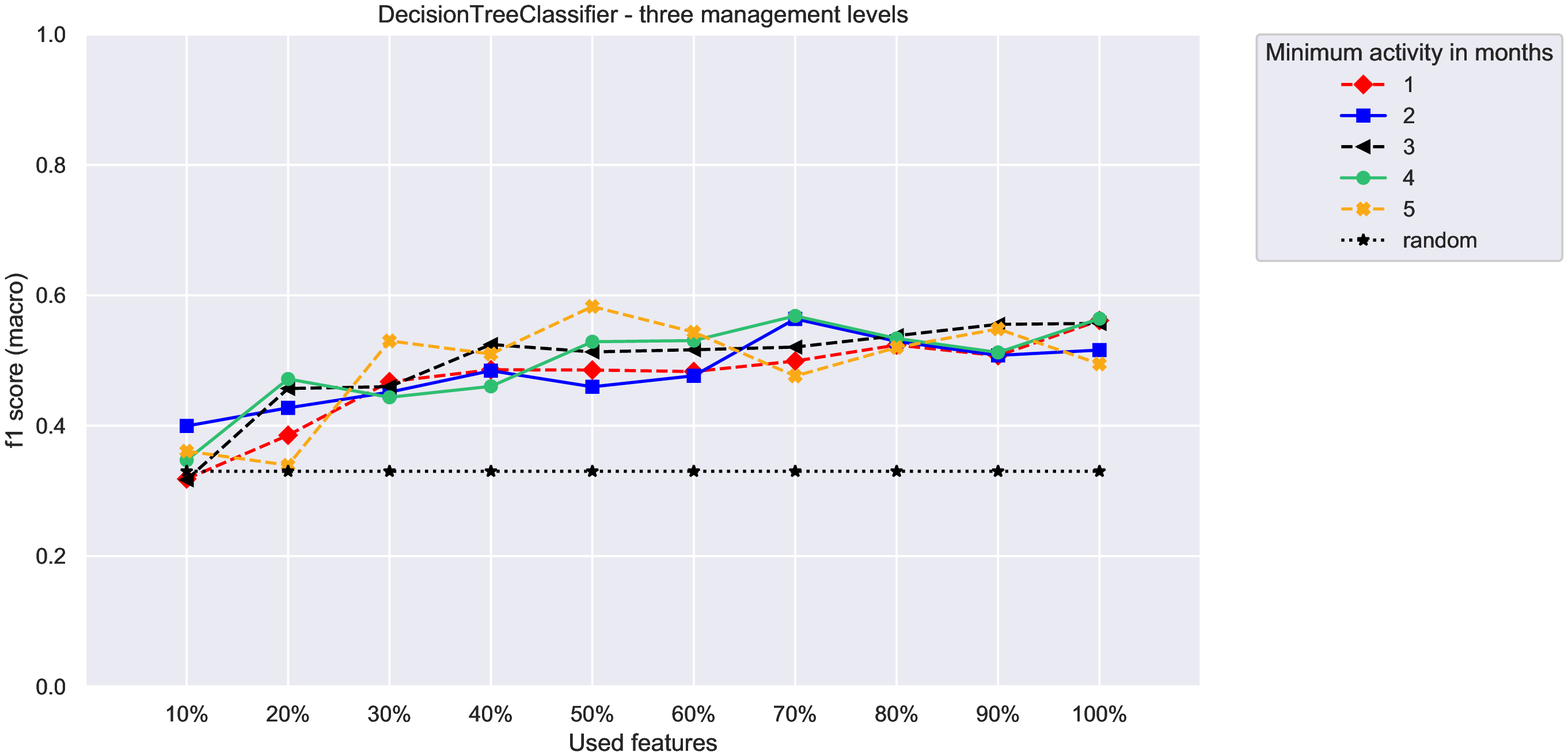}

\caption{\textit{Cont}.}
\end{figure}

\begin{figure}[H]\ContinuedFloat
\centering
\includegraphics[scale=0.45]{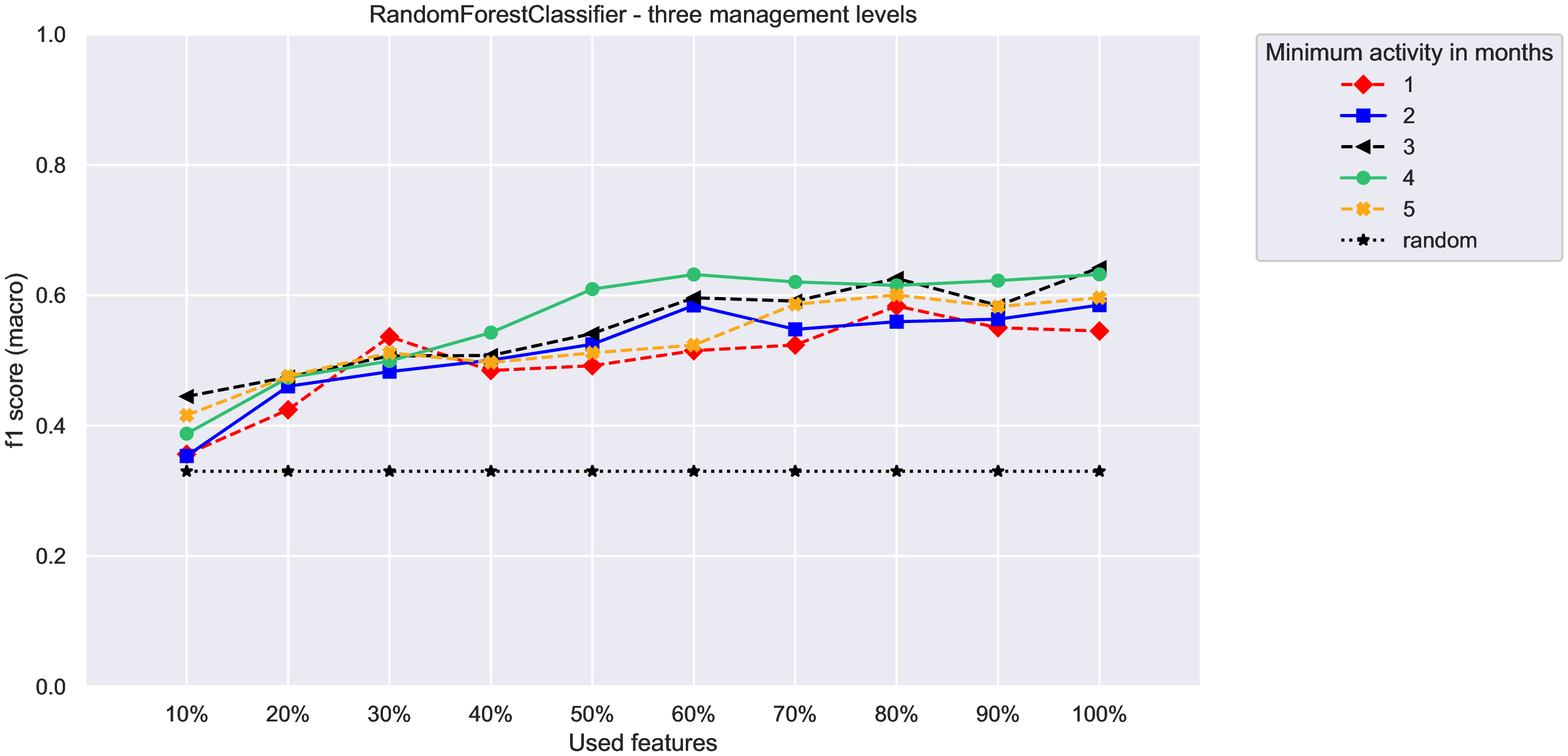}
\includegraphics[scale=0.45]{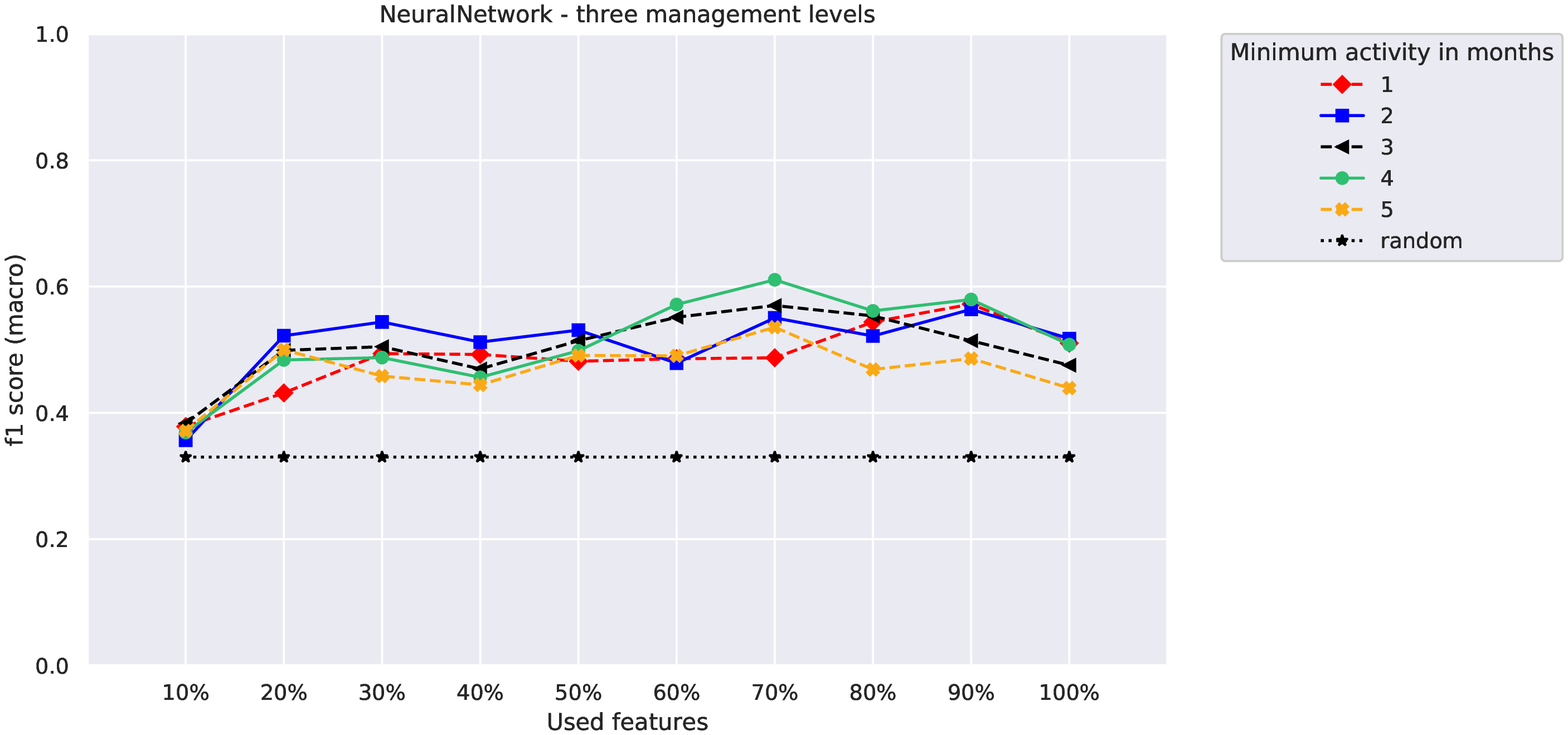}
\includegraphics[scale=0.45]{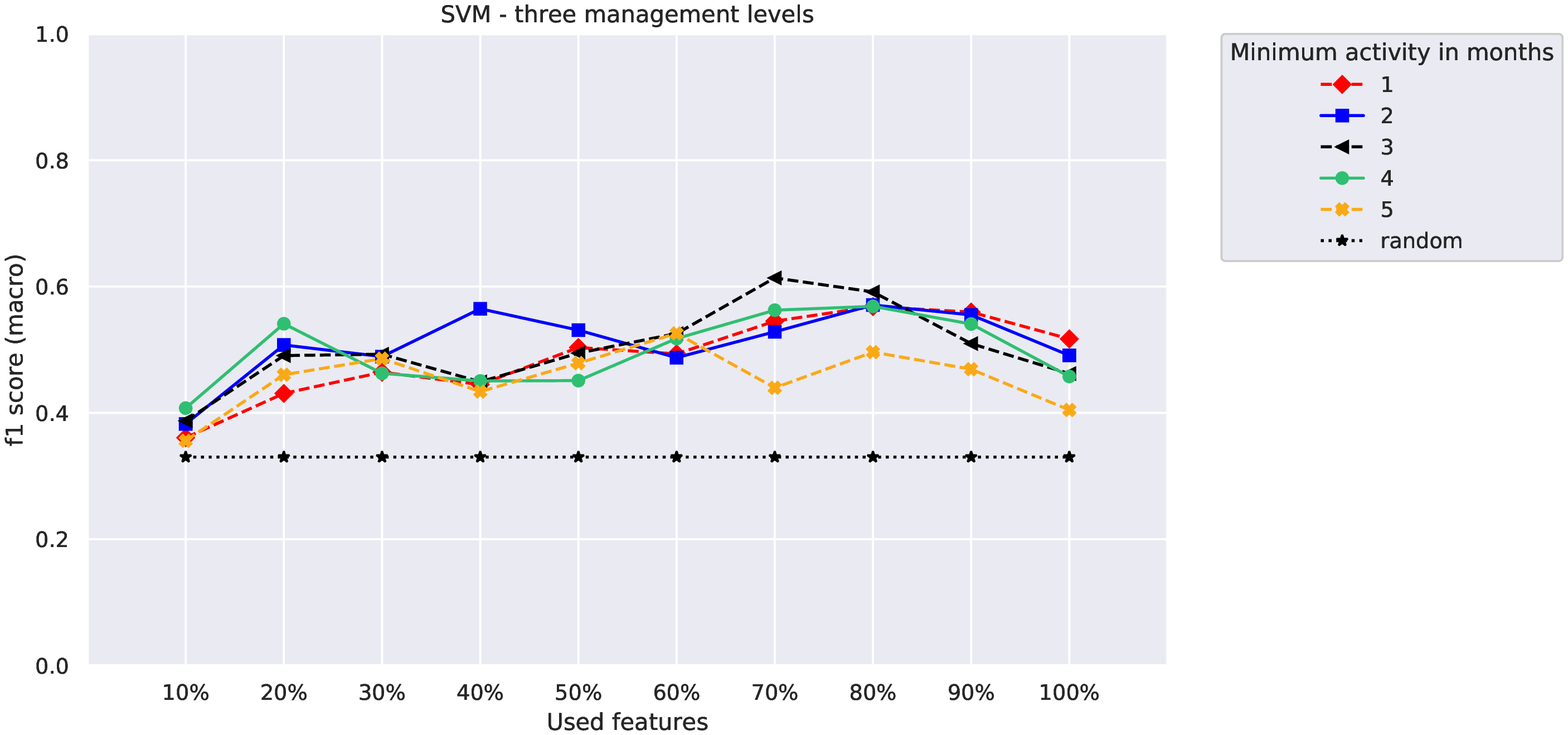}

\caption{The result of the classification of three groups for the Enron company~dataset.}
\label{fig:enron_classification_3_groups}
\end{figure}
\unskip

\unskip
\section{Discussion}

The comparison of the results for standard classification algorithms and collective classification show that the first of them cope better with balanced data such as the Enron datasets. {The conducted experiments also present that Decision Tree and Random Forest, as~well as Neural Network and SVM algorithm, have been able to obtain similar results. However, in~an organizational environment, the~possibility of the result interpretation could be highly appreciated, therefore the first two algorithms are a better solution if we would like to deeply understand the communication behavior on different organizational hierarchy levels.} Furthermore, the~presented collective classification algorithm obtained better results in the case of an unbalanced dataset.
These results indicate that the graph algorithm is able to reduce impact of majority classes and predict well contrary to the standard classification algorithms. In~addition, future research should examine if the above impact of unequal classes is also associated with some network hidden characteristics. Furthermore, it was presented that the minimum length of the period in which an employee was active influences the result, however the ideal value of the minimum activity depends on an analyzed dataset, as~well as other~parameters.

A network created from email communication may vary from organization to organization, especially regarding to the size of the company and different management models, for~example: a big international company with many employees and a complicated hierarchy could create a social network with totally different properties than a small startup with a bunch of employees and a simple structure. Moreover, in~some companies email communication could be one of many ways of passing messages, so a dataset of emails do not have to contain full information about the connection between employees in the company. These differences could cause some patterns of behavior assigned to a specific level of hierarchy, which does not have to appear in a constructed social~network.

Future work should focus on the study of communication coming from different types of companies, moreover, further research should discover which organizational structures can provide the best results of classification task. Therefore, better results could be an implication of some hidden graph properties corresponding to the way which an organization is managed, so future studies also should focus on the examination of a network structure and revealing its characteristics. The~biggest problem may be obtaining data for research due to the fact that internal communication of a company is confidential and has to be anonymized before being shared. {Another interesting approach is the attempt to use graph embeddings instead of conventional features provided to supervised learning algorithms. This way the properties of nodes will be encoded in a form of vectors making them more suitable as direct input to algorithms. Regarding the collective classification, instead of analysing particular features as the input for utility score, latent Dirichlet allocation could be used to create a utility score combining features. Lastly, the~reader would notice that the social network used in this study has been an aggregated one. This was mainly due to the fact that organizational structure of manufacturing company did not undergo any changes in the period covered by the dataset and in the case of Enron, the~structure was inferred from e-mails and no other information was known. However, for~applying proposed approach in organizations, it would be advised to verify the capabilities of temporal approach: both in the area of measures as well as networks.}



\vspace{6pt}

\supplementary{The manufacturing company email dataset alongside with corporate hierarchy has been published by the authors of this manuscript at Harvard Dataverse, see~\href{https://doi.org/10.7910/DVN/6Z3CGX}{https://doi.org/10.7910/DVN/6Z3CGX}. The~Enron dataset is available from Reference~\cite{McCallum2007Topic}.}


\authorcontributions{Conceptualization, R.M. and M.N.; methodology, R.M. and M.N.; software, M.N.; validation, R.M.; investigation, R.M. and M.N.; data curation, R.M. and M.N.; writing---original draft preparation, M.N. and R.M.; writing---review and editing, R.M.; visualization, M.N.; supervision, R.M.; project administration, R.M.; funding acquisition, R.M. All authors have read and agreed to the published version of the manuscript.}

\funding{This work was supported by the National Science Centre, Poland, project no. 2015/17/D/ST6/04046 as well as by the European Unions Horizon 2020 research and innovation programme under grant agreement no. 691152 (RENOIR) and the Polish Ministry of Science and Higher Education fund for supporting internationally co-financed projects in 2016–2019 (agreement no. 3628/H2020/2016/2).}

\acknowledgments{Authors of this manuscript would like to thank to Piotr Bródka for his valuable remarks. Moreover, we would like to express our gratitude to the Reviewers of the manuscript for their precious~feedback.}

\conflictsofinterest{The authors declare no conflict of~interest.}

\abbreviations{The following abbreviations are used in this manuscript:\\

\noindent
\begin{tabular}{@{}ll}
SNA & Social Network Analysis\\
IID & Independent and Identical Distribution \\
{SVM} & {Support Vector Machine}\\
{L-BFGS} & {limited memory Broyden-Fletcher-Goldfarb-Shanno algorithm}
\end{tabular}}


\appendixtitles{no} 
\appendix
\section{}
\unskip

\begin{table}[H]
\centering
\caption{F-score macro average for the Decision Tree classification of two levels of the hierarchy in the manufacturing company dataset. {The values which are bolded are the best results for a given minimum communication activity.}}
{
\begin{tabular}{cccccccccccc}
\toprule
& & \multicolumn{10}{c}{\textbf{Percentage of the Used Features}}\\ \midrule
& & \textbf{10\%} & \textbf{20\%} & \textbf{30\%} & \textbf{40\%} & \textbf{50\%} & \textbf{60\%} & \textbf{70\%} & \textbf{80\%} & \textbf{90\%} & \textbf{100\%} \\ \midrule
\multirow{5}{*}{\rotatebox[origin=c]{90}{Min. activity}}& 1 month & 0.5472 & 0.5270 & 0.5987 & 0.5368 & 0.5641 & 0.6130 & 0.5850 & 0.6422 & 0.6092 & \textbf{0.6568} \\ \cmidrule{2-12}
& 2 months & 0.5040 & 0.5360 & 0.5473 & 0.5938 & 0.6464 & 0.6417 & 0.6627 & \textbf{0.7039} & 0.6476 & 0.6642 \\ \cmidrule{2-12}
& 3 months & 0.5989 & 0.5336 & 0.5822 & 0.5312 & 0.5678 & 0.6391 & 0.5772 & 0.6230 & 0.6162 & \textbf{0.6642} \\ \cmidrule{2-12}
& 4 months & \textbf{0.6502} & 0.4818 & 0.5170 & 0.5315 & 0.5660 & 0.5848 & 0.5898 & 0.6234 & 0.6388 & 0.6398 \\ \cmidrule{2-12}
& 5 months & 0.6423 & 0.5622 & 0.6083 & 0.5826 & 0.6174 & 0.6387 & 0.6094 & 0.6231 & \textbf{0.6640} & 0.6388 \\ \bottomrule
\end{tabular}}
\label{tab:mc_dt_2l}
\end{table}
\unskip

\begin{table}[H]
\centering
\caption{F-score macro average for the Random Forest classification of two levels of the hierarchy in the manufacturing company dataset. {The values which are bolded are the best results for a given minimum communication activity.}}
{\begin{tabular}{cccccccccccc}
\toprule
& & \multicolumn{10}{c}{\textbf{Percentage of the Used Features}}\\ \midrule
& & \textbf{10\%} & \textbf{20\%} & \textbf{30\%} & \textbf{40\%} & \textbf{50\%} & \textbf{60\%} & \textbf{70\%} & \textbf{80\%} & \textbf{90\%} & \textbf{100\%} \\ \midrule
\multirow{5}{*}{\rotatebox[origin=c]{90}{Min. activity}}& 1 month & 0.4899 & 0.5789 & 0.5717 & 0.5786 & 0.5742 & 0.6426 & 0.5766 & 0.6290 & 0.6401 & \textbf{0.7146} \\ \cmidrule{2-12}
& 2 months & 0.6421 & 0.5311 & 0.5075 & 0.5857 & 0.6793 & 0.6542 & 0.6811 & 0.6464 & 0.6528 & \textbf{0.6963} \\ \cmidrule{2-12}
& 3 months & 0.4977 & 0.6228 & 0.5785 & 0.5849 & 0.5933 & 0.5718 & 0.6024 & 0.6054 & 0.6532 & \textbf{0.7111} \\ \cmidrule{2-12}
& 4 months & 0.5647 & 0.5104 & 0.5657 & 0.5928 & 0.6337 & 0.5848 & 0.6336 & 0.6164 & 0.6683 & \textbf{0.6895} \\ \cmidrule{2-12}
& 5 months & 0.6423 & 0.6000 & 0.5176 & 0.5923 & 0.6493 & 0.6533 & 0.6981 & 0.7425 & \textbf{0.7768} & 0.7057 \\ \bottomrule
\end{tabular}}
\label{tab:mc_rf_2l}
\end{table}
\unskip

\begin{table}[H]
\centering
\caption{F-score macro average for the Neural Network classification of two levels of the hierarchy in the manufacturing company dataset. {The values which are bolded are the best results for a given minimum communication activity.}}
{\begin{tabular}{cccccccccccc}
\toprule
& & \multicolumn{10}{c}{\textbf{Percentage of the Used Features}}\\ \midrule
& & \textbf{10\%} & \textbf{20\%} & \textbf{30\%} & \textbf{40\%} & \textbf{50\%} & \textbf{60\%} & \textbf{70\%} & \textbf{80\%} & \textbf{90\%} & \textbf{100\%} \\ \midrule
\multirow{5}{*}{\rotatebox[origin=c]{90}{Min. activity}}& 1 month & 0.4848 & 0.4635 & 0.5349 & 0.4842 & 0.5356 & 0.5595 & 0.5577 & \textbf{0.5760} & 0.5680 & 0.5627 \\ \cmidrule{2-12}
& 2 months & 0.5303 & 0.4910 & 0.5674 & 0.5532 & 0.5391 & 0.5814 & 0.5388 & 0.5986 & \textbf{0.6241} & 0.6187 \\ \cmidrule{2-12}
& 3 months & 0.5270 & 0.5210 & 0.5493 & 0.5544 & 0.5586 & 0.5798 & 0.5304 & 0.6144 & 0.6144 & \textbf{0.6247} \\ \cmidrule{2-12}
& 4 months & 0.4497 & 0.4725 & 0.4909 & 0.4845 & 0.5412 & 0.5906 & 0.5500 & 0.6013 & 0.5826 & \textbf{0.6037} \\ \cmidrule{2-12}
& 5 months & 0.4946 & 0.5275 & 0.5601 & \textbf{0.6112} & 0.5486 & 0.5431 & 0.5534 & 0.6042 & 0.5919 & 0.6000 \\ \bottomrule
\end{tabular}}
\label{tab:mc_nn_2l}
\end{table}
\unskip

\begin{table}[H]
\centering
\caption{F-score macro average for the SVM classification of two levels of the hierarchy in the manufacturing company dataset. {The values which are bolded are the best results for a given minimum communication activity.}}
{\begin{tabular}{cccccccccccc}
\toprule
& & \multicolumn{10}{c}{\textbf{Percentage of the Used Features}}\\ \midrule
& & \textbf{10\%} & \textbf{20\%} & \textbf{30\%} & \textbf{40\%} & \textbf{50\%} & \textbf{60\%} & \textbf{70\%} & \textbf{80\%} & \textbf{90\%} & \textbf{100\%} \\ \midrule
\multirow{5}{*}{\rotatebox[origin=c]{90}{Min. activity}}& 1 month & 0.4209 & 0.4703 & 0.5128 & 0.5236 & 0.4675 & 0.4998 & 0.4997 & 0.5711 & 0.5802 & \textbf{0.5977} \\ \cmidrule{2-12}
& 2 months & 0.4499 & 0.5461 & 0.5922 & 0.5609 & 0.5368 & 0.4896 & 0.5047 & 0.6087 & 0.6142 & \textbf{0.6245} \\ \cmidrule{2-12}
& 3 months & 0.4364 & 0.4896 & 0.5917 & 0.5186 & 0.5149 & 0.5027 & 0.4915 & 0.6018 & 0.6218 & \textbf{0.6517} \\ \cmidrule{2-12}
& 4 months & 0.4223 & 0.4338 & 0.5783 & 0.5416 & 0.5312 & 0.4873 & 0.5012 & 0.6001 & 0.6134 & \textbf{0.6250} \\ \cmidrule{2-12}
& 5 months & 0.4379 & 0.4493 & 0.5703 & 0.5722 & 0.5441 & 0.5363 & 0.5274 & 0.6053 & 0.6139 & \textbf{0.6501} \\ \bottomrule
\end{tabular}}
\label{tab:mc_svm_2l}
\end{table}
\unskip

\begin{table}[H]
\centering
\caption{F-score macro average for the collective classification of two levels of the hierarchy in the manufacturing company dataset. {The values which are bolded are the best results for a given minimum communication activity.}}
{\begin{tabular}{ccccccccccc}

\toprule
& & \multicolumn{9}{c}{\textbf{Percentage of the Known Nodes}}\\ \midrule
& & \textbf{10\%} & \textbf{20\%} & \textbf{30\%} & \textbf{40\%} & \textbf{50\%} & \textbf{60\%} & \textbf{70\%} & \textbf{80\%} & \textbf{90\%} \\ \midrule
\multirow{5}{*}{\rotatebox[origin=c]{90}{Min. activity}}& 1 month & 0.7389 & 0.7347 & 0.7876 & 0.8402 & 0.8899 & 0.8937 & 0.9005 & \textbf{1.0000} & \textbf{1.0000} \\ \cmidrule{2-11}
& 2 months & 0.7067 & 0.8172 & 0.8023 & 0.8613 & 0.8895 & 0.9253 & 0.9466 & \textbf{1.0000} & \textbf{1.0000} \\ \cmidrule{2-11}
& 3 months & 0.6834 & 0.8168 & 0.8290 & 0.8610 & 0.8895 & 0.9253 & 0.9463 & \textbf{1.0000} & \textbf{1.0000} \\ \cmidrule{2-11}
& 4 months & 0.7340 & 0.8163 & 0.8019 & 0.8790 & 0.8891 & 0.9253 & 0.9463 & \textbf{1.0000} & \textbf{1.0000} \\ \cmidrule{2-11}
& 5 months & 0.7233 & 0.8031 & 0.8010 & 0.8786 & 0.8887 & 0.9250 & 0.9463 & \textbf{1.0000} & \textbf{1.0000} \\ \bottomrule
\end{tabular}}
\label{tab:mc_cc_2l}
\end{table}
\unskip

\begin{table}[H]
\centering
\caption{F-score macro average for the Decision Tree classification of three levels of the hierarchy in the manufacturing company dataset. {The values which are bolded are the best results for a given minimum communication activity.}}
{\begin{tabular}{cccccccccccc}
\toprule
& & \multicolumn{10}{c}{\textbf{Percentage of the Used Features}}\\ \midrule
& & \textbf{10\%} & \textbf{20\%} & \textbf{30\%} & \textbf{40\%} & \textbf{50\%} & \textbf{60\%} & \textbf{70\%} & \textbf{80\%} & \textbf{90\%} & \textbf{100\%} \\ \midrule
\multirow{5}{*}{\rotatebox[origin=c]{90}{Min. activity}}& 1 month & 0.2910 & 0.2924 & 0.3685 & 0.3319 & 0.3992 & 0.3564 & 0.3835 & \textbf{0.4737} & 0.4071 & 0.4585 \\ \cmidrule{2-12}
& 2 months & 0.2766 & 0.2950 & 0.3541 & 0.3416 & 0.3670 & 0.3550 & 0.3981 & 0.4058 & 0.4057 & \textbf{0.4390} \\ \cmidrule{2-12}
& 3 months & 0.3425 & 0.3378 & 0.3239 & 0.3484 & 0.3404 & 0.3928 & 0.3903 & 0.3993 & \textbf{0.4136} & 0.4100 \\ \cmidrule{2-12}
& 4 months & 0.3024 & 0.2904 & 0.3041 & 0.3413 & 0.4014 & 0.3582 & 0.3755 & 0.3923 & 0.4156 & \textbf{0.4208} \\ \cmidrule{2-12}
& 5 months & 0.3373 & 0.4064 & 0.4089 & 0.3723 & 0.3674 & \textbf{0.4091} & 0.3790 & 0.3651 & 0.3852 & 0.3999 \\ \bottomrule
\end{tabular}}
\label{tab:mc_dt_3l}
\end{table}
\unskip

\begin{table}[H]
\centering
\caption{F-score macro average for the Random Forest classification of three levels of the hierarchy in the manufacturing company dataset. {The values which are bolded are the best results for a given minimum communication activity.}}
{\begin{tabular}{cccccccccccc}
\toprule
& & \multicolumn{10}{c}{\textbf{Percentage of the Used Features}}\\ \midrule
& & \textbf{10\%} & \textbf{20\%} & \textbf{30\%} & \textbf{40\%} & \textbf{50\%} & \textbf{60\%} & \textbf{70\%} & \textbf{80\%} & \textbf{90\%} & \textbf{100\%} \\ \midrule
\multirow{5}{*}{\rotatebox[origin=c]{90}{Min. activity}}& 1 month & 0.2872 & 0.3867 & 0.4075 & 0.4065 & 0.3954 & 0.3886 & 0.3780 & 0.4312 & \textbf{0.4380} & 0.4374 \\ \cmidrule{2-12}
& 2 months & 0.3137 & 0.3742 & 0.3389 & 0.2981 & 0.3834 & \textbf{0.4575} & 0.3510 & 0.4149 & 0.4420 & 0.4363 \\ \cmidrule{2-12}
& 3 months & 0.3122 & 0.3367 & 0.3705 & 0.3667 & 0.3839 & 0.3899 & 0.3935 & 0.3990 & \textbf{0.4435} & 0.4064 \\ \cmidrule{2-12}
& 4 months & 0.3520 & 0.3599 & 0.3347 & 0.3822 & 0.3917 & 0.3360 & 0.3407 & 0.3810 & 0.3852 & \textbf{0.4163} \\ \cmidrule{2-12}
& 5 months & 0.3720 & 0.3160 & 0.3778 & 0.3362 & 0.3925 & 0.3850 & 0.3571 & 0.4133 & \textbf{0.4249} & 0.4048 \\ \bottomrule
\end{tabular}}
\label{tab:mc_rf_3l}
\end{table}
\unskip

\begin{table}[H]
\centering
\caption{F-score macro average for the Neural Network classification of three levels of the hierarchy in the manufacturing company dataset. {The values which are bolded are the best results for a given minimum communication activity.}}
{\begin{tabular}{cccccccccccc}
\toprule
& & \multicolumn{10}{c}{\textbf{Percentage of the Used Features}}\\ \midrule
& & \textbf{10\%} & \textbf{20\%} & \textbf{30\%} & \textbf{40\%} & \textbf{50\%} & \textbf{60\%} & \textbf{70\%} & \textbf{80\%} & \textbf{90\%} & \textbf{100\%} \\ \midrule
\multirow{5}{*}{\rotatebox[origin=c]{90}{Min. activity}}& 1 month & 0.2838 & 0.3090 & 0.3273 & 0.3090 & 0.3494 & 0.3465 & 0.3090 & 0.4138 & 0.3520 & \textbf{0.4149} \\ \cmidrule{2-12}
& 2 months & 0.2957 & 0.3886 & 0.3197 & 0.3383 & 0.2818 & 0.3306 & 0.3097 & 0.3950 & \textbf{0.3959} & 0.3567 \\ \cmidrule{2-12}
& 3 months & 0.2662 & \textbf{0.3848} & 0.2898 & 0.3365 & 0.2814 & 0.3084 & 0.3084 & 0.3785 & 0.3506 & 0.3241 \\ \cmidrule{2-12}
& 4 months & 0.2660 & 0.3081 & 0.2836 & 0.3417 & 0.2965 & 0.3081 & 0.3081 & \textbf{0.3576} & 0.3517 & 0.3505 \\ \cmidrule{2-12}
& 5 months & 0.3093 & 0.3077 & 0.3419 & 0.3600 & 0.3291 & \textbf{0.4014} & 0.3574 & 0.3736 & 0.3077 & 0.3513 \\ \bottomrule
\end{tabular}}
\label{tab:mc_nn_3l}
\end{table}
\unskip

\begin{table}[H]
\centering
\caption{F-score macro average for the SVM classification of three levels of the hierarchy in the manufacturing company dataset. {The values which are bolded are the best results for a given minimum communication activity.}}
{\begin{tabular}{cccccccccccc}
\toprule
& & \multicolumn{10}{c}{\textbf{Percentage of the Used Features}}\\ \midrule
& & \textbf{10\%} & \textbf{20\%} & \textbf{30\%} & \textbf{40\%} & \textbf{50\%} & \textbf{60\%} & \textbf{70\%} & \textbf{80\%} & \textbf{90\%} & \textbf{100\%} \\ \midrule
\multirow{5}{*}{\rotatebox[origin=c]{90}{Min. activity}}& 1 month & 0.2449 & 0.3055 & \textbf{0.4085} & 0.0962 & 0.2835 & 0.2667 & 0.2941 & 0.3752 & 0.3669 & 0.2960 \\ \cmidrule{2-12}
& 2 months & 0.2493 & 0.3147 & 0.3279 & 0.3637 & \textbf{0.4203} & 0.2789 & 0.2681 & 0.3474 & 0.3286 & 0.3389 \\ \cmidrule{2-12}
& 3 months & 0.2495 & 0.3218 & 0.3666 & 0.3597 & \textbf{0.3984} & 0.2504 & 0.2588 & 0.3497 & 0.3881 & 0.3265 \\ \cmidrule{2-12}
& 4 months & 0.2615 & 0.3177 & 0.3715 & 0.3530 & \textbf{0.4622} & 0.2558 & 0.2499 & 0.3393 & 0.3437 & 0.3385 \\ \cmidrule{2-12}
& 5 months & 0.2688 & 0.3018 & 0.3615 & 0.3856 & \textbf{0.4241} & 0.4084 & 0.2666 & 0.3514 & 0.2399 & 0.3060 \\ \bottomrule
\end{tabular}}
\label{tab:mc_svm_3l}
\end{table}
\unskip

\begin{table}[H]
\centering
\caption{F-score macro average for the collective classification of three levels of the hierarchy in the manufacturing company dataset. {The values which are bolded are the best results for a given minimum communication activity.}}
{\begin{tabular}{ccccccccccc}

\toprule
& & \multicolumn{9}{c}{\textbf{Percentage of the Known Nodes}}\\ \midrule
& & \textbf{10\%} & \textbf{20\%} & \textbf{30\%} & \textbf{40\%} & \textbf{50\%} & \textbf{60\%} & \textbf{70\%} & \textbf{80\%} & \textbf{90\%} \\ \midrule
\multirow{5}{*}{\rotatebox[origin=c]{90}{Min. activity}}& 1 month & 0.3570 & 0.3735 & 0.3731 & 0.3095 & 0.4910 & 0.5328 & 0.5397 & 0.6000 & \textbf{0.8730} \\ \cmidrule{2-11}
& 2 months & 0.3564 & 0.4718 & 0.3640 & 0.5260 & 0.4969 & 0.5642 & 0.6344 & 0.6589 & \textbf{0.8730} \\ \cmidrule{2-11}
& 3 months & 0.3433 & 0.3665 & 0.3689 & 0.5258 & 0.3099 & 0.5642 & 0.5726 & 0.6589 & \textbf{0.8730} \\ \cmidrule{2-11}
& 4 months & 0.3472 & 0.3841 & 0.3757 & 0.5255 & 0.3095 & 0.5642 & 0.5846 & 0.6000 & \textbf{0.8713} \\ \cmidrule{2-11}
& 5 months & 0.3626 & 0.3972 & 0.3678 & 0.4795 & 0.5394 & 0.6496 & 0.5726 & 0.6000 & \textbf{0.8713} \\ \bottomrule
\end{tabular}}
\label{tab:mc_cc_3l}
\end{table}
\unskip

\begin{table}[H]
\centering
\caption{F-score macro average for the Decision Tree classification of two levels of the hierarchy in the Enron dataset. {The values which are bolded are the best results for a given minimum communication activity.}}
{\begin{tabular}{cccccccccccc}
\toprule
& & \multicolumn{10}{c}{\textbf{Percentage of the Used Features}}\\ \midrule
& & \textbf{10\%} & \textbf{20\%} & \textbf{30\%} & \textbf{40\%} & \textbf{50\%} & \textbf{60\%} & \textbf{70\%} & \textbf{80\%} & \textbf{90\%} & \textbf{100\%} \\ \midrule
\multirow{5}{*}{\rotatebox[origin=c]{90}{Min. activity}}& 1 month & 0.4880 & 0.5061 & 0.5874 & 0.6116 & 0.6369 & 0.6959 & 0.6602 & 0.7007 & 0.6672 & \textbf{0.7156} \\ \cmidrule{2-12}
& 2 months & 0.4997 & 0.5070 & 0.6003 & 0.6755 & 0.6297 & 0.6559 & 0.6814 & 0.6778 & 0.6912 & \textbf{0.7406} \\ \cmidrule{2-12}
& 3 months & 0.4332 & 0.5405 & 0.7040 & 0.7388 & 0.6710 & 0.6336 & 0.6979 & 0.6818 & 0.7328 & \textbf{0.7475} \\ \cmidrule{2-12}
& 4 months & 0.5138 & 0.4367 & 0.5903 & 0.6175 & 0.6702 & 0.7076 & 0.7168 & \textbf{0.7455} & 0.7421 & 0.7194 \\ \cmidrule{2-12}
& 5 months & 0.5538 & 0.5201 & 0.6829 & 0.7113 & 0.7275 & 0.7332 & 0.7294 & 0.7548 & 0.7506 & \textbf{0.7849} \\ \bottomrule
\end{tabular}}
\label{tab:enron_dt_2l}
\end{table}
\unskip

\begin{table}[H]
\centering
\caption{F-score macro average for the Random Forest classification of two levels of the hierarchy in the Enron dataset. {The values which are bolded are the best results for a given minimum communication~activity.}}
{\begin{tabular}{cccccccccccc}
\toprule
& & \multicolumn{10}{c}{\textbf{Percentage of the Used Features}}\\ \midrule
& & \textbf{10\%} & \textbf{20\%} & \textbf{30\%} & \textbf{40\%} & \textbf{50\%} & \textbf{60\%} & \textbf{70\%} & \textbf{80\%} & \textbf{90\%} & \textbf{100\%} \\ \midrule
\multirow{5}{*}{\rotatebox[origin=c]{90}{Min. activity}}& 1 month & 0.5162 & 0.4907 & 0.6073 & 0.6496 & 0.6493 & 0.6841 & 0.6784 & 0.7464 & \textbf{0.7507} & 0.7286 \\ \cmidrule{2-12}
& 2 months & 0.6281 & 0.6015 & 0.6747 & 0.6560 & 0.6709 & 0.7005 & 0.6987 & 0.6894 & \textbf{0.7310} & 0.7189 \\ \cmidrule{2-12}
& 3 months & 0.4397 & 0.6658 & 0.6812 & 0.7218 & 0.7096 & 0.7180 & 0.7724 & 0.7608 & 0.7609 & \textbf{0.7922} \\ \cmidrule{2-12}
& 4 months & 0.4604 & 0.5224 & 0.6759 & 0.6678 & 0.6908 & 0.6797 & 0.7307 & 0.7387 & 0.7535 & \textbf{0.7713} \\ \cmidrule{2-12}
& 5 months & 0.4897 & 0.6057 & 0.6567 & 0.7142 & 0.6998 & 0.7078 & 0.7188 & 0.8047 & 0.7911 & \textbf{0.8198} \\ \bottomrule
\end{tabular}}
\label{tab:enron_rf_2l}
\end{table}
\unskip

\begin{table}[H]
\centering
\caption{F-score macro average for the Neural Network classification of two levels of the hierarchy in the Enron dataset. {The values which are bolded are the best results for a given minimum communication~activity.}}
{\begin{tabular}{cccccccccccc}
\toprule
& & \multicolumn{10}{c}{\textbf{Percentage of the Used Features}}\\ \midrule
& & \textbf{10\%} & \textbf{20\%} & \textbf{30\%} & \textbf{40\%} & \textbf{50\%} & \textbf{60\%} & \textbf{70\%} & \textbf{80\%} & \textbf{90\%} & \textbf{100\%} \\ \midrule
\multirow{5}{*}{\rotatebox[origin=c]{90}{Min. activity}}& 1 month & 0.5032 & 0.5877 & 0.6263 & 0.6464 & 0.6100 & \textbf{0.7291} & 0.7260 & 0.7241 & 0.6985 & 0.6858 \\ \cmidrule{2-12}
& 2 months & 0.4938 & 0.6774 & 0.7053 & 0.7130 & 0.7089 & 0.6927 & \textbf{0.7443} & 0.7340 & 0.6892 & 0.6763 \\ \cmidrule{2-12}
& 3 months & 0.5135 & 0.6478 & 0.7019 & 0.6930 & 0.7165 & 0.7256 & \textbf{0.7402} & 0.7273 & 0.7318 & 0.6944 \\ \cmidrule{2-12}
& 4 months & 0.5184 & 0.6212 & 0.6831 & 0.6710 & 0.7299 & 0.7684 & 0.7456 & \textbf{0.7835} & 0.6937 & 0.6342 \\ \cmidrule{2-12}
& 5 months & 0.4684 & 0.6268 & 0.7305 & 0.6982 & 0.7697 & 0.7550 & 0.7663 & 0.7642 & \textbf{0.7790} & 0.6936 \\ \bottomrule
\end{tabular}}
\label{tab:enron_nn_2l}
\end{table}
\unskip

\begin{table}[H]
\centering
\caption{F-score macro average for the SVM classification of two levels of the hierarchy in the Enron dataset. {The values which are bolded are the best results for a given minimum communication activity.}}
{\begin{tabular}{cccccccccccc}
\toprule
& & \multicolumn{10}{c}{\textbf{Percentage of the Used Features}}\\ \midrule
& & \textbf{10\%} & \textbf{20\%} & \textbf{30\%} & \textbf{40\%} & \textbf{50\%} & \textbf{60\%} & \textbf{70\%} & \textbf{80\%} & \textbf{90\%} & \textbf{100\%} \\ \midrule
\multirow{5}{*}{\rotatebox[origin=c]{90}{Min. activity}}& 1 month & 0.5580 & 0.5615 & 0.6673 & 0.6423 & 0.6354 & 0.7230 & 0.7054 & \textbf{0.7249} & 0.6802 & 0.6733 \\ \cmidrule{2-12}
& 2 months & 0.5687 & 0.6350 & 0.6958 & 0.6824 & 0.7138 & 0.7300 & \textbf{0.7356} & 0.7329 & 0.6696 & 0.6249 \\ \cmidrule{2-12}
& 3 months & 0.5731 & 0.6430 & 0.7221 & \textbf{0.7519} & 0.7139 & 0.7082 & 0.7087 & 0.7475 & 0.6833 & 0.6290 \\ \cmidrule{2-12}
& 4 months & 0.4883 & 0.6367 & 0.6698 & 0.6483 & 0.7186 & \textbf{0.7565} & 0.7456 & 0.7534 & 0.6731 & 0.5987 \\ \cmidrule{2-12}
& 5 months & 0.5248 & 0.6167 & 0.6975 & 0.7146 & \textbf{0.7855} & 0.7430 & 0.7718 & 0.7757 & 0.7077 & 0.6200 \\ \bottomrule
\end{tabular}}
\label{tab:enron_svm_2l}
\end{table}
\unskip

\begin{table}[H]
\centering
\caption{F-score macro average for the collective classification of two levels of the hierarchy in the Enron dataset. {The values which are bolded are the best results for a given minimum communication~activity.}}
{\begin{tabular}{ccccccccccc}
\toprule
& & \multicolumn{9}{c}{\textbf{Percentage of the Known Nodes}}\\ \midrule
& & \textbf{10\%} & \textbf{20\%} & \textbf{30\%} & \textbf{40\%} & \textbf{50\%} & \textbf{60\%} & \textbf{70\%} & \textbf{80\%} & \textbf{90\%} \\ \midrule
\multirow{5}{*}{\rotatebox[origin=c]{90}{Min. activity}}& 1 month & 0.5285 & 0.5554 & 0.7629 & 0.6976 & 0.7812 & 0.7896 & 0.7896 & 0.7684 & \textbf{0.8286} \\ \cmidrule{2-11}
& 2 months & 0.5689 & 0.5920 & 0.6865 & 0.6727 & 0.6718 & 0.7350 & 0.7380 & 0.8120 & \textbf{0.8901} \\ \cmidrule{2-11}
& 3 months & 0.5863 & 0.5385 & 0.6334 & 0.6627 & 0.6573 & 0.6821 & 0.6641 & 0.6761 & \textbf{0.8000} \\ \cmidrule{2-11}
& 4 months & 0.5605 & 0.5359 & 0.5972 & 0.6117 & 0.6725 & 0.6824 & 0.7618 & \textbf{0.8750} & 0.8730 \\ \cmidrule{2-11}
& 5 months & 0.6141 & 0.5163 & 0.6003 & 0.6260 & 0.6581 & 0.5826 & 0.7429 & 0.7500 & \textbf{1.0000} \\ \bottomrule
\end{tabular}}
\label{tab:enron_cc_2l}
\end{table}
\unskip

\begin{table}[H]
\centering
\caption{F-score macro average for the Decision Tree classification of three levels of the hierarchy in the Enron dataset. {The values which are bolded are the best results for a given minimum communication~activity.}}
{\begin{tabular}{cccccccccccc}
\toprule
& & \multicolumn{10}{c}{\textbf{Percentage of the Used Features}}\\ \midrule
& & \textbf{10\%} & \textbf{20\%} & \textbf{30\%} & \textbf{40\%} & \textbf{50\%} & \textbf{60\%} & \textbf{70\%} & \textbf{80\%} & \textbf{90\%} & \textbf{100\%} \\ \midrule
\multirow{5}{*}{\rotatebox[origin=c]{90}{Min. activity}}& 1 month & 0.3183 & 0.3853 & 0.4674 & 0.4857 & 0.4853 & 0.4830 & 0.4991 & 0.5236 & 0.5074 & \textbf{0.5613} \\ \cmidrule{2-12}
& 2 months & 0.3995 & 0.4273 & 0.4514 & 0.4843 & 0.4596 & 0.4765 & \textbf{0.5637} & 0.5293 & 0.5078 & 0.5159 \\ \cmidrule{2-12}
& 3 months & 0.3170 & 0.4570 & 0.4598 & 0.5245 & 0.5131 & 0.5164 & 0.5205 & 0.5380 & 0.5554 & \textbf{0.5568} \\ \cmidrule{2-12}
& 4 months & 0.3471 & 0.4716 & 0.4435 & 0.4603 & 0.5287 & 0.5305 & \textbf{0.5683} & 0.5337 & 0.5123 & 0.5641 \\ \cmidrule{2-12}
& 5 months & 0.3609 & 0.3393 & 0.5300 & 0.5099 & \textbf{0.5827} & 0.5434 & 0.4761 & 0.5199 & 0.5484 & 0.4946 \\ \bottomrule
\end{tabular}}
\label{tab:enron_dt_3l}
\end{table}
\unskip

\begin{table}[H]
\centering
\caption{F-score macro average for the Random Forest classification of three levels of the hierarchy in the Enron dataset. {The values which are bolded are the best results for a given minimum communication~activity.}}
{\begin{tabular}{cccccccccccc}
\toprule
& & \multicolumn{10}{c}{\textbf{Percentage of the Used Features}}\\ \midrule
& & \textbf{10\%} & \textbf{20\%} & \textbf{30\%} & \textbf{40\%} & \textbf{50\%} & \textbf{60\%} & \textbf{70\%} & \textbf{80\%} & \textbf{90\%} & \textbf{100\%} \\ \midrule
\multirow{5}{*}{\rotatebox[origin=c]{90}{Min. activity}}& 1 month & 0.3562 & 0.4244 & 0.5364 & 0.4846 & 0.4919 & 0.5150 & 0.5237 & \textbf{0.5835} & 0.5502 & 0.5451 \\ \cmidrule{2-12}
& 2 months & 0.3535 & 0.4604 & 0.4828 & 0.5005 & 0.5248 & 0.5843 & 0.5477 & 0.5594 & 0.5633 & \textbf{0.5849} \\ \cmidrule{2-12}
& 3 months & 0.4447 & 0.4745 & 0.5070 & 0.5076 & 0.5413 & 0.5961 & 0.5910 & 0.6258 & 0.5842 & \textbf{0.6423} \\ \cmidrule{2-12}
& 4 months & 0.3877 & 0.4735 & 0.4991 & 0.5427 & 0.6094 & 0.6318 & 0.6204 & 0.6151 & 0.6224 & \textbf{0.6320} \\ \cmidrule{2-12}
& 5 months & 0.4158 & 0.4758 & 0.5117 & 0.4970 & 0.5116 & 0.5235 & 0.5862 & \textbf{0.6004} & 0.5826 & 0.5963 \\ \bottomrule
\end{tabular}}
\label{tab:enron_rf_3l}
\end{table}
\unskip

\begin{table}[H]
\centering
\caption{F-score macro average for the Neural Network classification of three levels of the hierarchy in the Enron dataset. {The values which are bolded are the best results for a given minimum communication activity.}}
{\begin{tabular}{cccccccccccc}
\toprule
& & \multicolumn{10}{c}{\textbf{Percentage of the Used Features}}\\ \midrule
& & \textbf{10\%} & \textbf{20\%} & \textbf{30\%} & \textbf{40\%} & \textbf{50\%} & \textbf{60\%} & \textbf{70\%} & \textbf{80\%} & \textbf{90\%} & \textbf{100\%} \\ \midrule
\multirow{5}{*}{\rotatebox[origin=c]{90}{Min. activity}}& 1 month & 0.3783 & 0.4316 & 0.4938 & 0.4925 & 0.4817 & 0.4857 & 0.4872 & 0.5439 & \textbf{0.5721} & 0.5104 \\ \cmidrule{2-12}
& 2 months & 0.3566 & 0.5221 & 0.5439 & 0.5121 & 0.5310 & 0.4787 & 0.5504 & 0.5217 & \textbf{0.5639} & 0.5176 \\ \cmidrule{2-12}
& 3 months & 0.3835 & 0.4990 & 0.5048 & 0.4696 & 0.5142 & 0.5514 & \textbf{0.5700} & 0.5535 & 0.5142 & 0.4754 \\ \cmidrule{2-12}
& 4 months & 0.3688 & 0.4838 & 0.4875 & 0.4563 & 0.4983 & 0.5716 & \textbf{0.6107} & 0.5615 & 0.5795 & 0.5080 \\ \cmidrule{2-12}
& 5 months & 0.3714 & 0.4998 & 0.4583 & 0.4443 & 0.4909 & 0.4904 & \textbf{0.5356} & 0.4687 & 0.4860 & 0.4393 \\ \bottomrule
\end{tabular}}
\label{tab:enron_nn_3l}
\end{table}
\unskip

\begin{table}[H]
\centering
\caption{F-score macro average for the SVM classification of three levels of the hierarchy in the Enron dataset. {The values which are bolded are the best results for a given minimum communication~activity.}}
{\begin{tabular}{cccccccccccc}
\toprule
& & \multicolumn{10}{c}{\textbf{Percentage of the Used Features}}\\ \midrule
& & \textbf{10\%} & \textbf{20\%} & \textbf{30\%} & \textbf{40\%} & \textbf{50\%} & \textbf{60\%} & \textbf{70\%} & \textbf{80\%} & \textbf{90\%} & \textbf{100\%} \\ \midrule
\multirow{5}{*}{\rotatebox[origin=c]{90}{Min. activity}}& 1 month & 0.3607 & 0.4308 & 0.4644 & 0.4448 & 0.5035 & 0.4934 & 0.5452 & \textbf{0.5682} & 0.5596 & 0.5171 \\ \cmidrule{2-12}
& 2 months & 0.3825 & 0.5075 & 0.4891 & 0.5648 & 0.5310 & 0.4874 & 0.5284 & \textbf{0.5708} & 0.5551 & 0.4910 \\ \cmidrule{2-12}
& 3 months & 0.3874 & 0.4907 & 0.4929 & 0.4492 & 0.4946 & 0.5256 & \textbf{0.6137} & 0.5915 & 0.5097 & 0.4622 \\ \cmidrule{2-12}
& 4 months & 0.4075 & 0.5411 & 0.4626 & 0.4505 & 0.4511 & 0.5177 & 0.5626 & \textbf{0.5687} & 0.5407 & 0.4576 \\ \cmidrule{2-12}
& 5 months & 0.3559 & 0.4604 & 0.4859 & 0.4337 & 0.4787 & \textbf{0.5259} & 0.4397 & 0.4962 & 0.4691 & 0.4046 \\ \bottomrule
\end{tabular}}
\label{tab:enron_svm_3l}
\end{table}
\unskip

\begin{table}[H]
\centering
\caption{F-score macro average for the collective classification of three levels of the hierarchy in the Enron dataset. {The values which are bolded are the best results for a given minimum communication~activity.}}
{\begin{tabular}{ccccccccccc}
\toprule
& & \multicolumn{9}{c}{\textbf{Percentage of the Known Nodes}}\\ \midrule
& & \textbf{10\%} & \textbf{20\%} & \textbf{30\%} & \textbf{40\%} & \textbf{50\%} & \textbf{60\%} & \textbf{70\%} & \textbf{80\%} & \textbf{90\%} \\ \midrule
\multirow{5}{*}{\rotatebox[origin=c]{90}{Min. activity}}& 1 month & 0.2826 & 0.3044 & 0.3739 & 0.4514 & 0.4730 & 0.5517 & 0.5160 & \textbf{0.5654} & 0.5596 \\ \cmidrule{2-11}
& 2 months & 0.2874 & 0.3193 & 0.3708 & 0.3933 & 0.4393 & 0.5084 & 0.5426 & 0.5333 & \textbf{0.6746} \\ \cmidrule{2-11}
& 3 months & 0.2593 & 0.3368 & 0.3681 & 0.4007 & 0.4284 & \textbf{0.5915} & 0.5753 & 0.5166 & 0.5000 \\ \cmidrule{2-11}
& 4 months & 0.2687 & 0.3077 & 0.3628 & 0.4031 & 0.5094 & 0.5105 & 0.6427 & 0.6476 & \textbf{0.7302} \\ \cmidrule{2-11}
& 5 months & 0.2725 & 0.2774 & 0.3542 & 0.3489 & 0.3954 & 0.4832 & 0.4593 & 0.4551 & \textbf{0.6111} \\ \bottomrule
\end{tabular}}
\label{tab:enron_cc_3l}
\end{table}
\unskip

\begin{table}[H]
\centering
\caption{Hyperparameter search space for supervised learning~algorithms.}
{\scalebox{0.8}[0.8]{
\begin{tabular}{ccc}
\toprule
& & \textbf{Values}\\ \midrule
\multirow{2}{*}{Decision Tree\vspace{-5pt}
}& max\_depth & 1, 2, 3, ..., 20 \\ \cmidrule{2-3}
& max\_features & 1, 2, 3, ..., 16 (manufacturing company); 1, 2, 3, ..., 15 (Enron)  \\ \midrule
\multirow{3}{*}{Random Forest\vspace{-10pt}}& max\_depth & 1, 2, 3, ..., 20 \\ \cmidrule{2-3}
& max\_features & 1, 2, 3, ..., 16 (manufacturing company); 1, 2, 3, ..., 15 (Enron)  \\ \cmidrule(lr){2-3}
& n\_estimators & 1, 2, 4, 8, 16, 32, 64, 100, 200 \\ \midrule
\multirow{2}{*}{Neural Network\vspace{-5pt}}& alpha & 0.0001, 0.001, 0.01, 1 \\ \cmidrule{2-3}
& hidden\_layer\_sizes & (13), (9), (4), (13, 9), (13, 4), (9, 4), (13, 9, 4), (4, 9, 4), (9, 13, 9), (9, 13, 4), (4, 9, 13, 9, 4)  \\  \midrule
\multirow{2}{*}{SVM\vspace{-5pt}}& degree & 3, 4, 5, 6, 7, 8, 9, 10, 15, 20 \\ \cmidrule{2-3}
& C & 1, 2, 3, 4, 5, 10, 15, 20, 25, 50, 100, 150, 200  \\ \bottomrule
\end{tabular}}}
\label{tab:hp_space}
\end{table}
\unskip

\begin{table}[H]
\centering
\caption{The best values of the hyperparameters for the models which uses the full set of features for the manufacturing company dataset with two hierarchy~levels.}
{\begin{tabular}{ccccccc}
\toprule
& & \multicolumn{5}{c}{\textbf{Min. Activity}}\\ \midrule
& & \textbf{1 month} & \textbf{2 months} & \textbf{3 months} & \textbf{4 months} & \textbf{5 months} \\ \midrule
\multirow{2}{*}{Decision Tree\vspace{-5pt}
} & max\_depth & 4 & 8 & 8 & 10 & 5 \\ \cmidrule{2-7}
& max\_features  & 3 & 2 & 1 & 9 & 1 \\ \midrule
\multirow{3}{*}{Random Forest\vspace{-10pt}} & max\_depth & 8 & 6 & 6 & 10 & 6 \\ \cmidrule{2-7}
& max\_features  & 2 & 2 & 12 & 3 & 2 \\ \cmidrule{2-7}
& n\_estimators  & 32 & 16 & 16 & 16 & 8 \\ \midrule
\multirow{2}{*}{Neural Network\vspace{-5pt}} & alpha & 0.0001 & 0.01 & 0.0001 & 0.001 & 0.01 \\ \cmidrule{2-7}
& hidden\_layer\_sizes  & (13, 4) & (13) & (13, 9, 4) & (13, 9) & (13, 9) \\ \midrule
\multirow{2}{*}{SVM\vspace{-5pt}} & degree & 20 & 20 & 20 & 15 & 20 \\ \cmidrule{2-7}
& C  & 150 & 200 & 200 & 200 & 200 \\ \bottomrule
\end{tabular}}
\label{tab:mc_best_hp_2l}
\end{table}
\unskip

\begin{table}[H]
\centering
\caption{The best values of the hyperparameters for the models which uses the full set of features for the manufacturing company dataset with three hierarchy~levels.}
{\begin{tabular}{ccccccc}
\toprule
& & \multicolumn{5}{c}{\textbf{Min. Activity}}\\ \midrule
& & \textbf{1 Month} & \textbf{2 Months} & \textbf{3 Months} & \textbf{4 Months} & \textbf{5 Months} \\ \midrule
\multirow{2}{*}{Decision Tree\vspace{-5pt}} & max\_depth & 10 & 7 & 12 & 13 & 8 \\ \cmidrule{2-7}
& max\_features  & 1 & 8 & 2 & 1 & 8 \\ \midrule
\multirow{3}{*}{Random Forest\vspace{-10pt}} & max\_depth & 7 & 5 & 7 & 10 & 10 \\ \cmidrule{2-7}
& max\_features  & 3 & 11 & 14 & 6 & 7 \\ \cmidrule{2-7}
& n\_estimators  & 2 & 4 & 16 & 32 & 8 \\ \midrule
\multirow{2}{*}{Neural Network\vspace{-5pt}} & alpha & 0.001 & 0.0001 & 0.01 & 0.01 & 0.001 \\ \cmidrule{2-7}
& hidden\_layer\_sizes  & (9, 13, 4) & (4, 9, 13, 9, 4) & (4, 9, 13, 9, 4) & (13, 9, 4) & (13, 9, 4) \\ \midrule
\multirow{2}{*}{SVM\vspace{-5pt}} & degree & 15 & 20 & 20 & 20 & 20 \\ \cmidrule{2-7}
& C  & 200 & 200 & 200 & 200 & 150 \\ \bottomrule
\end{tabular}}
\label{tab:mc_best_hp_3l}
\end{table}
\unskip

\begin{table}[H]
\centering
\caption{The best values of the hyperparameters for the models which uses the full set of features for the Enron dataset with two hierarchy~levels.}
{\begin{tabular}{ccccccc}
\toprule
& & \multicolumn{5}{c}{\textbf{Min. Activity}}\\ \midrule
& & \textbf{1 Month} & \textbf{2 Months} & \textbf{3 Months} & \textbf{4 Months} & \textbf{5 Months} \\ \midrule
\multirow{2}{*}{Decision Tree\vspace{-5pt}} & max\_depth & 2 & 8 & 7 & 3 & 2 \\ \cmidrule{2-7}
& max\_features  & 1 & 9 & 3 & 4 & 12 \\ \midrule
\multirow{3}{*}{Random Forest\vspace{-10pt}} & max\_depth & 2 & 5 & 3 & 9 & 4 \\ \cmidrule{2-7}
& max\_features  & 3 & 7 & 5 & 2 & 2 \\ \cmidrule{2-7}
& n\_estimators  & 4 & 2 & 2 & 16 & 2 \\ \midrule
\multirow{2}{*}{Neural Network\vspace{-5pt}} & alpha & 0.01 & 0.001 & 0.0001 & 0.001 & 0.0001 \\ \cmidrule{2-7}
& hidden\_layer\_sizes  & (13, 9) & (4, 9, 4) & (4, 9, 4) & (4, 9, 13, 9, 4) & (9) \\ \midrule
\multirow{2}{*}{SVM\vspace{-5pt}} & degree & 6 & 3 & 7 & 3 & 6 \\ \cmidrule{2-7}
& C  & 20 & 10 & 20 & 100 & 50 \\ \bottomrule
\end{tabular}}
\label{tab:enron_best_hp_2l}
\end{table}
\unskip

\begin{table}[H]
\centering
\caption{The best values of the hyperparameters for the models which uses the full set of features for the Enron dataset with three hierarchy~levels.}
{\begin{tabular}{ccccccc}
\toprule
& & \multicolumn{5}{c}{\textbf{Min. Activity}}\\ \midrule
& & \textbf{1 Month} & \textbf{2 Months} & \textbf{3 Months} & \textbf{4 Months} & \textbf{5 Months} \\ \midrule
\multirow{2}{*}{Decision Tree\vspace{-5pt}} & max\_depth & 5 & 4 & 5 & 8 & 4 \\ \cmidrule{2-7}
& max\_features  & 7 & 8 & 11 & 9 & 13 \\ \midrule
\multirow{3}{*}{Random Forest\vspace{-10pt}} & max\_depth & 2 & 4 & 7 & 2 & 3 \\ \cmidrule{2-7}
& max\_features  & 2 & 12 & 2 & 14 & 14 \\ \cmidrule{2-7}
& n\_estimators  & 2 & 4 & 8 & 2 & 16 \\ \midrule
\multirow{2}{*}{Neural Network\vspace{-5pt}} & alpha & 0.001 & 0.01 & 0.01 & 0.0001 & 0.001 \\ \cmidrule{2-7}
& hidden\_layer\_sizes  & (4, 3) & (13) & (13) & (9) & (13, 9, 4) \\ \midrule
\multirow{2}{*}{SVM\vspace{-5pt}} & degree & 9 & 3 & 3 & 6 & 3 \\ \cmidrule{2-7}
& C  & 5 & 20 & 200 & 150 & 4 \\ \bottomrule
\end{tabular}}
\label{tab:enron_best_hp_3l}
\end{table}
\unskip

\begin{table}[H]
\centering
\caption{Hyperparameter search space for collective~classification.}
{\begin{tabular}{cc}
\toprule
\textbf{Parameter} & \textbf{Values} \\ \midrule
utility score & all features from Table~\ref{tab:features} \\
threshold & 1, 2, 3, 4, 5, 6, 7, 8, 9, 10 \\
Jaccard value & 0.7, 0.8, 0.9, 0.99 \\
\bottomrule
\end{tabular}}
\label{tab:hp_space_cc}
\end{table}
\unskip

\begin{table}[H]
\centering
\caption{The best values of the hyperparameters for the collective {classification} algorithm for the manufacturing company dataset with two hierarchy levels (1 to 3 months of minimum activity).}
{\begin{tabular}{ccccc}
\toprule
& & \multicolumn{3}{c}{\textbf{Hyperparameters}}\\ \midrule
\textbf{Min. Activity} & \textbf{\% of the Known Nodes} & \textbf{Utility Score} & \textbf{Threshold} & \textbf{Jaccard Value}  \\ \midrule
\multirow{17}{*}{1 month\vspace{-40pt}
} & \multirow{3}{*}{90\%} & hubs centrality & 4 & 0.7 \\
& & clustering coefficient & 5 & 0.7 \\
& & overtime & 4 & 0.7 \\ \cmidrule{2-5}
& \multirow{1}{*}{80\%} & hubs centrality & 4 & 0.9 \\ \cmidrule(lr){2-5}
&\multirow{3}{*}{70\%} & indegree centrality & 4 & 0.7 \\
& & the total numbers of cliques & 4 & 0.7 \\
\bottomrule
\end{tabular}}
\label{tab:mc_best_hp_cc_1_2l}
\end{table}

\begin{table}[H]\ContinuedFloat
\centering
\caption{\textit{Cont}.}
{\begin{tabular}{ccccc}
\toprule
& & \multicolumn{3}{c}{\textbf{Hyperparameters}}\\ \midrule
\textbf{Min. Activity} & \textbf{\% of the Known Nodes} & \textbf{Utility Score} & \textbf{Threshold} & \textbf{Jaccard Value}  \\ \midrule
\multirow{16}{*}{3 months\vspace{-30pt}} & \multirow{3}{*}{90\%} & closeness centrality & 4 & 0.7 \\
& & hubs centrality & 4 & 0.7 \\
& & clustering coefficient & 5 & 0.7 \\ \cmidrule{2-5}
& \multirow{2}{*}{80\%} & hubs centrality & 4 & 0.9 \\
& & the biggest clique & 4 & 0.7 \\ \cmidrule{2-5}
&\multirow{3}{*}{70\%} & indegree centrality & 4 & 0.7 \\
& & the total numbers of cliques & 4 & 0.7 \\
& & the biggest clique & 4 & 0.7 \\ \cmidrule{2-5}
&\multirow{3}{*}{60\%} & indegree centrality & 4 & 0.7 \\
& & outdegree centrality & 4 & 0.7 \\
& & the total numbers of cliques & 4 & 0.7 \\ \cmidrule{2-5}
&\multirow{1}{*}{50\%} & received neighborhood variability & 4 & 0.7 \\ \cmidrule(lr){2-5}
& \multirow{1}{*}{40\%} & the total numbers of cliques & 4 & 0.7 \\ \cmidrule(lr){2-5}
& \multirow{1}{*}{30\%} & eigenvector centrality & 4 & 0.7 \\ \cmidrule(lr){2-5}
& \multirow{1}{*}{20\%} & betweenness centrality & 5 & 0.7 \\ \cmidrule(lr){2-5}
& \multirow{1}{*}{10\%} & eigenvector centrality & 4 & 0.7 \\
\bottomrule
\end{tabular}}
\label{tab:mc_best_hp_cc_1_2l}
\end{table}
\unskip

\begin{table}[H]
\centering
\caption{The best values of the hyperparameters for the collective {classification} algorithm for the manufacturing company dataset with two hierarchy levels (4 to 5 months of minimum activity).}
{\begin{tabular}{ccccc}
\toprule
& & \multicolumn{3}{c}{\textbf{Hyperparameters}}\\ \midrule
\textbf{Min. Activity} & \textbf{\% of the Known Nodes} & \textbf{Utility Score} & \textbf{Threshold} & \textbf{Jaccard Value}  \\ \midrule
\multirow{17}{*}{4 month\vspace{-30pt}
} & \multirow{5}{*}{90\%} & outdegree centrality & 3 & 0.7 \\
& & eigenvector centrality & 3 & 0.7 \\
& & hubs centrality & 4 & 0.7 \\
& & clustering coefficient & 5 & 0.7 \\ \cmidrule{2-5}
& \multirow{2}{*}{80\%} & hubs centrality & 4 & 0.9 \\
& & the biggest clique & 4 & 0.7 \\
\bottomrule
\end{tabular}}
\label{tab:mc_best_hp_cc_2_2l}
\end{table}

\begin{table}[H]\ContinuedFloat
\centering
\caption{\textit{Cont}.}
{\begin{tabular}{ccccc}
\toprule
& & \multicolumn{3}{c}{\textbf{Hyperparameters}}\\ \midrule
\textbf{Min. Activity} & \textbf{\% of the Known Nodes} & \textbf{Utility Score} & \textbf{Threshold} & \textbf{Jaccard Value}  \\
\midrule
\multirow{16}{*}{5 months\vspace{-50pt}} & \multirow{5}{*}{90\%} & outdegree centrality & 3 & 0.7 \\
& & betweenness centrality & 3 & 0.7 \\
& & closeness centrality & 4 & 0.7 \\
& & hubs centrality & 4 & 0.7 \\
& & clustering coefficient & 5 & 0.7 \\ \cmidrule{2-5}
& \multirow{2}{*}{80\%} & hubs centrality & 4 & 0.9 \\
& & the biggest clique & 4 & 0.7 \\ \cmidrule{2-5}
&\multirow{3}{*}{70\%} & indegree centrality & 4 & 0.7 \\
& & the total numbers of cliques & 4 & 0.7 \\
& & the biggest clique & 4 & 0.7 \\ \cmidrule{2-5}
&\multirow{1}{*}{60\%} & the total numbers of cliques & 4 & 0.7 \\ \cmidrule(lr){2-5}
&\multirow{2}{*}{50\%} & the biggest clique & 4 & 0.99 \\
& & received neighborhood variability & 4 & 0.7 \\ \cmidrule{2-5}
& \multirow{1}{*}{40\%} & the total numbers of cliques & 4 & 0.7 \\ \cmidrule(lr){2-5}
& \multirow{1}{*}{30\%} & authorities centrality & 4 & 0.7 \\ \cmidrule(lr){2-5}
& \multirow{1}{*}{20\%} & betweenness centrality & 5 & 0.7 \\ \cmidrule(lr){2-5}
& \multirow{1}{*}{10\%} & the number of weekends worked & 4 & 0.7 \\
\bottomrule
\end{tabular}}
\label{tab:mc_best_hp_cc_2_2l}
\end{table}
\unskip

\begin{table}[H]
\centering
\caption{The best values of the hyperparameters for the collective {classification} algorithm for the manufacturing company dataset with three hierarchy~levels.}
{\begin{tabular}{ccccc}
\toprule
& & \multicolumn{3}{c}{\textbf{Hyperparameters}}\\ \midrule
\textbf{Min. Activity} & \textbf{\% of the Known Nodes} & \textbf{Utility Score} & \textbf{Threshold} & \textbf{Jaccard Value}  \\ \midrule
\multirow{10}{*}{1 month \vspace{-40pt}
} & \multirow{1}{*}{90\%} & the number of weekends worked & 7 & 0.7 \\ \cmidrule(lr){2-5}
& \multirow{1}{*}{80\%} & overtime & 4 & 0.7 \\ \cmidrule(lr){2-5}
&\multirow{1}{*}{70\%} & overtime & 5 & 0.7 \\ \cmidrule(lr){2-5}
&\multirow{1}{*}{60\%} & hubs centrality & 5 & 0.7 \\ \cmidrule(lr){2-5}
&\multirow{1}{*}{50\%} & the number of weekends worked & 5 & 0.7 \\ \cmidrule(lr){2-5}
& \multirow{3}{*}{40\%} & closeness centrality & 5 & 0.7 \\
& & clustering coefficient & 2 & 0.7 \\
& & the number of weekends worked & 4 & 0.7 \\ \cmidrule{2-5}
& \multirow{1}{*}{30\%} & closeness centrality & 2 & 0.7 \\ \cmidrule(lr){2-5}
& \multirow{1}{*}{20\%} & received neighborhood variability & 3 & 0.7 \\ \cmidrule(lr){2-5}
& \multirow{1}{*}{10\%} & the biggest clique & 2 & 0.7 \\
\bottomrule
\end{tabular}}
\label{tab:mc_best_hp_cc_3l}
\end{table}

\begin{table}[H]\ContinuedFloat
\centering
\caption{\textit{Cont}.}
{\begin{tabular}{ccccc}
\toprule
& & \multicolumn{3}{c}{\textbf{Hyperparameters}}\\ \midrule
\textbf{Min. Activity} & \textbf{\% of the Known Nodes} & \textbf{Utility Score} & \textbf{Threshold} & \textbf{Jaccard Value}  \\ \midrule
\multirow{8}{*}{2 months\vspace{-40pt}} & \multirow{1}{*}{90\%} & the number of weekends worked & 7 & 0.7 \\ \cmidrule(lr){2-5}
& \multirow{1}{*}{80\%} & overtime & 5 & 0.7 \\ \cmidrule(lr){2-5}
&\multirow{1}{*}{70\%} & betweenness centrality & 5 & 0.7 \\ \cmidrule(lr){2-5}
&\multirow{1}{*}{60\%} & hubs centrality & 5 & 0.7 \\ \cmidrule(lr){2-5}
&\multirow{1}{*}{50\%} & the number of weekends worked & 5 & 0.7 \\ \cmidrule(lr){2-5}
& \multirow{1}{*}{40\%} & general neighborhood variability & 5 & 0.7 \\
\bottomrule
\end{tabular}}
\label{tab:mc_best_hp_cc_3l}
\end{table}
\unskip

\begin{table}[H]
\centering
\caption{The best values of the hyperparameters for the collective {classification} algorithm for the Enron dataset with two hierarchy~levels.}
{\begin{tabular}{ccccc}
\toprule
& & \multicolumn{3}{c}{\textbf{Hyperparameters}}\\ \midrule
\textbf{Min. Activity} & \textbf{\% of the Known Nodes} & \textbf{Utility Score} & \textbf{Threshold} & \textbf{Jaccard Value}  \\ \midrule
\multirow{9}{*}{1 month\vspace{-30pt}
} & \multirow{1}{*}{90\%} & page rank centrality & 1 & 0.7 \\ \cmidrule(lr){2-5}
& \multirow{1}{*}{80\%} & authorities centrality & 1 & 0.7 \\ \cmidrule(lr){2-5}
&\multirow{1}{*}{70\%} & authorities centrality & 1 & 0.7 \\ \cmidrule(lr){2-5}
&\multirow{1}{*}{60\%} & indegree centrality & 1 & 0.8 \\ \cmidrule(lr){2-5}
&\multirow{1}{*}{50\%} & the biggest clique & 1 & 0.7 \\ \cmidrule(lr){2-5}
& \multirow{1}{*}{40\%} & indegree centrality & 1 & 0.7 \\
\bottomrule
\end{tabular}}
\label{tab:enron_best_hp_cc_2l}
\end{table}

\begin{table}[H]\ContinuedFloat
\centering
\caption{\textit{Cont}.}
{\begin{tabular}{ccccc}
\toprule
& & \multicolumn{3}{c}{\textbf{Hyperparameters}}\\ \midrule
\textbf{Min. Activity} & \textbf{\% of the Known Nodes} & \textbf{Utility Score} & \textbf{Threshold} & \textbf{Jaccard Value}  \\ \midrule
\multirow{10}{*}{5 months\vspace{-30pt}} & \multirow{2}{*}{90\%} & authorities centrality & 1 & 0.7 \\
& & hubs centrality & 1 & 0.7 \\ \cmidrule{2-5}
& \multirow{1}{*}{80\%} & indegree centrality & 1 & 0.7 \\ \cmidrule(lr){2-5}
&\multirow{1}{*}{70\%} & outdegree centrality & 1 & 0.7 \\ \cmidrule(lr){2-5}
&\multirow{1}{*}{60\%} & outdegree centrality & 1 & 0.7 \\ \cmidrule(lr){2-5}
&\multirow{1}{*}{50\%} & betweenness centrality & 1 & 0.7 \\ \cmidrule(lr){2-5}
& \multirow{1}{*}{40\%} & betweenness centrality & 1 & 0.7 \\ \cmidrule(lr){2-5}
& \multirow{1}{*}{30\%} & betweenness centrality & 1 & 0.7 \\ \cmidrule(lr){2-5}
& \multirow{1}{*}{20\%} & betweenness centrality & 1 & 0.7 \\ \cmidrule(lr){2-5}
& \multirow{1}{*}{10\%} & the biggest clique & 1 & 0.7 \\
\bottomrule
\end{tabular}}
\label{tab:enron_best_hp_cc_2l}
\end{table}
\unskip

\begin{table}[H]
\centering
\caption{The best values of the hyperparameters for the collective {classification} algorithm for the Enron dataset with three hierarchy~levels.}
{\begin{tabular}{ccccc}
\toprule
& & \multicolumn{3}{c}{\textbf{Hyperparameters}}\\ \midrule
\textbf{Min. Activity} & \textbf{\% of the Known Nodes} & \textbf{Utility Score} & \textbf{Threshold} & \textbf{Jaccard Value}  \\ \midrule
\multirow{9}{*}{1 month\vspace{-30pt}} & \multirow{1}{*}{90\%} & the biggest clique & 1 & 0.7 \\ \cmidrule(lr){2-5}
& \multirow{1}{*}{80\%} & hubs centrality & 1 & 0.7 \\ \cmidrule(lr){2-5}
&\multirow{1}{*}{70\%} & page rank centrality & 1 & 0.7 \\ \cmidrule(lr){2-5}
&\multirow{1}{*}{60\%} & outdegree centrality & 1 & 0.7 \\ \cmidrule(lr){2-5}
&\multirow{1}{*}{50\%} & eigenvector centrality & 1 & 0.7 \\ \cmidrule(lr){2-5}
& \multirow{1}{*}{40\%} & the total numbers of cliques & 1 & 0.7 \\ \cmidrule(lr){2-5}
& \multirow{1}{*}{30\%} & indegree centrality & 1 & 0.7 \\ \cmidrule(lr){2-5}
& \multirow{1}{*}{20\%} & eigenvector centrality & 1 & 0.7 \\ \cmidrule(lr){2-5}
& \multirow{1}{*}{10\%} & general neighborhood variability & 1 & 0.7 \\ \midrule
\multirow{9}{*}{2 months\vspace{-30pt}} & \multirow{1}{*}{90\%} & received neighborhood variability & 1 & 0.7 \\ \cmidrule(lr){2-5}
& \multirow{1}{*}{80\%} & hubs centrality & 1 & 0.7 \\ \cmidrule(lr){2-5}
&\multirow{1}{*}{70\%} & eigenvector centrality & 1 & 0.7 \\ \cmidrule(lr){2-5}
&\multirow{1}{*}{60\%} & outdegree centrality & 1 & 0.7 \\ \cmidrule(lr){2-5}
&\multirow{1}{*}{50\%} & outdegree centrality & 1 & 0.7 \\ \cmidrule(lr){2-5}
& \multirow{1}{*}{40\%} & hubs centrality & 1 & 0.7 \\ \cmidrule(lr){2-5}
& \multirow{1}{*}{30\%} & indegree centrality & 1 & 0.7 \\ \cmidrule(lr){2-5}
& \multirow{1}{*}{20\%} & authorities centrality & 1 & 0.7 \\ \cmidrule(lr){2-5}
& \multirow{1}{*}{10\%} & the total numbers of cliques & 1 & 0.99 \\
\bottomrule
\end{tabular}}
\label{tab:enron_best_hp_cc_3l}
\end{table}

\begin{table}[H]\ContinuedFloat
\centering
\caption{\textit{Cont}.}
{\begin{tabular}{ccccc}
\toprule
& & \multicolumn{3}{c}{\textbf{Hyperparameters}}\\ \midrule
\textbf{Min. Activity} & \textbf{\% of the Known Nodes} & \textbf{Utility Score} & \textbf{Threshold} & \textbf{Jaccard Value}  \\ \midrule
\multirow{9}{*}{3 months\vspace{-30pt}} & \multirow{1}{*}{90\%} & the biggest clique & 1 & 0.7 \\ \cmidrule(lr){2-5}
& \multirow{1}{*}{80\%} & hubs centrality & 1 & 0.7 \\ \cmidrule(lr){2-5}
&\multirow{1}{*}{70\%} & closeness centrality & 1 & 0.7 \\ \cmidrule(lr){2-5}
&\multirow{1}{*}{60\%} & outdegree centrality & 1 & 0.7 \\ \cmidrule(lr){2-5}
&\multirow{1}{*}{50\%} & the biggest clique & 1 & 0.7 \\ \cmidrule(lr){2-5}
& \multirow{1}{*}{40\%} & authorities centrality & 1 & 0.7 \\
\bottomrule
\end{tabular}}
\label{tab:enron_best_hp_cc_3l}
\end{table}
\unskip

\begin{figure}[H]
\centering
\includegraphics[scale=0.35]{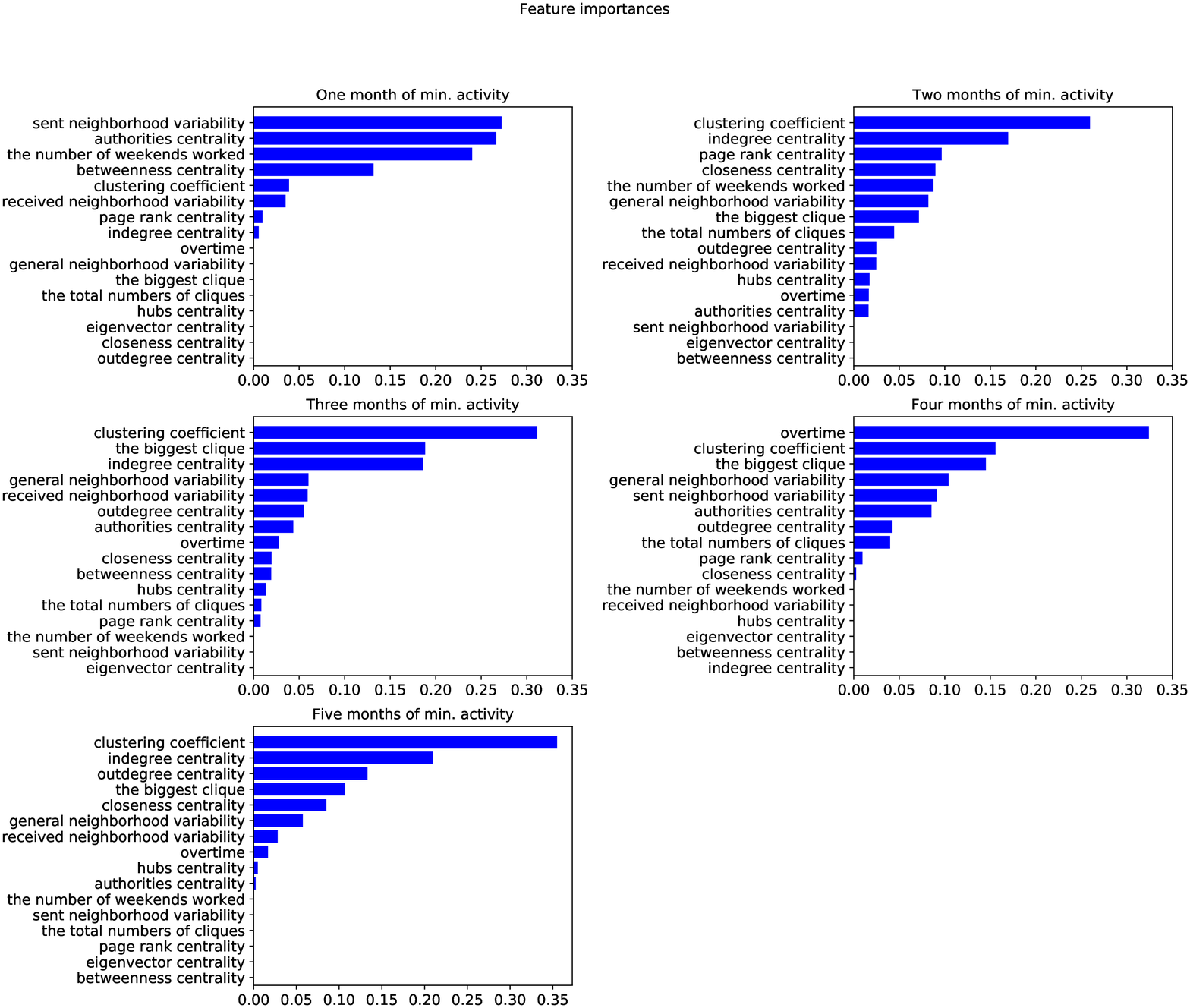}
\caption{{Features importance (Gini importance) for the Decision Tree which uses the full set of features for the manufacturing company dataset with two levels of the hierarchy.}}
\label{fig:mc_fi_dt_2l}
\end{figure}
\unskip

\begin{figure}[H]
\centering
\includegraphics[scale=0.35]{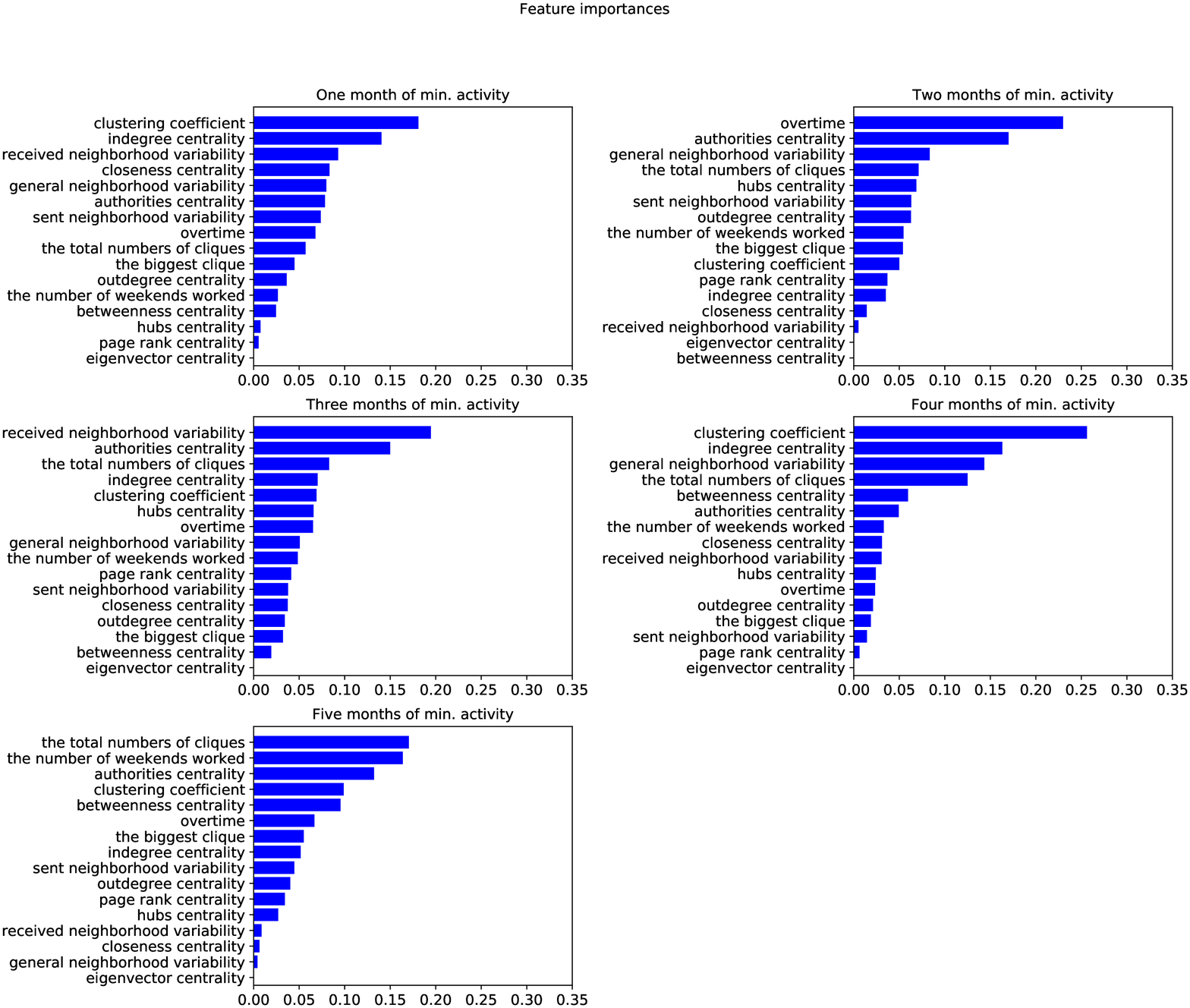}
\caption{{Features importance (Gini importance) for the Decision Tree which uses the full set of features for the manufacturing company dataset with three levels of the hierarchy.}}
\label{fig:mc_fi_dt_3l}
\end{figure}
\unskip

\begin{figure}[H]
\centering
\includegraphics[scale=0.35]{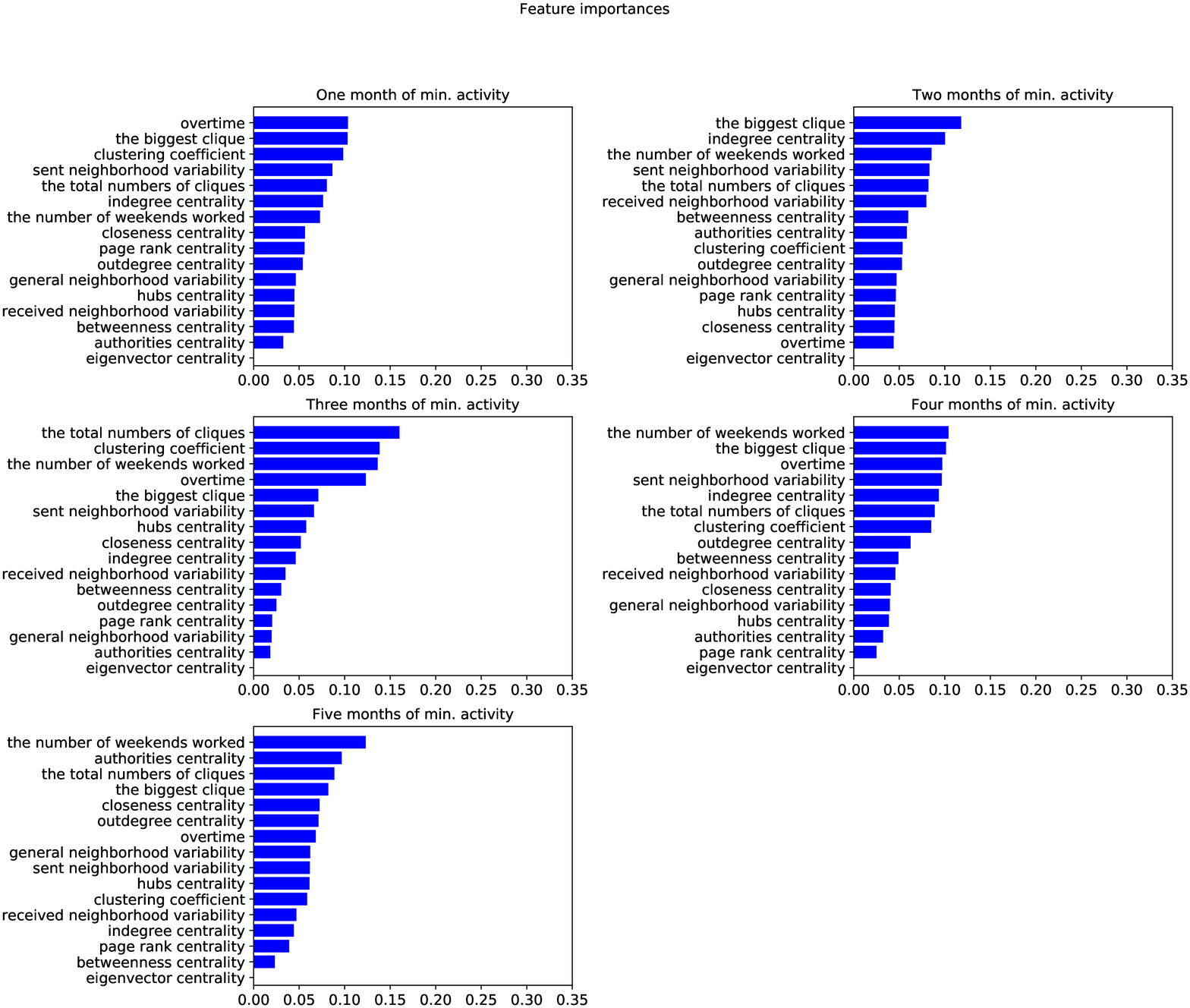}
\caption{{Features importance (Gini importance) for the Random Forest which uses the full set of features for the manufacturing company dataset with two levels of the hierarchy.}}
\label{fig:mc_fi_rf_2l}
\end{figure}
\unskip

\begin{figure}[H]
\centering
\includegraphics[scale=0.35]{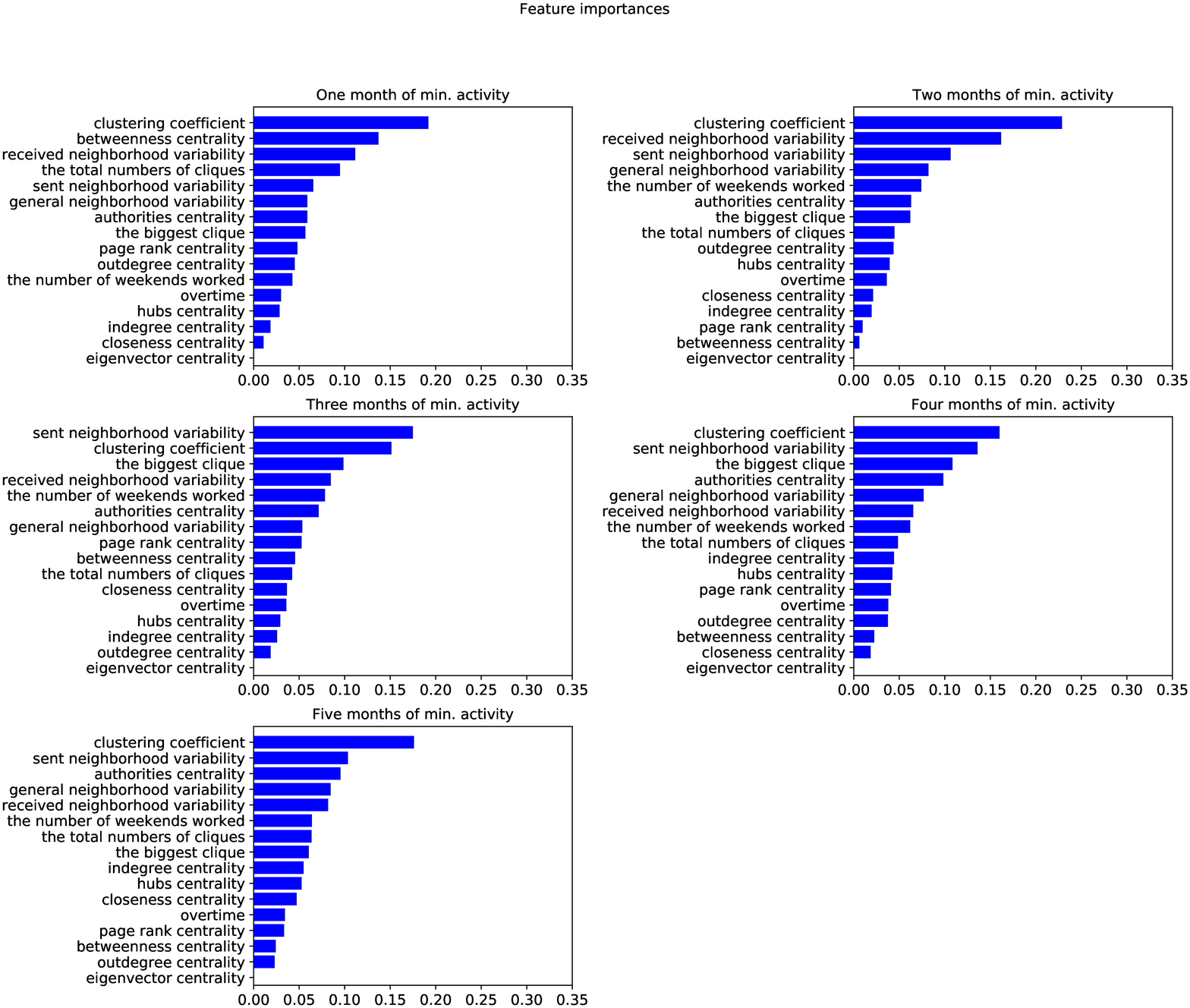}
\caption{{Features importance (Gini importance) for the Random Forest which uses the full set of features for the manufacturing company dataset with three levels of the hierarchy.}}
\label{fig:mc_fi_rf_3l}
\end{figure}
\unskip

\begin{figure}[H]
\centering
\includegraphics[scale=0.33]{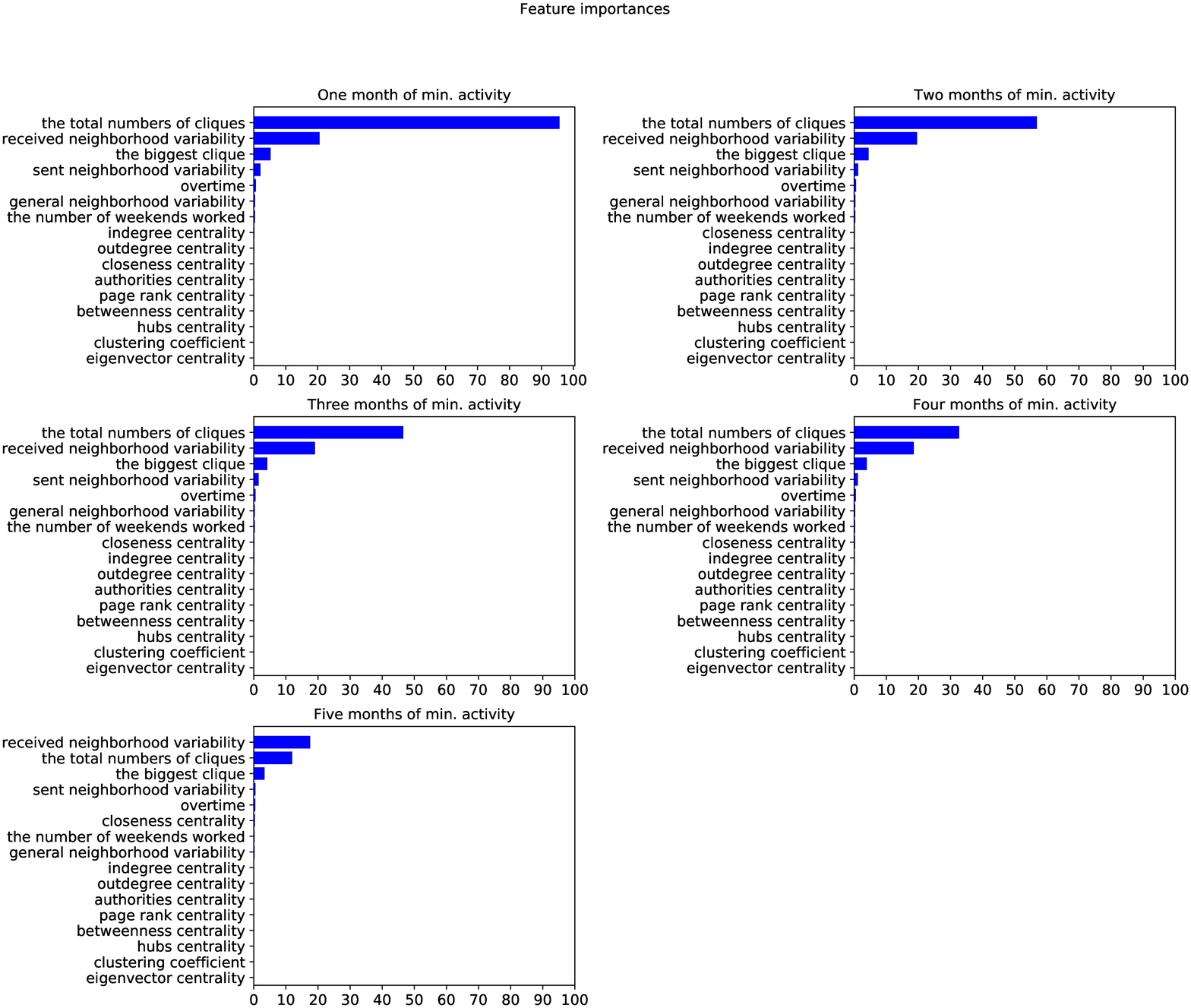}
\caption{{Features importance (based on univariate feature selection method which uses chi-squared test) for the Neural Network and SVM which use the full set of features for the manufacturing company dataset with two levels of the hierarchy.}}
\label{fig:mc_fi_chi_2l}
\end{figure}
\unskip

\begin{figure}[H]
\centering
\includegraphics[scale=0.33]{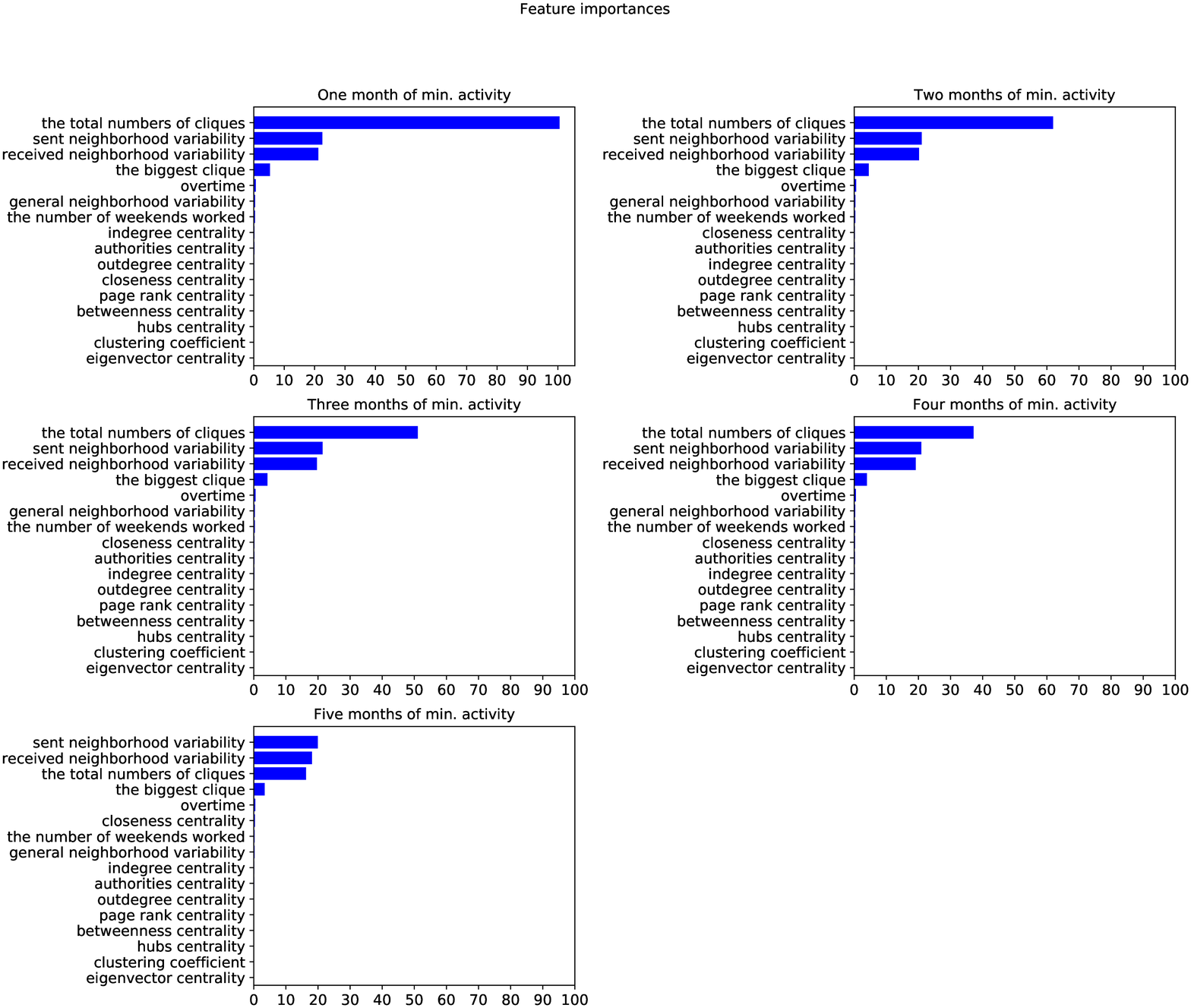}
\caption{{Features importance (based on univariate feature selection method which uses chi-squared test) for the Neural Network and SVM which use the full set of features for the manufacturing company dataset with three levels of the hierarchy.}}
\label{fig:mc_fi_chi_3l}
\end{figure}
\unskip

\begin{figure}[H]
\centering
\includegraphics[scale=0.35]{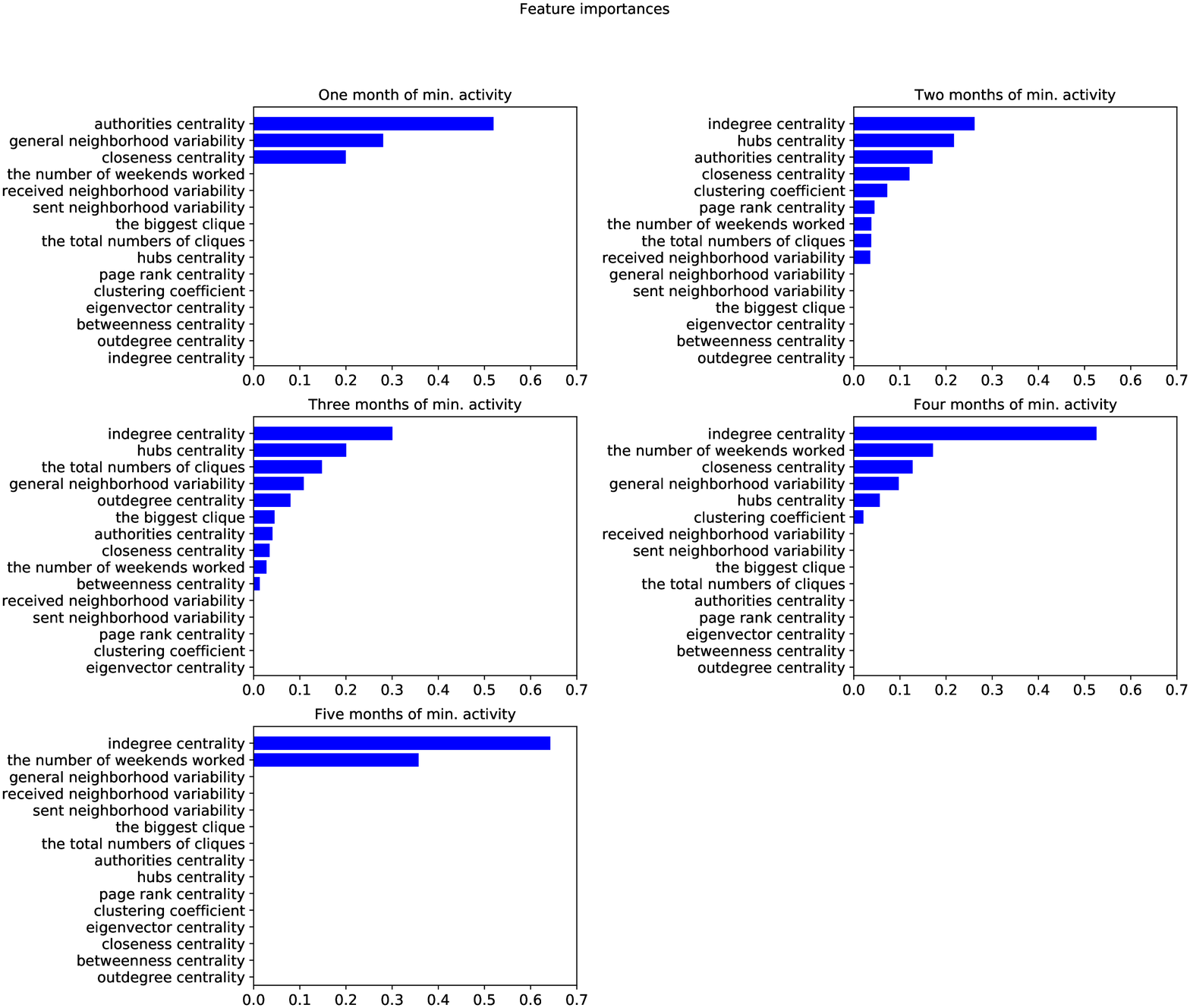}
\caption{{Features importance (Gini importance) for the Decision Tree which uses the full set of features for the Enron dataset with two levels of the hierarchy.}}
\label{fig:enron_fi_dt_2l}
\end{figure}
\unskip

\begin{figure}[H]
\centering
\includegraphics[scale=0.35]{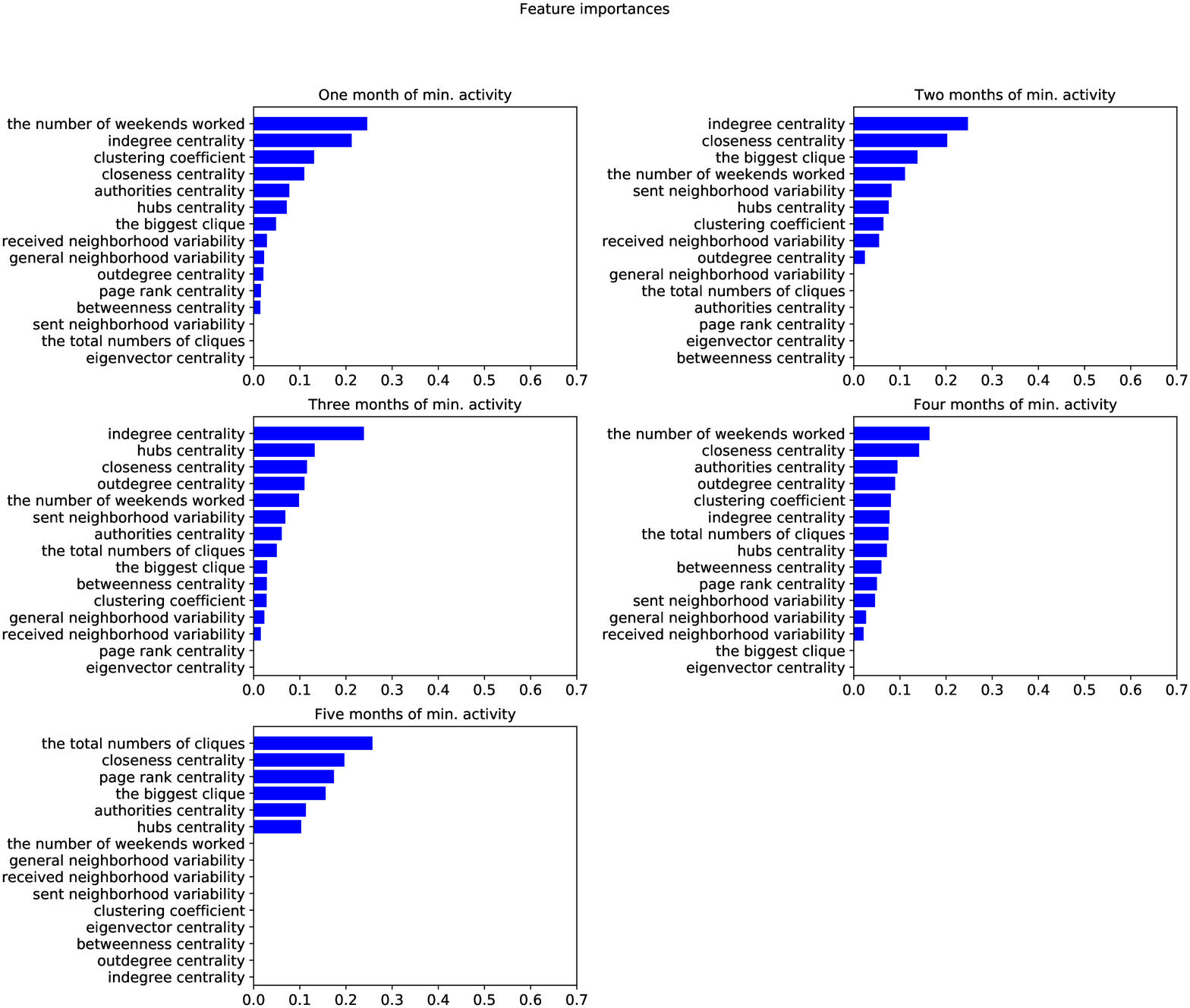}
\caption{{Features importance (Gini importance) for the Decision Tree which uses the full set of features for the Enron dataset with three levels of the hierarchy.}}
\label{fig:enron_fi_dt_3l}
\end{figure}
\unskip

\begin{figure}[H]
\centering
\includegraphics[scale=0.35]{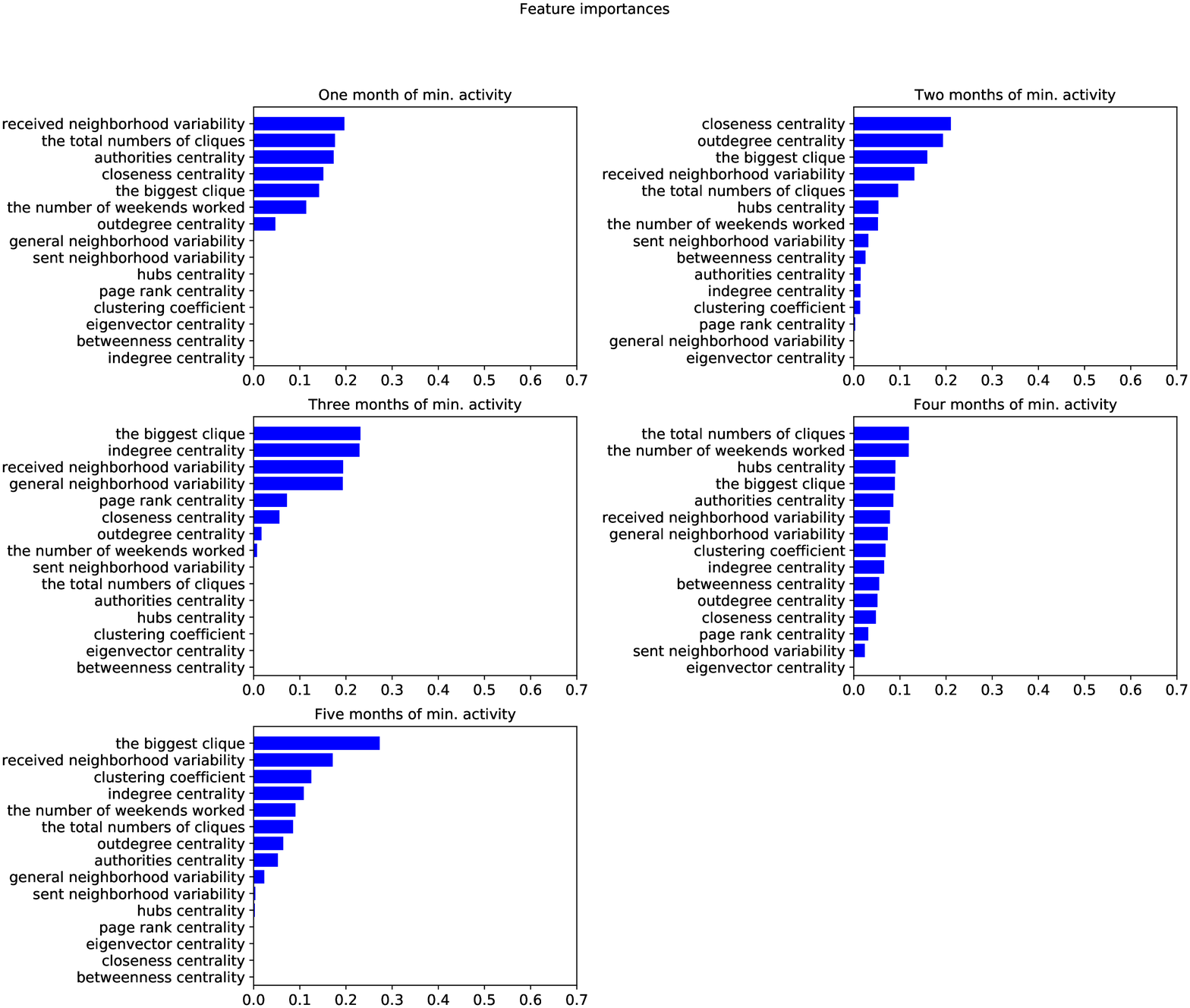}
\caption{{Features importance (Gini importance) for the Random Forest which uses the full set of features for the Enron dataset with two levels of the hierarchy.}}
\label{fig:enron_fi_rf_2l}
\end{figure}
\unskip

\begin{figure}[H]
\centering
\includegraphics[scale=0.35]{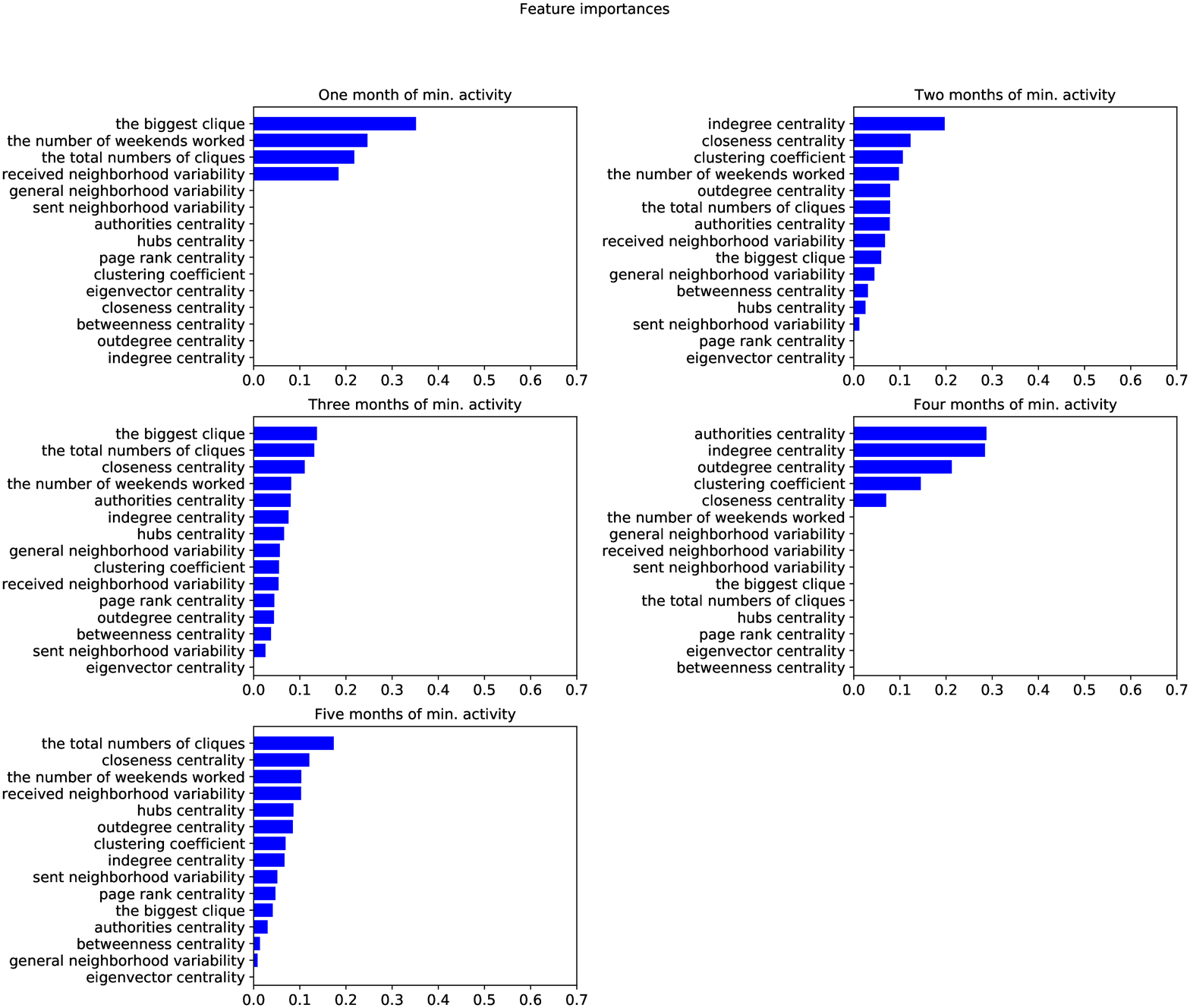}
\caption{{Features importance (Gini importance) for the Random Forest which uses the full set of features for the Enron dataset with three levels of the hierarchy.}}
\label{fig:enron_fi_rf_3l}
\end{figure}
\unskip

\begin{figure}[H]
\centering
\includegraphics[scale=0.33]{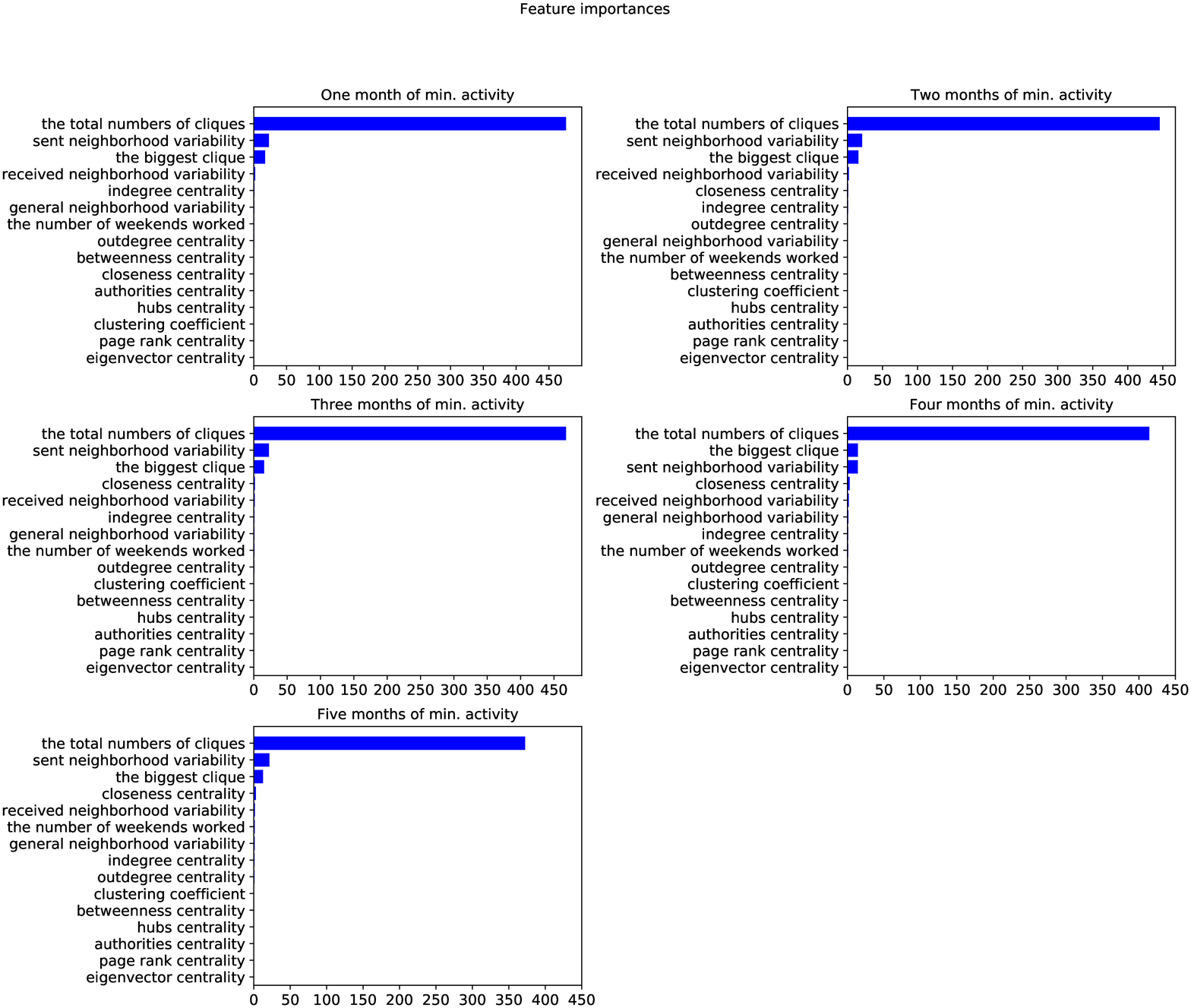}
\caption{{Features importance (based on univariate feature selection method which uses chi-squared test) for the Neural Network and SVM which use the full set of features for the Enron dataset with two levels of the hierarchy.}}
\label{fig:enron_fi_chi_2l}
\end{figure}
\unskip

\begin{figure}[H]
\centering
\includegraphics[scale=0.33]{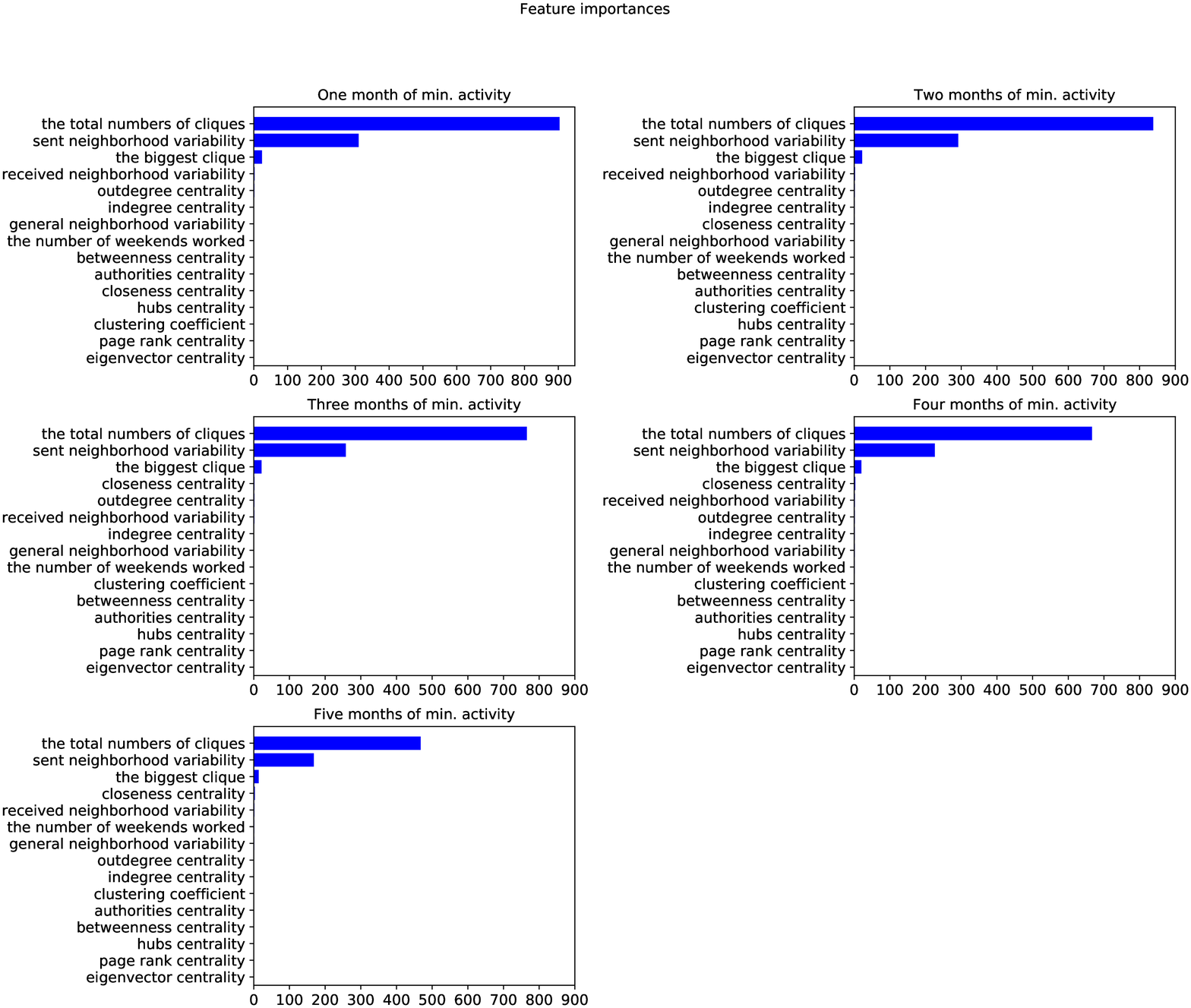}
\caption{{Features importance (based on univariate feature selection method which uses chi-squared test) for the Neural Network and SVM which use the full set of features for the Enron dataset with three levels of the hierarchy.}}
\label{fig:enron_fi_chi_3l}
\end{figure}





\reftitle{References}





\end{document}